\documentclass[12pt]{article}
\usepackage{scicite}
\usepackage{times}
\usepackage{bm}
\usepackage{graphicx}
\usepackage{booktabs}
\usepackage{longtable} 

\def\farcm{\hbox{.\kern -0.7ex\raisebox{.9ex}{\scriptsize$\prime$}}}
\def\farcs{\hbox{\kern 0.13ex.\kern -0.95ex%
\raisebox{.9ex}{\scriptsize$\prime\prime$}\kern -0.1ex}}

\def\micron{\hbox{$\mu$m}}
\def\kms{km s${}^{-1}$}

\def\lesssim{\mathrel{\hbox{\rlap{\hbox{\lower4pt\hbox{$\sim$}}}\hbox{$<$}}}}
\def\gtrsim{\mathrel{\hbox{\rlap{\hbox{\lower4pt\hbox{$\sim$}}}\hbox{$>$}}}}

\def\farcm{\hbox{.\kern -0.7ex\raisebox{.9ex}{\scriptsize$\prime$}}}
\def\farcs{\hbox{\kern 0.13ex.\kern -0.95ex%
\raisebox{.9ex}{\scriptsize$\prime\prime$}\kern -0.1ex}}

\def\kms{km s${}^{-1}$}

\def\deg{\hbox{$^\circ$}}
\def\la{\mathrel{\hbox{\rlap{\hbox{\lower4pt\hbox{$\sim$}}}\hbox{$<$}}}}
\def\ga{\mathrel{\hbox{\rlap{\hbox{\lower4pt\hbox{$\sim$}}}\hbox{$>$}}}}

\def\farcm{\hbox{$.\mkern-4mu^\prime$}}
\def\farcs{\hbox{$.\!\!^{\prime\prime}$}}

\def\fnum@figure{{\rmfamily Fig.\space\thefigure.---}}%
\def\fnum@table{{\rmfamily Table \thetable:}}%
\def\fnum@plate{{\bfseries Plate \theplate.}}%
\def\fps@figure{bp}%
\def\fps@table{bp}%
\def\fps@plate{bp}%

\usepackage{hyperref}
\newenvironment{sciabstract}{%
\begin{quote} \bf}
{\end{quote}}

\title{Relativistic redshift of the star S0-2 orbiting the Galactic center supermassive black hole} 

\author
{Tuan Do,$^{1\ast}$ Aurelien Hees,$^{2,1}$ Andrea Ghez,$^{1}$ Gregory D. Martinez,$^{1}$\\
Devin S. Chu,$^{1}$ Siyao Jia,$^{3}$  Shoko Sakai,$^{1}$ Jessica R. Lu,$^{3}$ \\
Abhimat K. Gautam,$^{1}$ Kelly Kosmo O'Neil,$^{1}$ \\
Eric E. Becklin,$^{1,4}$Mark R. Morris,$^{1}$ Keith Matthews,$^{5}$ Shogo Nishiyama,$^{6}$\\
Randy Campbell,$^{7}$ Samantha Chappell,$^{1}$ Zhuo Chen,$^{1}$ Anna Ciurlo,$^{1}$ \\
Arezu Dehghanfar,$^{1,8}$ Eulalia Gallego-Cano,$^{9}$  Wolfgang E. Kerzendorf,$^{10,11,12,13}$ \\
James E. Lyke,$^{7}$ Smadar Naoz,$^{1,14}$ Hiromi Saida,$^{15}$ Rainer Sch\"odel,$^{9}$\\
Masaaki Takahashi,$^{16}$, Yohsuke Takamori,$^{17}$ Gunther Witzel,$^{1,18}$ Peter Wizinowich,$^{7}$\\
\footnotesize{$^{1}$Department of Physics and Astronomy, University of California, Los Angeles, California 90095, USA}\\
\footnotesize{$^{2}$Syt\`emes de R\'ef\'erence Temps Espace, Observatoire de Paris, Universit\'e Paris-Sciences-et-Lettres,}\\
\footnotesize{Centre national de la Recherche Scientifique, Sorbonne Universit\'e,} \\
\footnotesize{Laboratoire National de m\'etrologie et d'Essais, 61 avenue de l'Observatoire, 75014 Paris, France}\\
\footnotesize{$^{3}$Department of Astronomy, University of California, Berkeley, CA 94720-3411, USA}\\
\footnotesize{$^{4}$Universities Space Research Association/Stratospheric Observatory for Infrared Astronomy,}\\
\footnotesize{NASA Ames Research Center, Mail Stop N232-12, Moffet Field, CA 94035}\\
\footnotesize{$^{5}$Division of Physics, Mathematics, and Astronomy, California Institute of Technology,}\\
\footnotesize{MC 301-17, Pasadena, California 91125, USA}\\
\footnotesize{$^{6}$ Miyagi University of Education, 149 Aramaki-aza-aoba, Aoba-ku, Sendai, Miyagi 980-0845, Japan}\\
\footnotesize{$^{7}$ W. M. Keck Observatory, 65-1120 Mamalahoa Highway, Kamuela, HI 96743, USA}\\
\footnotesize{$^{8}$ Institut de Plan\'etologie et d'Astrophysique de Grenoble,} \\
\footnotesize{414 Rue de la Piscine, 38400 Saint-Martin-d'H\'eres, France.}\\
\footnotesize{$^{9}$Instituto de Astrof\'isica de Andaluc\'ia, Consejo Superior de Investigaciones Cient\'ificas,}\\
\footnotesize{Glorieta de la Astronom\'ia S/N, 18008 Granada, Spain}\\
\footnotesize{$^{10}$European Southern Observatory, Karl-Schwarzschild-Stra\ss e 2,85748 Garching bei M\"unchen, Germany}\\
\footnotesize{$^{11}$Center for Cosmology and Particle Physics, New York University,}\\
\footnotesize{726 Broadway, New York, NY 10003, USA}\\
\footnotesize{$^{12}$Department of Physics and Astronomy, Michigan State University, East Lansing, MI 48824, USA}\\
\footnotesize{$^{13}$Department of Computational Mathematics, Science, and Engineering, Michigan State University,}\\
\footnotesize{East Lansing, MI 48824, USA}\\
\footnotesize{$^{14}$Mani L. Bhaumik Institute for Theoretical Physics, Department of Physics and Astronomy,}\\ 
\footnotesize{University of California, Los Angeles, CA 90095, USA}\\
\footnotesize{$^{15}$Daido University,  10-3 Takiharu-cho, Minami-ku, Nagoya, Aichi 457-8530, Japan}\\
\footnotesize{$^{16}$Aichi University of Education, 1 Hirosawa, Igaya-cho, Kariya, Aichi 448-8542, Japan}\\
\footnotesize{$^{17}$National Institute of Technology, Wakayama College,}\\
\footnotesize{77 Noshima, Nada-cho, Gobo, Wakayama 644-0023, Japan}\\
\footnotesize{$^{18}$Max Planck Institute for Radio Astronomy, Auf dem H\"ugel 69, D-53121 Bonn (Endenich), Germany}\\
\footnotesize{$^\ast$Corresponding author e-mail: tdo@astro.ucla.edu}
}

\date{}

\begin{document} 

\baselineskip24pt

\maketitle 

\begin{sciabstract}

General Relativity predicts that a star passing close to a supermassive black hole should exhibit a relativistic redshift. We test this using observations of the Galactic center star S0-2. We combine existing spectroscopic and astrometric measurements from 1995-2017, which cover S0-2's 16-year orbit, with measurements in 2018 March to September which cover three events during its closest approach to the black hole.  We detect the combination of special relativistic- and gravitational-redshift, quantified using a redshift parameter, $\Upsilon$.  Our result, $\Upsilon=0.88 \pm 0.17$, is consistent with General Relativity ($\Upsilon=1$) and excludes a Newtonian model ($\Upsilon=0$ ) with a statistical significance of 5 $\sigma$.
\end{sciabstract}

\begin{figure}
\begin{center}
\includegraphics[scale=0.35]{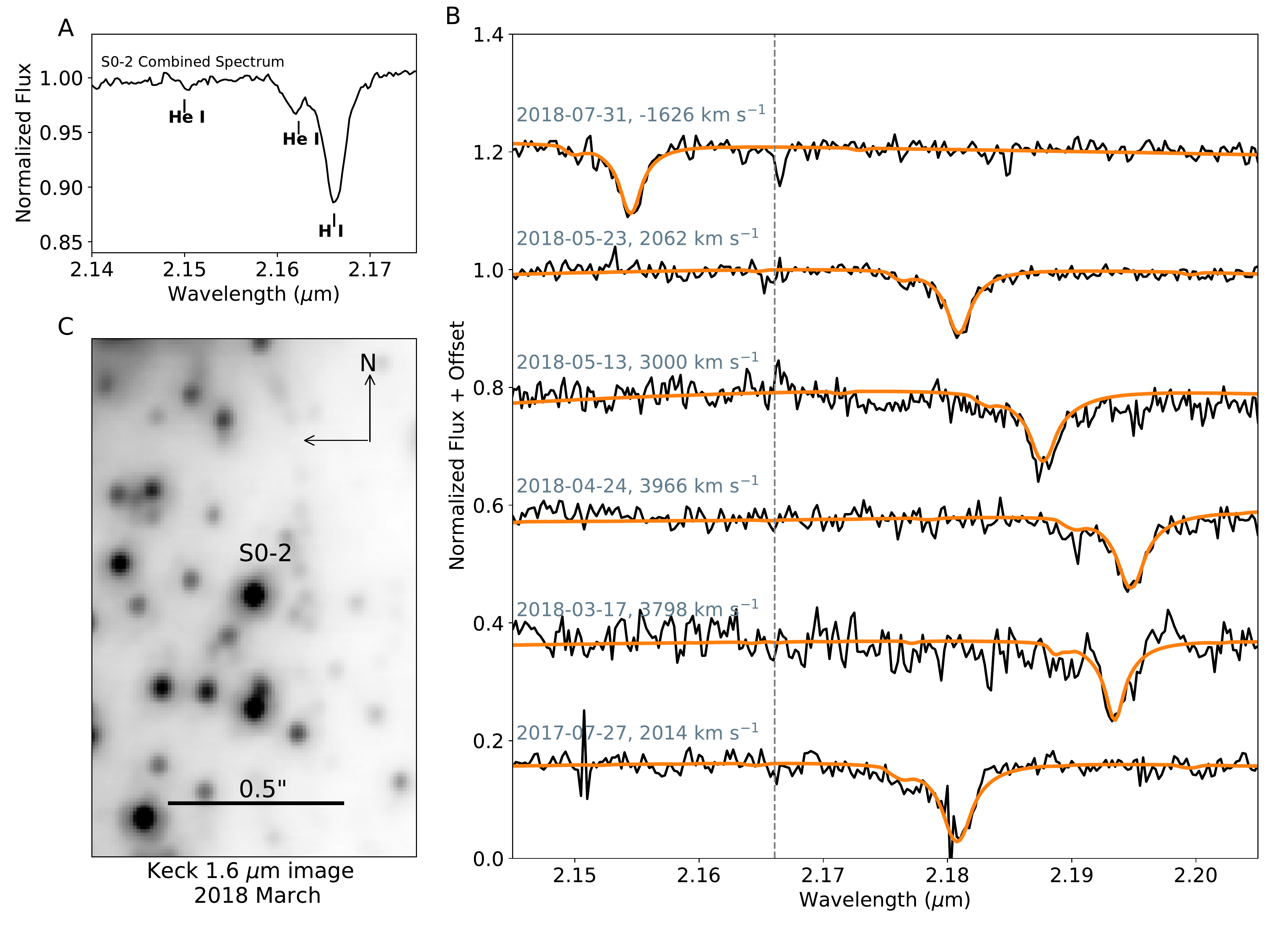}
\end{center}
\caption{\textbf{Spectroscopy and imaging of the star S0-2.} \textbf{A:} A weighted-average spectrum of S0-2 from data obtained from 2006-2018 using Keck data. The strongest feature, which provides most of the RV constraint, is from the H~\textsc{i} line at 2.1661 $\mu$m. \textbf{B:} A sequence of S0-2 spectra observed in 2017 and 2018 (black lines). The RV of the star changes by over 6000 km s$^{-1}$ throughout 2018. The dashed line shows the rest wavelength of the H~\textsc{i} line. We fit a model to the spectrum that simultaneously constrains the star's physical properties such as effective temperature and rotation along with the RV of the star (orange). This model accounts for the asymmetries in the H~\textsc{i} feature. \textbf{C:} An inverted Keck AO image of S0-2 (center of image) from March 2018 with the H-band filter (1.6 $\micron$). 
\label{fig:data}
}
\end{figure}

\begin{figure}
\begin{center}
\includegraphics[scale=0.6]{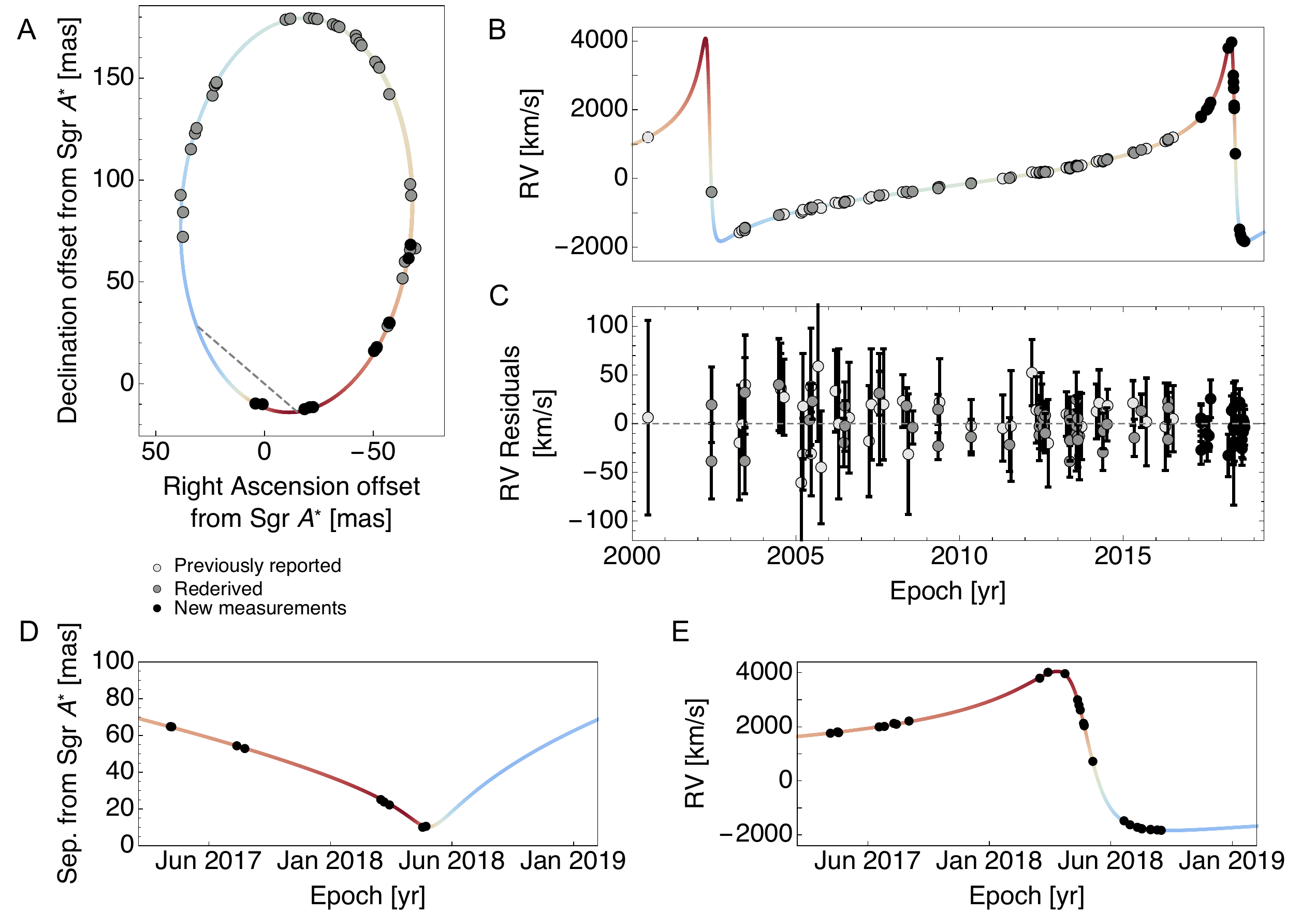}
\end{center}
\caption{\textbf{GR orbit modeling of S0-2.} \textbf{A:} Astrometric measurements of the short-period star S0-2 in orbit around the SMBH (Sgr A*) overlain with our best-fitting projected orbit in the plane of the sky. The origin of the coordinate system coincides with the fitted SMBH center of mass \cite{supplements}. The x and y axes corresponds to offsets in right ascension and declination from the SMBH. 
45 astrometric measurements from 1995-2018 of which 11 are new observations (black) and 34 rederived measurements (grey). The best-fitting SMBH linear drift has been removed from the measurements. The line of nodes (dashed line) shows the intersection of the orbital plane with the plane of the sky (this line also passes through the position of the black hole). S0-2 moves clockwise in this projection; the star is behind the black hole below the line of nodes and in front of the black hole above the line of nodes. The color and intensity used in the best-fitting orbital plot represent the direction and magnitude of the line-of-sight velocity with colors corresponding to panel B.
\textbf{B:} RV measurements and the best-fitting RV model (colored line) using 115 RV measurements from 2000-2018. 42 measurements were previously reported (empty circles), 45 were rederived for this work with improved methodology (grey dots), and 28 are new observations (black dots).  The color of the best-fitting orbit represents the value and sign of the line-of-sight velocity.
\textbf{C:}  residuals from the best-fitting RV model. \textbf{D \& E}: Observations around the three turning points, 1 at the closest approach to Sgr A* in the plane of the sky (\textbf{D}) and 2 RV turning points  (maximum and minimum RV, \textbf{E}) provide the greatest sensitivity to the relativistic redshift. 
\label{fig:res}
}
\end{figure}

General Relativity (GR) has been thoroughly tested in weak gravitational fields in the Solar System \cite{Will:2014}, with binary pulsars \cite{Kramer:2016} and with measurements of gravitational waves from  stellar-mass black-hole binaries \cite{Virgo:2016,Virgo:2017}. Observations of short-period stars in our Galactic center (GC) \cite{Genzel:1996,Ghez:1998,Boehle:2016,Gillessen:2017} allow GR to be tested in a different regime \cite{Gravity:2018}: the strong field near a supermassive black hole (SMBH)  
\cite{Psaltis:2004,Baker:2015}. 
The star S0-2 (also known as S2) has a 16 year orbit around Sagittarius A* (Sgr A*), the SMBH at the center of the Milky Way. In 2018 May, it reached its point of closest approach, at a distance of 120 astronomical units (au) with a velocity reaching 2.7\% of the speed of light. 
Within a 6 months interval of that date, the star also passed through its maximum (March) and minimum velocity (September) along the line-of-sight,  spanning a range of 6000 \kms~ in radial velocity (RV - Fig. \ref{fig:data}).  We present observations of all three events and combine them with data from 1995-2017 (Fig. \ref{fig:res}).

\begin{figure}[h!]
    \centering
    
    \includegraphics[width=6.0in]{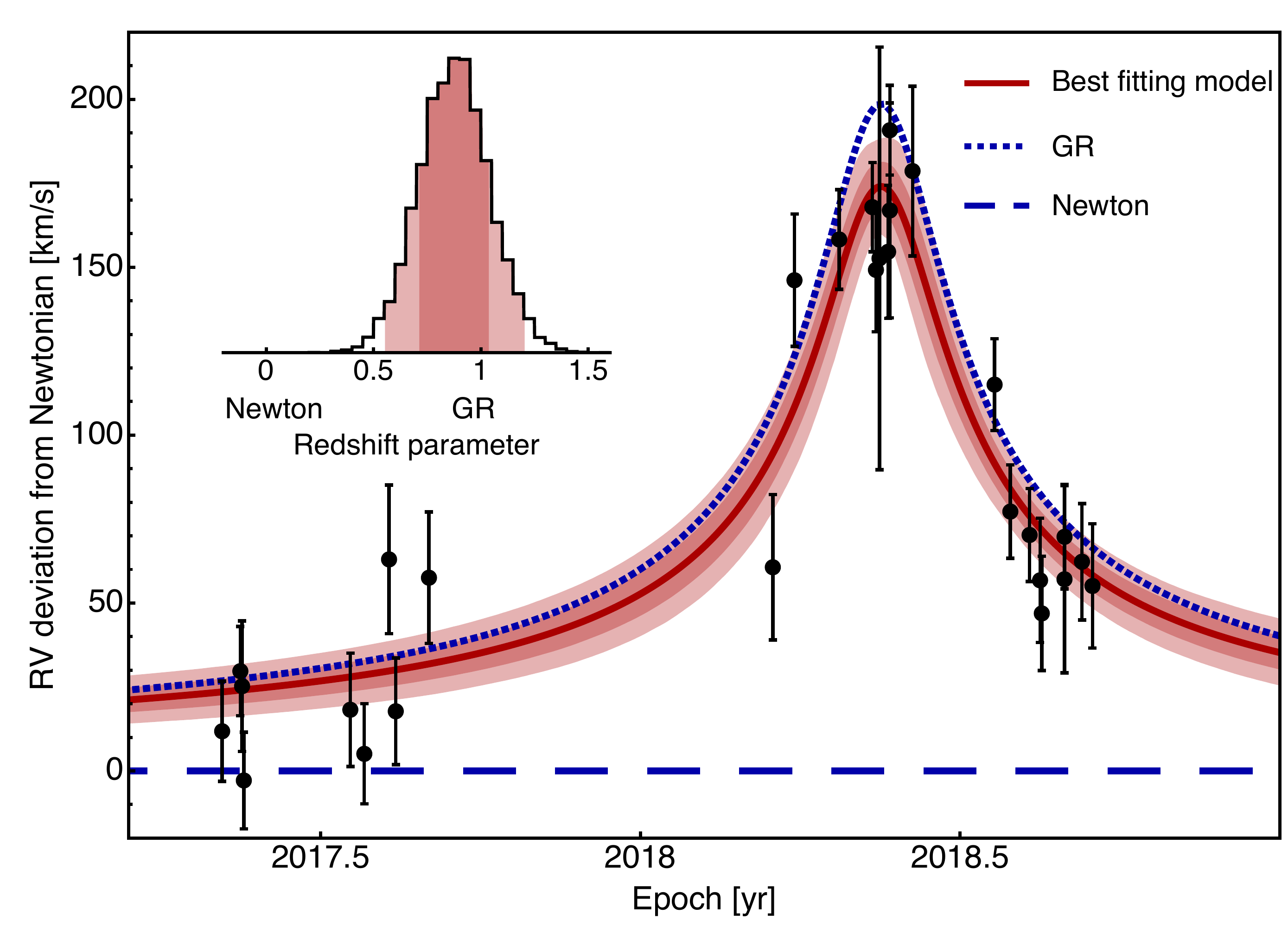}
    \caption{\textbf{Measured deviation from Newtonian predictions.} 
The fitted deviation from  Newtonian prediction, overlaid with the best-fitting orbit model (red line) corresponding to $\Upsilon=0.88$. The inset shows the posterior probability distribution for $\Upsilon$; 0.88 is the median value. The  red shaded areas show the model 68\% and 95\% confidence intervals. The observed RVs are shown as black points after removing the Newtonian part of the model. For comparison, we show the RV deviation expected for a purely relativistic signal ($\Upsilon= 1$, dotted blue line) and for a purely Newtonian model ($\Upsilon=0$, dashed blue line) for an orbit with the same orbital parameters. Our measurement is consistent with the GR model at the 1$\sigma$ confidence level while the Newtonian model is excluded at $>5\sigma$ confidence.}
\label{fig:signal}
\end{figure}

During 2018, the close proximity of S0-2 to the SMBH causes the relativistic redshift, which is the combination of the transverse Doppler shift from special relativity and the gravitational redshift from GR. This deviation from a Keplerian orbit was predicted to reach 200 \kms~ (Fig.~\ref{fig:signal}) and is detectable with current telescopes. The GRAVITY collaboration \cite{Gravity:2018} previously reported a similar measurement. Our measurements are complementary:
i) we present a complete set of independent measurements with 3 additional months of data, doubling the time baseline for the year of closest approach, and including the third turning point (RV minimum) in September 2018, ii) we use three different spectroscopic instruments in 2018, which allows us to probe the presence of instrumental biases, iii) we perform an analysis of the systematic errors that may arise from an experiment spanning over 20 years to test for bias in the result, and iv) we publicly release the stellar measurements and the posterior probability distributions.

We use a total of 45 astrometric positional measurements (spanning 24 years) and 115 RVs (18 years) to fit the orbit of S0-2. Of these, 11 are new astrometric measurements of S0-2 from 2016 to 2018 and 28 are new RV measurements from 2017 and 2018 (Fig \ref{fig:data}). Astrometric measurements were obtained at the W. M. Keck Observatory using speckle imaging (a technique to overcome blurring from the atmosphere by taking very short exposures and combining the images with software) from 1995-2005 and adaptive optics (AO) imaging \cite{Wizinowich:2006} from 2005-2018. RV measurements were obtained from the W. M. Keck Observatory, Gemini North Telescope, and Subaru Telescope. All our RV observations were taken using AO. We supplement our observations with previously reported RVs from Keck from 2000 \cite{Boehle:2016} and the Very Large Telescope (VLT) from 2003-2016 \cite{Gillessen:2017}. This work includes data from a total of 2 imaging instruments and 6 spectroscopic instruments \cite{supplements}. 

We scheduled our 2018 observations using a tool designed to maximize the sensitivity of the experiment to the redshift signal \cite{supplements}.  Using this tool, we predicted that, given the existing data (1995-2017), spectroscopic measurements at the RV maximum and minimum in 2018 would provide the most sensitivity to detect the relativistic redshift (see Fig.~\ref{fig:signal}).  While they are less sensitive to the effect, imaging observations of the sky position of S0-2 in 2018 also slightly improve the measurement of the relativistic redshift.

The RVs of S0-2 are measured by fitting a physical model (which includes properties of the star such as its effective temperature, surface gravity, and rotational velocity in addition to RV), to its observed spectrum \cite{supplements}. The same procedure is applied to the new and archival observations; in the latter case this
spectroscopic method improves the precision by a factor of 1.7 compared to previous analyses \cite{Ghez:2008,Chu:2018}.

We also characterized additional sources of uncertainties beyond the uncertainties in the fitted model: i) the wavelength solution, which transforms locations on the detector to vacuum wavelengths, was characterized by comparing the observed wavelengths of atmospheric OH emission lines in the spectra of S0-2 and in observations of blank sky to their known vacuum wavelengths. This comparison shows the uncertainty of the wavelength solution of the spectroscopic instruments to be about 2 \kms, with some observations from 2002-2004 with lower accuracy between 2-26 \kms. ii) Re-examination of the spectroscopic data showed that one spectroscopic instrument had additional systematic bias from its optical system, which resulted in a systematic offset in RV compared to other instruments. We include an RV offset parameter in the orbit fit to account for this systematic offset. 
iii) We assessed systematic uncertainties by observations of bright RV standards stars of the same spectral-type as S0-2 (Table \ref{tab:rv_standards}). This systematic error is 1.3$\pm1.2$ km s$^{-1}$, smaller than the statistical uncertainties and about 6 times smaller than previous RV observations of S0-2 \cite{Chu:2018}. When these sources of systematic error are included in the analysis, the average RV uncertainty of S0-2 is found to be 20 km s$^{-1}$ for the Keck and Gemini observations.

The astrometric positions of S0-2 with respect to Sgr A* are placed into a common absolute astrometric reference frame using a multi-step cross-matching and transformation process. 
We adopted an improved methodology for obtaining precise astrometry and a more accurate absolute reference frame compared to previous work \cite{Boehle:2016}. This resulted in an average astrometric uncertainty for S0-2 of 1.1 milliarcsecond (mas) for speckle imaging, and 0.26 mas for AO imaging. 

The astrometric and RV measurements are combined in a global orbital model fitting using a standard post-Newtonian approximation which includes the first-order GR corrections on the Newtonian equations of motion, the R\"omer time delay due to variations in the light propagation time between S0-2 and the observer, and the relativistic redshift. For the astrometric observables, we ignore the negligible effect of light deflection by the SMBH but include a 2D linear drift of the gravitational center of mass. This drift accounts for systematic uncertainties in the construction of the astrometric reference frame. The RV observable is fully derived in \cite{supplements}. To our level of accuracy, it is:
\begin{equation}
    RV = v_{z_0} + V_{Z,\mathrm{S0-2}} + \Upsilon \left[\frac{V^2_\mathrm{S0-2}}{2c} + \frac{GM}{cR_\mathrm{S0-2}}\right] \, ,
\end{equation}
where $c$ is the speed of light in a vacuum, $v_{z_0}$ is a constant offset introduced to account for systematic uncertainties within our RV reduction, $V_{Z,\mathrm{S0-2}}$ is the Newtonian line-of-sight velocity of S0-2, $V^2_\mathrm{S0-2}/2c$ is the transverse Doppler shift predicted by special relativity depending on S0-2's velocity $\bm V_\mathrm{S0-2}$ 
and $GM/cR_\mathrm{S0-2}$ is the gravitational redshift predicted by GR incoporating the SMBH gravitational parameter $GM$ (the graviational constant G, and SMBH mass M) and on the distance, $R_\mathrm{S0-2}$, between S0-2 and the SMBH. $\Upsilon$ is a scale parameter introduced to characterize deviations from GR: its value is 0 in a purely Newtonian model and 1 in GR \cite{supplements} (for more details see S\ref{sec:model} in supplementary material). The model has 14 parameters: 6 orbital parameters for S0-2, the gravitational parameter of the SMBH ($GM$), the distance to the Galactic center $R_0$, a 2-D linear drift of the SMBH parametrized by the 2-D position ($x_0$,$y_0$) and velocity ($v_{x_0},v_{y_0}$) of the black hole from the center of the reference frame, an offset for the RV $v_{z_0}$, and the redshift parameter $\Upsilon$.

Several statistical tests are performed to assess systematic effects, using two different information criteria estimators to compare models: the Bayesian evidence and the expected logarithm predicted density \cite{supplements}. We examine several sources of systematic uncertainties in the orbital fit: (i) potential offsets in RVs and astrometric positions from different instruments  and (ii) potentially correlated uncertainties in astrometric measurements. 
Based on Bayesian model selection, we find that NIRC2 spectroscopy requires a RV offset with respect to other instruments (likely due to optical fringing). No other instruments require an RV or astrometric positional offset. We include a parameter for the NIRC2 RV offset in the model so it is fitted simultaneously. Based on the model selection criteria, we also find spatial correlation in the astrometric uncertainties. The correlated uncertainties are modeled with a multivariate likelihood characterized by a covariance matrix. The correlation matrix introduces a characteristic correlation length scale $\Lambda$ and a mixing parameter $p$, both of which are simultaneously fitted with the model parameters (see section \ref{sec:dom} in supplementary materials). We validated this approach by a Monte Carlo analysis, by randomly choosing one astrometric measurement per length scale to empirically estimate the effect of correlation scales. 
While the inclusion of these systematic effects does not significantly affect the best-fitting $\Upsilon$ value, it increases the uncertainties, affecting the precision of the results.

We developed an orbit modeling software package to model the orbits. The software uses Bayesian inference for model fitting, using nested sampling to estimate the posterior probability distribution via the multinest package \cite{Feroz:2008,Feroz:2009}. We also perform Monte Carlo simulations to evaluate our fitting methodology and to show that the statistical uncertainties are robust (see supplementary materials).

We initially compare a purely Newtonian model with a purely relativistic ($\Upsilon$ fixed to 1) model. We use the Bayes factor model selection criterion to show that the relativistic model is preferred by the data with high confidence. The difference of the logarithm of the Bayesian evidence between these two models is 10.68. Expressed as an odds ratio, the GR model is 43,000 times more likely than the Newtonian model in explaining the observations. 

We then fitted the more general model that includes the $\Upsilon$ redshift parameter as a free parameter. The estimated values for the 17 fitted parameters are in Table \ref{tab:fit_results} (the posterior distributions are shown in Figs. \ref{fig:corner_M_R}-\ref{fig:redshift_correlations}). The estimation $\Upsilon=0.88\pm 0.16$ and its marginal posterior is shown in Fig.~\ref{fig:signal}c. We estimate the systematic uncertainties due to the astrometric reference frame construction by performing a jackknife analysis on stars used to construct the reference frame. This adds a systematic uncertainty on the redshift parameter of $\sim0.047$, which when added in quadrature with the statistical uncertainties, results in a total uncertainty $\sigma_{\Upsilon} = 0.17$. The measured redshift parameter is therefore $0.88\pm 0.17$, consistent with GR at the 1$\sigma$ level while the Newtonian value $\Upsilon=0$ is excluded by $>5\sigma$. 
Our estimation also agrees at the 1$\sigma$ level with the measurement by the GRAVITY collaboration \cite{Gravity:2018}. Our experiment is independent from theirs, using a different set of measurements that includes the third turning point. We examined additional sources of systematic error that were previously not considered. The best-fitting model to the RV and the fit residuals is presented in Fig.~\ref{fig:res}. A fit using a parameter encoding deviations from GR only at the level of the gravitational redshift gives $\alpha = -0.24 \pm 0.32$, where $\alpha=2\left(\Upsilon-1\right)$ is the standard gravitational redshift parameter \cite{supplements} (see Supplements and Section 2.1.3. from \cite{Will:2014}). 

Our observations also constrain two other parameters: the mass of the black hole ($M_\mathrm{BH}$) and the distance to the Galactic center ($R_0$). From our model with $\Upsilon$ as free parameter, the 68\% marginalized confidence interval for $M_\mathrm{BH}=\left(3.984 \pm 0.058 \pm 0.026 \right) \times 10^{6} M_\odot$ and $R_0= 7,971\pm 59 \pm 32$ pc, where the first uncertainty is the statistical uncertainty and the second uncertainty is the systematic error $\sigma$ from the jackknife analysis (see Table \ref{tab:fit_results}). If we assume GR is true, then  $M_\mathrm{BH}=\left(3.964 \pm 0.047 \pm 0.026 \right) \times 10^{6} M_\odot$ and $R_0= 7,946\pm 50 \pm 32$ pc (see Supplemental texts for discussion). The nested sampling chains are provided in the Data Supplements. 

The gravitational redshift is a direct consequence of the universality of free fall and of special relativity \cite{Schiff1960}, hence of the Einstein equivalence principle, a fundamental principle of GR, which provides a geometric interpretation for gravitational interactions. Violations of the equivalence principle are predicted by some theories of modified gravity motivated by the development of a quantum theory of gravitation, unification theories, and some models of dark energy \cite{Damour:1994}. While the gravitational redshift has been measured with higher precision within in the Solar System \cite{herrmann:2018aa,delva:2018aa}, our results and those of the GRAVITY collaboration \cite{Gravity:2018} extend the measurements to higher gravitational redshift and around a massive compact object, a SMBH. Sgr A* has a mass $\sim4\times10^6$ times larger than that of the Sun. This constrains modified theories of gravitation that exhibit large non-perturbative effects around black holes, but not around non-compact objects like those in the Solar System (see \cite{antoniou:2018aa,silva:2018aa,doneva:2018aa} and Supplemental Text). This redshift test is also performed in a different environment than in the Solar System, where some theories predict modifications of GR to be screened or hidden (e.g. \cite{Khoury:2004}).



\bibliographystyle{Science}
\bibliography{scibib}

\section*{Acknowledgments}
We thank the staff and astronomers at Keck Observatory and Gemini Observatory, especially Gary Puniwai, Jason McIlroy, Sherry Yeh, John Pelletier, Joel Hicock, Greg Doppmann, Julie Renaud-Kim, Tony Ridenour, Alan Hatakeyama, Josh Walawender, Carolyn Jordan, Cynthia Wilburn, Terry Stickel, Heather Hershey, Jason Macilroy, John Pelletierm, Julie Renauld-Kim, Alessandro Rettura, Luca Rizzi, Carlos Alvarez, Marie Lemoine-Busserolle, Matthew Taylor, Trent Dupuy, Meg Schwamb, for all their help in obtaining the new data. The W.M. Keck Observatory is operated as a scientific partnership among the California Institute of Technology, the University of California, and the National Aeronautics and Space Administration. The authors wish to recognize that the summit of Maunakea has always held a very significant cultural role for the indigenous Hawaiian community. We are most fortunate to have the opportunity to observe from this mountain. We thank the Subaru telescope staff, especially
Y. Minowa, T.-S. Pyo, J.-H. Kim, and E. Mieda,
for their support for the Subaru observations. The Subaru Telescope is operated by the National Astronomical Observatory of Japan.

\section*{Funding}
Support for this work was provided by NSF AAG grant AST-1412615, the W. M. Keck Foundation, the Heising-Simons Foundation, and the Gordon and Betty Moore Foundation. S. J. and J.R.L. acknowledge support from NSF AAG (AST-1518273). The W. M. Keck Observatory was made possible by the generous financial support of the W. M. Keck Foundation. S. N. acknowledges financial support by JSPS KAKENHI,
Grant Number JP25707012, JP15K13463, JP18K18760, and JP19H00695.
H. S. was supported by JSPS KAKENHI Grant Number JP26610050 and JP19H01900.
Y. T. was supported by JSPS KAKENHI Grant Number JP26800150.
M. T. was supported by JSPS KAKENHI Grant Number JP17K05439, 
and DAIKO FOUNDATION. W. E. K. was
supported by an ESO Fellowship and the Excellence Cluster Universe,
Technische Universitat M\"unchen. R. S and E. G. have received funding from the European Research Council under the European Union's Seventh Framework Programme (FP7/2007-2013) / ERC grant agreement no [614922]. R. S. acknowledges financial support from the State Agency for Research of the Spanish MCIU through the "Center of Excellence Severo Ochoa" award for the Instituto de Astrof\`isica de Andaluc\`i�a (SEV-2017-0709).

\section*{Author contributions}
A.M.G, T.D., J.R.L, M.R.M, E.E.B, K.M, A.H. contributed to conceptualization and design of the experiment.  A.M.G, T.D., J.R.L, M.R.M, E.E.B, K.M, D.C., S.J., S.S., A.K.G., K.K.O., S.N., H.S., M.T., Y.T., R.C., Z.C., A.C., J.E.L, G.W.,S.C., performed observations. T.D., D.C., S.N., S.C., A.C., participated in reducing spectroscopic data and making RV measurements. J.R.L., S.J., S.S., A.K.G, Z.C., G.W.,R.S., E.G-O., reduced imaging data and and made astrometric measurements. A.M.G.,T.D., A.H, G.D.M, J.R.L., D.C., S.J., R.S., E.G-O., S.S., A.K.G, W.E.K., G.W., A.Z., participated in methodology development for improving astrometric and RV measurements. G.D.M, A.H., T.D. participated in statistical modeling and model comparisons. K.M., R.C., P.W., J.E.L., participated in building and improving instrumentation. All authors participated in writing and discussions of the paper.

\section*{Competing Interests} 
The authors declare no competing interests.

\section*{Data and material availability:} 
All (other) data needed to evaluate the conclusions in the paper are present in the paper or the Supplementary Materials.
Astrometric, RV measurements, and nested sampling chains are presented in the supplementary materials. Data from Keck Observatory are available at \url{https://www2.keck.hawaii.edu/koa/public/koa.php}, Gemini Telescope data available at \url{https://archive.gemini.edu/searchform}, and Subaru data available at \url{https://smoka.nao.ac.jp/}

\section*{Supplementary materials}
Materials and Methods\\
Supplementary Text\\
Figs. S1 to S18\\
Tables S1 to S13\\
Data S1 to S4 (Astrometry Measurements, RV Measurements, Nested Sampling Chains, S0-2 points in jackknife analysis)\\
References (26-92)

\begin{table}[htb]
 \caption{\textbf{Estimation of the model parameters.} Col 3: the maximum of the likelihood. Col 4: the median of the marginalized 1D posterior. Col 5: the half width of the 68\% confidence interval centered around the median. Col 6: the 1$\sigma$ systematics uncertainty from the reference frame estimated from the jackknife analysis \cite{supplements}. }
	\label{tab:fit_results} 
	\footnotesize
	\centering
	\begin{tabular}{c c c c c c}
	\hline
	    Parameter &  Description & Max of  & Estimation& Statistical & Systematic $\sigma$\\
      &  & likelihood  &  &uncertainty & from jackknife \\
	\hline
$M_\mathrm{BH}$ [$10^6 M_\odot$] & Black Hole Mass  & 3.984      & $   3.975 $ & 0.058    & 0.026  \\
$R_0$ [kpc]                       &  Distance to GC & 7.971      & $   7.959$ & 0.059 & 0.032\\
$\Upsilon$                       & Redshift Parameter & 0.80      & $   0.88 $    & 0.16    & 0.047 \\
$x_0$ [mas]                      & $x$ Dynamical Center &     0.99   & $   1.22$  & 0.32  & 0.51 \\
$y_0$ [mas]                      & $y$ Dynamical Center &   -0.85   & $  -0.88$  & 0.34  & 1.16 \\
$v_{x_0}$ [mas.yr$^{-1}$]               & $x$ Velocity &    -0.060   & $  -0.077$  & 0.018   & 0.14 \\
$v_{y_0}$ [mas.yr$^{-1}$]               & $y$ Velocity &     0.221   & $   0.226$  & 0.019   & 0.066 \\     
$v_{z_0}$ [km/s]                 & $z$ Velocity &    -3.6   & $  -6.2$  & 3.7     & 0.79 \\ \hline
$P$ [yr]                         &  Period &   16.041   & $  16.042 $  & 0.0016  & 7.8 $\times 10^{-5}$\\
$T_0$ [yr]                       &  Closest Approach &  2018.3765   & $2018.3763$ & 0.0004  & 1.9 $\times 10^{-5}$\\
$e$                              & Eccentricity &     0.886   &    0.8858 & 0.0004   & 2.8 $\times 10^{-5}$\\
$i$ [deg]                        &  Inclination &  133.88   & $ 133.82 $ & 0.18   & 0.13 \\
$\omega$ [deg]                   &  Argument of Periapsis &   66.03   & $  66.11 $ & 0.24   & 0.077\\     
$\Omega$ [deg]                   &  Angle to the Ascending Node &  227.40   & $ 227.49 $ & 0.29   & 0.11 \\ \hline
NIRC2 offset [km.s$^{-1}$]              &  RV Offset &  80    & $  81 $ & 19     & 0.8 \\ \hline
$\Lambda$ [mas]                  &  Astrometric Correlation Length &  21    & $  28 $ & $^{\phantom{-}24.6}_{-13.6}$ & 11.8 \\
$p$                              &  Astrometric Mixing Coefficient &  0.47     & $   0.55 $ & 0.13  & 0.11 \\ \hline
	\end{tabular}
\end{table}

\clearpage
\newpage

\setcounter{page}{1}
\renewcommand{\thetable}{S\arabic{table}}   
\renewcommand{\thefigure}{S\arabic{figure}}
\renewcommand{\theequation}{S\arabic{equation}}

\setcounter{figure}{0}
\setcounter{table}{0}
\setcounter{equation}{0}
\begin{center}
\section*{Supplementary Materials for}

\large
Relativistic redshift of the star S0-2 orbiting the Galactic center supermassive black hole \\

\bigskip

\footnotesize
Tuan Do, Aurelien Hees, Andrea Ghez, Gregory D. Martinez, Devin S. Chu, Siyao Jia, Shoko Sakai, Jessica R. Lu, Abhimat K. Gautam, Kelly Kosmo O'Neil, Eric E. Becklin, Mark R. Morris, Keith Matthews, Shogo Nishiyama,
Randy Campbell, Samantha Chappell, Zhuo Chen, Anna Ciurlo, Arezu Dehghanfar, Eulalia Gallego-Cano, Wolfgang E. Kerzendorf, James E. Lyke, Smadar Naoz, Hiromi Saida, Rainer Sch\"odel, Masaaki Takahashi, Yohsuke Takamori, Gunther Witzel, Peter Wizinowich\\

\bigskip
Correspondence to: tdo@astro.ucla.edu
\end{center}

\noindent\textbf{This PDF file includes}\\
\indent Materials and Methods\\
\indent Supplementary Text\\
\indent Figs. S1 to S18\\
\indent Tables S1 to S13\\

\noindent\textbf{Other Supplementary Materials for this manuscript include the following:}\\
\indent Data S1 to S4 (Astrometry Measurements, RV Measurements, Nested Sampling Chains, S0-2 points in jackknife analysis)\\

\clearpage

\section{Materials and Methods}

\subsection{Spectroscopy and Radial Velocity Measurements}

We use RV measurements from 6 spectrographs: NIRSPEC (Near-Infrared Spectrograph), NIRC2 (Near-Infrared Camera 2), OSIRIS (OH-Suppressing Infra-Red Imaging Spectrograph) on Keck, NIFS (Near-infrared Integral Field Spectrometer) on Gemini, IRCS (Infrared Camera and Spectrograph) on Subaru, and SINFONI (SINgle Faint Object Near-IR Investigation) on VLT. We use published values for NIRSPEC \cite{Boehle:2016} and SINFONI \cite{Gillessen:2017}, rederive the RVs from NIRC2, OSIRIS, and IRCS and performed new observations from OSIRIS, NIFS, and IRCS in 2017 and 2018 (Table \ref{tab:new_spec_obs}). 
Table \ref{tab:rv_measurements} presents all the S0-2 RV measurements used in this work.

\begin{longtable}{llllllr}
\caption{\textbf{New Spectroscopic Observations.} Col 1: date(s) of observations, Col 2: instrument name, Col 3: number of frames combined, Col 4: integration time per frame, Col 5: signal-to-noise ratio, Col 6: filter name, Col 7: plate scale of observations}\\
\label{tab:new_spec_obs}\\
\toprule
Date (UT)	&	Instrument	&	$N_{\textrm{frames}}$	&	Int. Time (s) &	SNR	&	Filter	&	Scale (mas)	\\
\midrule
\endhead
\midrule
\multicolumn{3}{r}{{Continued on next page}} \\
\midrule
\endfoot

\bottomrule
\endlastfoot

2017-05-05 - 2017-05-08 & IRCS & 98 & 300 & 48 & K & 55 \\
2017-05-17 &     OSIRIS	&	 11	   &	900	 &	101      &	Kn3	       &	35      \\
2017-05-18 &     OSIRIS	&	 9 	   &	900	 &	49       &	Kn3	       &	35      \\
2017-05-19 &     OSIRIS	&	 6 	   &	900	 &	77       &	Kn3		   &	35      \\
2017-07-19 &     OSIRIS	&	 12	   &	900	 &	55       &	Kn3		   &	35      \\
2017-07-27 &     OSIRIS	&	 13	   &	900	 &	76       &	Kn3		   &	35      \\
2017-08-09 - 2017-08-11 &  IRCS & 57 & 300 & 23 & K & 55 \\
2017-08-14 &     OSIRIS	&	 8 	   &	900	 &	71       &	Kn3		   &	35      \\
2017-09-02 &     OSIRIS	&	 4 	   &	900	 &	41       &	Kn3		   &	35      \\
2018-03-17 &     OSIRIS	&	 2 	   &	900	 &	41       &	Kn3		   &	35      \\
2018-03-29 - 2018-03-30 & IRCS & 39 & 300 & 21 & K & 55 \\
2018-04-24 &     OSIRIS	&	 7 	   &	900	 &	67       &	Kn3		   &	35      \\
2018-05-13 &     NIFS  	&	 12	   &	600	 &	84       &	K		   &	50 $\times$ 100       \\
2018-05-15 &     NIFS  	&	 7 	   &	600	 &	41       &	K		   &	50 $\times$ 100        \\
2018-05-17 &     OSIRIS	&	 4 	   &	900	 &	23       &	Kn3		   &	50      \\
2018-05-22 &     NIFS  	&	 12	   &	600	 &	66       &	K		   &	50 $\times$ 100        \\
2018-05-23 &     NIFS	&	 14	   &	600	 &	31       &	K		   &	50 $\times$ 100        \\
2018-05-23 &     OSIRIS	&	 14	   &	900	 &	97       &	Kn3		   &	35      \\
2018-06-05 &     OSIRIS	&	 10	   &	900	 &	44       &	Kn3		   &	35      \\
2018-07-22 &     OSIRIS	&	 11	   &	900	 &	121      &	Kn3		   &	35      \\
2018-07-31 &     OSIRIS	&	 11	   &	900	 &	125      &	Kn3		   &	35      \\
2018-08-11 &     OSIRIS	&	 9 	   &	900	 &	101      &	Kn3		   &	35      \\
2018-08-17 &     NIFS  	&	 8 	   &	600	 &	54       &	K		   &	50 $\times$ 100        \\
2018-08-18 &     NIFS  	&	 6 	   &	600	 &	58       &	K		   &	50 $\times$ 100        \\
2018-08-31 &     NIFS  	&	 3 	   &	600	 &	28       &	K		   &	50 $\times$ 100        \\
2018-08-31 &     OSIRIS	&	 3 	   &	900	 &	67       &	Kn3		   &	35      \\
2018-09-10 &     NIFS  	&	 3 	   &	600	 &	42       &	K	   &	50 $\times$ 100        \\
2018-09-16 &     NIFS  	&	 4 	   &	600	 &	63       &	K	   &	50 $\times$ 100

\end{longtable}

\begin{longtable}{lrrrrrr}
\caption{S0-2 RV measurements. Col. 1: Date of observation, Col. 2: Julian Year, Col. 3 Modified Julian Date, Col 4: Measured radial velocity, Col. 5: Radial velocity uncertainty, Col. 6: Radial velocity corrected for local standard of rest velocity. Full table (115 entries) is provided in Data S1}\\
\label{tab:rv_measurements}\\
\toprule
Date & Epoch & MJD Date & $v_z$&   $\sigma_{v_z}$  & $v_{lsr}$ &Source \\
(UT) & & (day) &  (km s$^{-1}$) &  (km s$^{-1}$) & (km s$^{-1}$) &  \\
\midrule
\endhead
\midrule
\multicolumn{3}{r}{{Continued on next page}} \\
\midrule
\endfoot

\bottomrule
\endlastfoot
2000-06-23 & 2000.4764 & 51718.50000 & 1192 &  100 & 1199 & NIRSPEC \\
2002-06-02 & 2002.4175 & 52427.50000 & -491 &   39 & -473 & NIRC2 \\
2002-06-03 & 2002.4203 & 52428.50000 & -494 &   39 & -476 & NIRC2 \\
2003-04-10 & 2003.2710 & 52739.23275 &  &   59 & -1571 & VLT \\
2003-05-10 & 2003.3530 & 52769.18325 &  &   40 & -1512 & VLT \\
 ... & ... &  ... & ... & ... & ... & ... \\
\midrule
\end{longtable}

\subsubsection{Keck NIRC2 Spectroscopy}\label{sec:NIRC2RV}

We used the NIRC2 instrument in slit spectroscopy mode for observations from 2002-2005 with the Keck 2 Natural Guide Star AO  system\cite{Wizinowich:2000}. 
The data and instrument have been reported in \cite{Ghez:2003}. We use the same spectra as in \cite{Ghez:2008}, except for 2003 Jun 08 and 2003 Jun 09. In previous publications, the spectra for those two nights were combined. In re-examining the data, we found that the wavelength solution varied between the two nights. By not combining the nights, we reduce interpolation errors from shifting the spectra and thereby better capture intrinsic orbital variations in the RV. 

\subsubsection{Keck OSIRIS Spectroscopy}

The Keck OSIRIS instrument is an integral-field spectrograph that can sample two spatial dimensions and one spectral dimension simultaneously. It has a resolving power of $ R = \lambda/\delta\lambda = 4000$ (wavelength divided by the resolution element). Data cubes were produced using the standard OSIRIS Data Reduction Pipeline \cite{Lyke:2017}. 
We utilize the Kn3 (2.121 -- 2.229 $\mu$m) and Kbb (1.965 -- 2.381 $\mu$m) filters at the 35 and 20 mas plate scales at various times from 2005 to 2018 (see Table \ref{tab:new_spec_obs}) with the Keck I Laser Guide Star AO system\cite{Wizinowich:2006,vanDam:2006}. 
The spectra of S0-2 and other stars in the data cubes are extracted using a circular aperture centered on the star on each spectral channel, with an annulus around the star to estimate sky background. Blank sky and standard stars are observed during the night to correct for sky emission and atmospheric absorption lines.  Further details are available in \cite{Do:2013}. We produce a combined spectrum from all the data cubes taken each night. 

\subsubsection{Gemini NIFS Spectroscopy}

The Gemini NIFS instrument (R = 5000) is also an integral-field spectrograph, which produces data similar to OSIRIS. All the NIFS data were obtained in 2018 using the natural guide star adaptive optics system Altair\cite{Herriot:2000} and with the K grating and the HK filter. The NIFS data were reduced using the data reduction package Nifty4Gemini\cite{nifty4gemini}, following standard methods (e.g. \cite{Stostad:2015}). Similar to OSIRIS, blank sky and standard stars are observed during the night to correct for sky emission and atmospheric absorption lines. We use the software package molecfit\cite{2015A&A...576A..77S} to fit and remove the atmospheric features.  We use the same method for spectral extraction as with OSIRIS. 

\subsubsection{Subaru IRCS Spectroscopy}

We carried out Echelle spectroscopic observations of S0-2 using the IRCS \cite{Kobayashi:2000} instrument (R = 20000) on Subaru Telescope \cite{Iye:2004}.
We took spectra in the K+ setting, with the correction of the Subaru AO system AO188 \cite{Hayano:2010}.
The reduction procedure includes dark subtraction, flat-fielding, bad pixel correction, cosmic-ray removal,
spectrum extraction, wavelength calibration, telluric correction, and spectrum continuum fitting.
A sky field and standard stars were observed during the nights for the correction of 
atmospheric emission and absorption lines.
The sky OH lines were used for the wavelength calibration.
The details of the observations before 2017 and data reduction procedure are presented in \cite{Nishiyama:2018}.
Spectra on four nights (2017 May 05 - 08), three nights (2017 Aug 09 - 11) and two nights (2018 Mar 29 - 30)
were combined to produce the spectra of the three epochs, 2017.346, 2017.607, and 2018.241, respectively.

\subsubsection{New and Re-derived RV Measurements}
We present new radial velocity measurements from spectra with NIRC2, OSIRIS, NIFS, and IRCS instruments. While the RV measurements from NIRC2, OSIRIS, and IRCS taken before 2017 were presented previously \cite{Ghez:2008,Boehle:2016,Nishiyama:2018}, here we re-derive all NIRC2, OSIRIS, and IRCS RVs using an improved method. We use a synthetic spectral grid and Bayesian inference to model the spectra using a physical model that includes the physical properties of the star (e.g. effective temperature) along with its rotational velocity and radial velocity. We use the BOSZ spectral grid \cite{Bohlin:2017} which has synthetic spectra calculated over the range of wavelength of the observations as well as the reported physical properties of S0-2. This grid reaches high effective temperatures (up to 35,000 K), which covers previously reported temperatures for S0-2 and other stars within 0.04 pc of the SMBH \cite{Habibi:2017}.  We use the StarKit spectral fitting software package to perform the parameter estimation \cite{Kerzendorf2015}. StarKit simultaneously models the physical properties of the star (effective temperature, surface gravity, metallicity, alpha-elemental abundance), the continuum (modeled as a second-order polynomial), rotational velocity, radial velocity, instrumental broadening, and wavelength sampling in order to compare the model with the data directly. We use a Gaussian likelihood to compare the model to the observed spectrum and the uncertainty on the flux. The flux uncertainty was estimated using the standard deviation of the flux in the continuum of the star. To account for potential mis-estimation of the flux uncertainties, we also included an additive flux uncertainties term in the spectroscopic fit. We find that this additive term is smaller than the flux uncertainties. Parameter estimation is done via Bayesian inference with sampling of the posterior using the nested-sampling algorithm MultiNest \cite{Feroz:2009}. More details on StarKit and its application to Galactic center data are given in \cite{Do:2015,Do:2018,Feldmeier-Krause:2017}. We use the median and 1 sigma central credible interval of the marginalized posterior as the radial velocity and its uncertainty. The spectroscopic observable $\mathcal V$ is defined as: 

\begin{equation}
\mathcal V = \frac{\lambda_\mathrm{obs}-\lambda_\mathrm{em}}{\lambda_\mathrm{em}}c
\label{eq:spectro}
\end{equation}

where $\lambda_\mathrm{obs}$ is the observed wavelength, $\lambda_\mathrm{em}$ is the emitted wavelength of the spectral features in vacuum, and $c$ is the speed of light. The radial velocity measurements are corrected for the local standard of rest with respect to the Galactic center. We use the IRAF procedure \textit{rvcorrect}. This correction uses a velocity of 20 \kms for the solar motion with respect to the local standard of rest in the direction $\alpha = 18^{h}, \delta = +30\deg$ for epoch 1900 \cite{1986MNRAS.221.1023K}, corresponding to $(u,v,w) = (10, 15.4, 7.8)$ \kms. The result is the Newtonian radial velocity of the star. We fitted the relativistic corrections as part of the orbit model. 

\begin{figure}
\begin{center}
\includegraphics[scale=0.5]{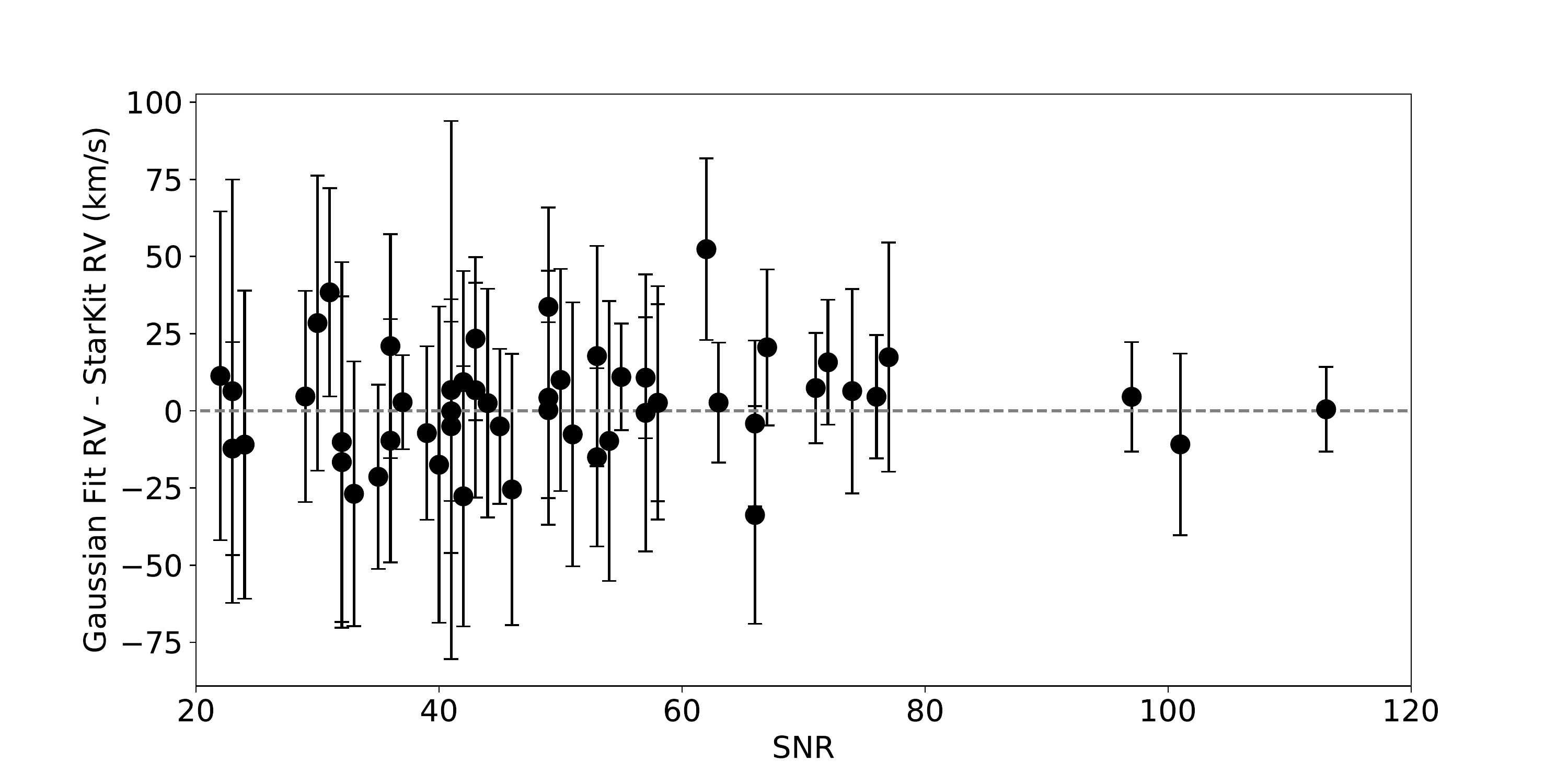}
\end{center}
\caption{
The difference between the S0-2 RV derived using Gaussian fit to the Bracket gamma line and using the full-spectrum fitter StarKit as function of SNR of the spectra. The two methods result in consistent RV values within the individual uncertainties. The weighted average RV offset between a Gaussian fit and StarKit is about 4 km s$^{-1}$. 
\label{fig:gaussian_starkit}
}
\end{figure}

\begin{figure}
\begin{center}
\includegraphics[scale=0.5]{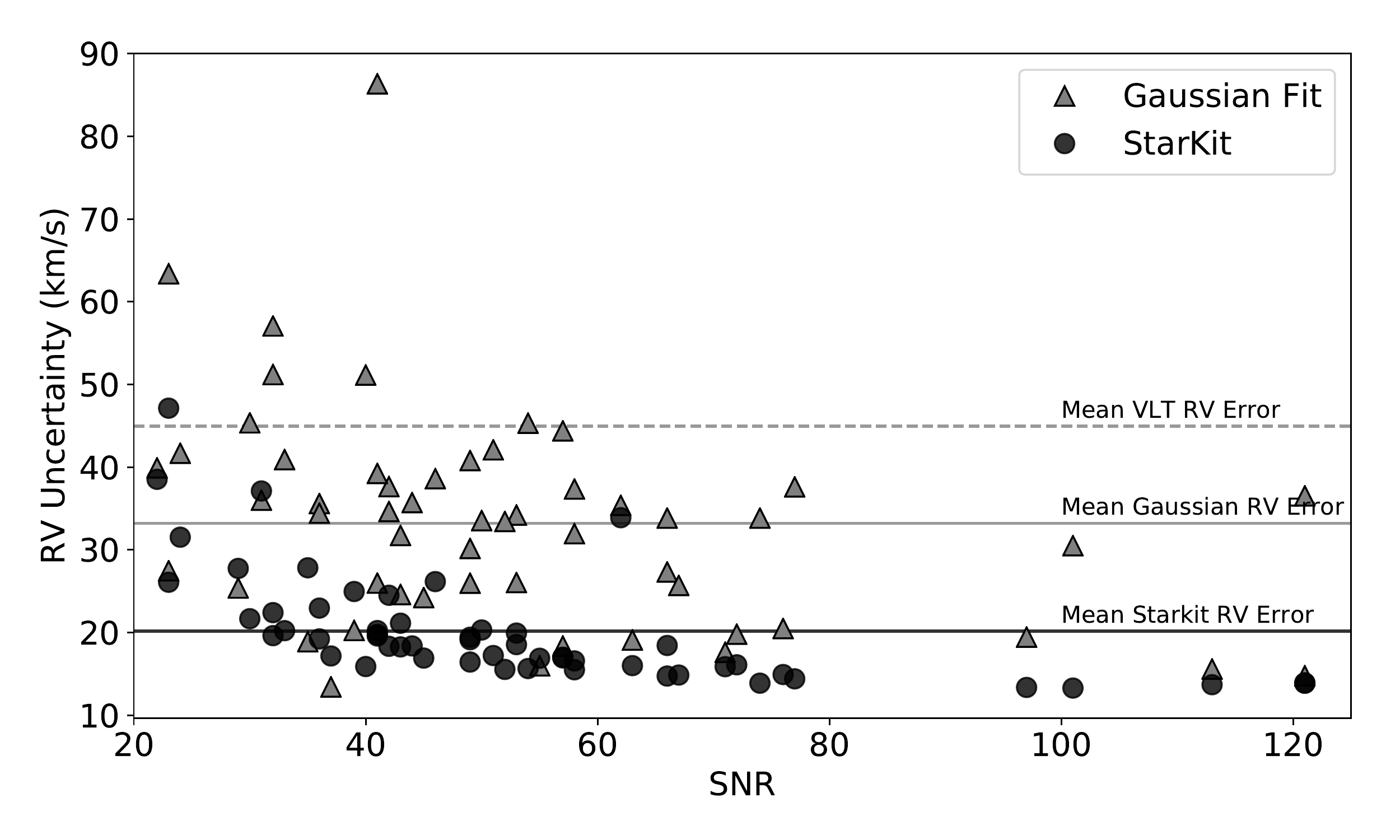}
\end{center}
\caption{
Comparisons of the RV uncertainty for S0-2 RV measured using a Gaussian fit to the Bracket Gamma line (grey triangles) and using the full-spectrum fitter StarKit (black circles). RV uncertainties are shown as a function of SNR. On average, StarKit uncertainty estimates are about 1.7 times more precise than using a Gaussian. These measurements are also on average about 2.2 times more precise than average RV values reported by \cite{Gillessen:2017} using SINFONI on VLT.
\label{fig:gaussian_starkit_err}
}
\end{figure}

We find that our new method of measuring the RV of S0-2 is consistent with, more precise and more accurate than the method used previously to extract S0-2's RV \cite{Ghez:2008,Gillessen:2009}(Fig. \ref{fig:gaussian_starkit}). 
Previously, a Gaussian profile was fitted to the hydrogen absorption line at 2.1661 micron (Bracket gamma), the strongest spectral feature in the K-band. 
While a Gaussian fit can determine the centroid of the line, at high SNR ratio, the intrinsic line shape becomes more important, thus limiting the precision of a Gaussian fit. In addition, there is a weak helium line at 2.1617 micron, which is not well resolved from the hydrogen line, resulting in an asymmetric line profile, potentially biasing the Gaussian fit.  By modeling the spectrum with a physically motivated model with the appropriate atomic line data, we use more information than with a Gaussian fit alone resulting in more precise measurements. The radial velocity measurements using StarKit have an average uncertainty of 17 km s$^{-1}$ compared to about 30 km s$^{-1}$ with the Gaussian fit, an improvement of a factor of 1.7 (Fig. \ref{fig:gaussian_starkit_err}).

We find that when the SNR in the continuum of spectrum falls below 20, the Bracket gamma feature becomes too noisy to be robustly measured by this technique; we therefore only include measurements with SNR $>$ 20. 
This criteria excludes the RV measurement from 2007 July 21 (SNR = 16) that was previously included in \cite{Boehle:2016}.

Using observations of stars that are RV standards on the same night as observations of S0-2, we can also evaluate the accuracy of the radial velocity measurements. We selected radial velocity standards to be stars that have spectral type similar to that of S0-2 and which have been previously observed as radial velocity standards. We extract these stars using the Set of Identifications, Measurements and Bibliography for Astronomical Data (SIMBAD) database, selecting ones that do not have significant variations in radial velocities between multiple previous measurements. Measurements of these stars are included in Table \ref{tab:rv_standards}.

\begin{figure}
\begin{center}
\includegraphics[scale=0.5]{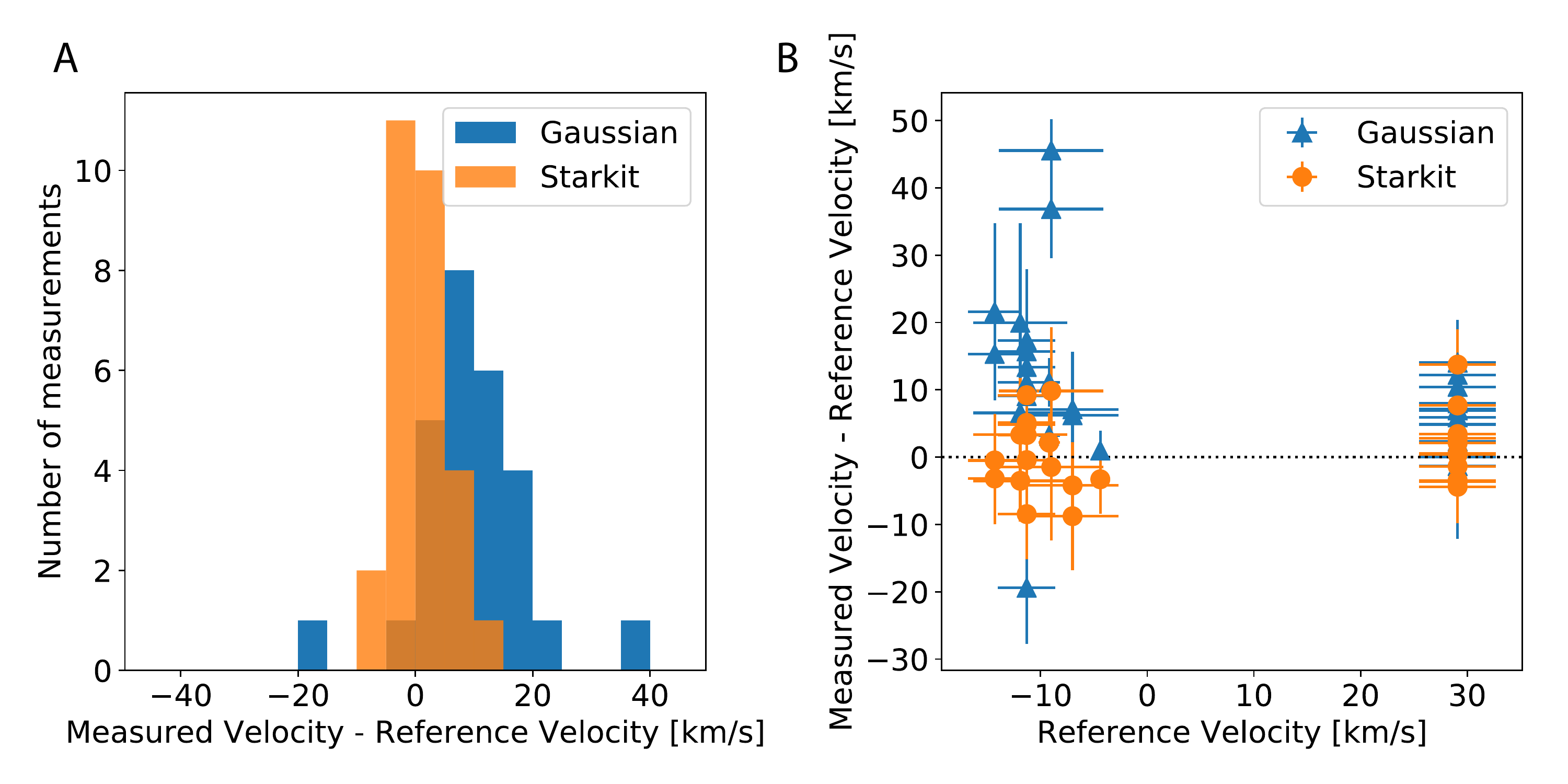}
\end{center}
\caption{
\textbf{A:} Histogram of the differences between radial velocity measurements and reference values of standard stars given in Table \ref{tab:rv_standards}. \textbf{B:} Differences between radial velocity measurements and reference values compared to their reference velocity for each individual measurement.
\label{fig:standard_stars}
}
\end{figure}

When the Gaussian fitting method was used, the weighted average difference from the reported SIMBAD values was 8.3 $\pm$ 1.2 km s$^{-1}$, but when using StarKit, radial velocity measurements of the standard stars had a weighted average difference from the reported SIMBAD velocities of only 1.3 $\pm$ 1.2 km s$^{-1}$ (Fig. \ref{fig:standard_stars}). 
We attribute this improvement to StarKit's ability to fit the non-Gaussian absorption lines. This result shows the robustness of the StarKit method and a reduction in systematic uncertainty. 

{\footnotesize
\begin{longtable}{lllllllll}
 \caption{\textbf{RV Standard Stars Measurements}. Col. 1: Date of observation, Col. 2: Star name, Col. 3: Filter Name, Col. 4: Radial velocity using Gaussian fit, Col 5. Radial velocity using StarKit, Col. 6: Reference velocity from SIMBAD, Col. 7: Radial velocity offset from reference value using a Gaussin fit, Col. 8: Radial velocity offset from reference value using StarKit, Col. 9: Reference for SIMBAD velocity}\\
	\toprule
	Date &  Star	&	Filter	&	RV$_{\textrm{Gauss}}$ 	&	RV$_{\textrm{StarKit}}$ 	&	SIMBAD 	&	$\Delta$ RV$_{\textrm{Guass}}$ 	&	$\Delta$ RV$_{\textrm{StarKit}}$ &   Reference\\
   \textrm{(UT)} &	&	&	(\kms)	&	(\kms)	&	(\kms)	&	(\kms)	&	(\kms)  &\\
	\midrule
\endhead
\midrule
\multicolumn{3}{r}{{Continued on next page}} \\
\midrule
\endfoot

\bottomrule
\endlastfoot

2015-08-07 &  HD 217811 &    Kn3 &          -30.7$\pm$8.4 	&           -19.8$\pm$6.7 &   -11.3$\pm$2.7 &            -19.4 &                 -8.5 	&	\cite{pulkovo2006}\\
2016-05-15 &  HD 172488 &    Kbb &           43.1$\pm$6.4 	&            42.8$\pm$5.2 &    29.1$\pm$3.6 &             14.0 &                 13.7 	&	\cite{pulkovo2006,Bobylev2008}\\
2016-05-16 &  HD 172488 &    Kbb &           41.3$\pm$7.3 	&            36.8$\pm$5.3 &    29.1$\pm$3.6 &             12.2 &                  7.7 	&	\cite{pulkovo2006,Bobylev2008}\\
2016-07-11 &  HD 172488 &    Kbb &           37.1$\pm$4.3 	&            31.2$\pm$4.6 &    29.1$\pm$3.6 &              8.0 &                  2.1 	&	\cite{pulkovo2006,Bobylev2008}\\
2016-07-12 &  HD 172488 &    Kbb &           34.0$\pm$4.3 	&            32.5$\pm$4.4 &    29.1$\pm$3.6 &              4.9 &                  3.4 	&	\cite{pulkovo2006,Bobylev2008}\\
2017-05-17 &  HD 172488 &    Kn3 &           36.0$\pm$3.6 	&            27.7$\pm$5.1 &    29.1$\pm$3.6 &              6.9 &                 -1.4 	&	\cite{pulkovo2006,Bobylev2008}\\
2017-08-14 &  HD 217811 &    Kn3 &           -0.2$\pm$6.8 	&            -8.0$\pm$4.5 &   -11.3$\pm$2.7 &             11.1 &                  3.3 	&	\cite{pulkovo2006}\\
2017-08-14 &  HD 215191 &    Kbb &            1.0$\pm$6.5 	&           -17.5$\pm$6.8 &   -14.3$\pm$2.5 &             15.3 &                 -3.2 	&	\cite{pulkovo2006,huang2010}\\
2017-08-14 &  HD 215191 &    Kn3 &            7.3$\pm$13.1 	&           -14.8$\pm$6.9 &   -14.3$\pm$2.5 &             21.6 &                 -0.5 	&	
\cite{pulkovo2006,huang2010}\\
2017-08-14 &  HD 191639 &    Kbb &           -0.8$\pm$9.4 	&           -11.2$\pm$6.4 &    -7.0$\pm$4.3 &              6.2 &                 -4.2 	&	\cite{pulkovo2006}\\
2017-08-14 &  HD 191639 &    Kn3 &            0.1$\pm$6.3 	&           -15.8$\pm$8.0 &    -7.0$\pm$4.3 &              7.1 &                 -8.8 	&	\cite{pulkovo2006}\\
2017-08-14 &  HD 217811 &    Kbb &            4.4$\pm$5.0 	&            -6.5$\pm$4.3 &   -11.3$\pm$2.7 &             15.7 &                  4.8 	&	\cite{pulkovo2006}\\
2017-09-02 &  HD 217811 &    Kn3 &            6.0$\pm$10.6 	&            -2.1$\pm$6.1 &   -11.3$\pm$2.7 &             17.3 &                  9.2 	&	\cite{pulkovo2006}\\
2017-09-02 &  HD 214652 &    Kbb &           -5.3$\pm$4.7 	&           -15.4$\pm$6.1 &   -11.9$\pm$4.4 &              6.6 &                 -3.5 	&	\cite{pulkovo2006}\\
2017-09-02 &  HD 214652 &    Kn3 &            8.0$\pm$14.8 	&            -8.6$\pm$8.4 &   -11.9$\pm$4.4 &             20.0 &                  3.3 	&	\cite{pulkovo2006}\\
2017-09-02 &  HD 186568 &    Kbb &           -6.0$\pm$2.8 	&            -7.0$\pm$3.6 &    -9.2$\pm$1.0 &              3.2 &                  2.2 	&	\cite{pulkovo2006,khalack2015}\\
2017-09-02 &  HD 186568 &    Kn3 &            1.9$\pm$3.6 	&            -7.0$\pm$4.3 &    -9.2$\pm$1.0 &             11.1 &                  2.2 	&	\cite{pulkovo2006,khalack2015}\\
2017-09-02 &  HD 217811 &    Kbb &           -2.2$\pm$7.7 	&            -6.2$\pm$5.2 &   -11.3$\pm$2.7 &              9.1 &                  5.1 	&	\cite{pulkovo2006}\\
2018-04-27 &  HD 172488 &    Kn3 &           35.0$\pm$5.1 	&            25.6$\pm$6.3 &    29.1$\pm$3.6 &              5.9 &                 -3.5 	&	\cite{pulkovo2006,Bobylev2008}\\
2018-04-27 &  HD 170783 &    Kn3 &           -3.4$\pm$3.0 	&            -7.7$\pm$5.1 &   -4.4$\pm$0.3 &              1.0 &                 -3.3 	&	\cite{pulkovo2006}\\
2018-04-27 &  HD 146416 &    Kbb &           27.8$\pm$7.3 	&           -10.5$\pm$10.9 &    -9.0$\pm$4.9 &             36.8 &                 -1.5 	&	\cite{pulkovo2006}\\
2018-04-27 &  HD 146416 &    Kn3 &           36.5$\pm$4.7 	&             0.8$\pm$9.4 &    -9.0$\pm$4.9 &             45.5 &                  9.8 	&	\cite{pulkovo2006}\\
2018-05-23 &  HD 172488 &    Kn3 &           39.5$\pm$6.4 	&            29.7$\pm$5.7 &    29.1$\pm$3.6 &             10.4 &                  0.6 	&	\cite{pulkovo2006,Bobylev2008}\\
2018-06-05 &  HD 172488 &    Kn3 &           29.2$\pm$9.9 	&            31.9$\pm$5.6 &    29.1$\pm$3.6 &              0.0 &                  2.8 	&	\cite{pulkovo2006,Bobylev2008}\\
2018-06-05 &  HD 164900 &    Kbb &           16.2$\pm$7.4 	&           -11.7$\pm$8.5 &   -36.0$\pm$3.7 &             52.2 &                 24.3 	&	
\cite{Kharchenko2007}\\
2018-06-05 &  HD 164900 &    Kn3 &           23.2$\pm$10.9 	&            -8.0$\pm$7.0 &   -36.0$\pm$3.7 &             59.2 &                 28.0 	&	\cite{Kharchenko2007}\\
2018-07-22 &  HD 172488 &    Kn3 &           36.2$\pm$5.9 	&            29.5$\pm$5.1 &    29.1$\pm$3.6 &              7.1 &                  0.4 	&	\cite{pulkovo2006,Bobylev2008}\\
2018-07-31 &  HD 172488 &    Kn3 &           31.4$\pm$10.2 	&            24.7$\pm$5.4 &    29.1$\pm$3.6 &              2.3 &                 -4.4 	&	\cite{pulkovo2006,Bobylev2008}\\
2018-08-11 &  HD 172488 &    Kn3 &           27.8$\pm$10.8 	&            25.5$\pm$5.7 &    29.1$\pm$3.6 &             -1.3 &                 -3.6 	&	\cite{pulkovo2006,Bobylev2008}\\
2018-08-11 &  HD 217811 &    Kn3 &            2.1$\pm$6.7 	&           -11.7$\pm$4.9 &   -11.3$\pm$2.7 &             13.4 &                 -0.4	&	\cite{pulkovo2006}
\label{tab:rv_standards} 
\end{longtable}}

A potential source of systematic uncertainty arises from subtracting emission lines from the night sky and gas emission at the Galactic center. In the near-infrared, strong OH line emission from the Earth's atmosphere are superimposed on the observed spectrum of S0-2. These lines are removed by observing a sky location free of stars and subtracting the sky spectrum. While we account for variations in the OH line strengths by scaling the reference sky observation, the remaining residuals after sky subtraction that can affect the intrinsic stellar lines if the velocity of the star places them near the OH lines. In addition, the Galactic center has emission from hydrogen gas from the Bracket gamma line (Br gamma) in the vicinity of Sgr A* \cite{Paumard:2004}. We remove this gas emission by extracting the spectrum in an annulus around S0-2. 
However, spatial variations in this line can create residuals that affect the fit quality. When residuals from Br gamma subtraction are strong, we mask two spectral channels on both sides of the line from the fit to reduce the impact of the residual. To quantitatively assess the impact of sky and gas subtraction residuals, we perform a series of Monte Carlo simulations. Using a high SNR spectrum of S0-2 observed when its velocity is very far from features we want to test, we plant S0-2's stellar features in the spectrum at specific locations near OH lines and the Br gamma line. Based on these simulations, we find that S0-2's radial velocity uncertainty can be underestimated by about 14 km s$^{-1}$ (added in quadrature). There is a velocity bias from the presence of these lines, but the bias is smaller than the uncertainty.

The simulations of the effect of sky and gas subtraction suggest there should be an additive error to the RVs. 
To better assess this, we fit the RVs of 3 stars near S0-2, which have similar brightness and are of similar spectral-type so that their spectral features are comparable to S0-2. These three stars, S0-9, S0-14, and S0-15, only show a linear trend in RV so we include three parameters in the fit: a baseline RV value, an acceleration in RV, and an additive error to be added in quadrature that is simultaneously fitted with the two model parameters.
The posterior probability distribution function (PDF) of the systematic uncertainty resulting from independent fits of these three stars are presented in Fig.~\ref{fig:rv_uncertainty} and are consistent with each other. A combined analysis in which we fit the three stars simultaneously is shown in Fig.~\ref{fig:rv_uncertainty}. The 68\% confidence interval for this systematic uncertainty is $11^{+4.8}_{-4.1}$~km~s$^{-1}$. We also check for an additive error for S0-2 RVs by including an additive error parameter in the S0-2 orbit fit. This fit is similar to that from the three other stars and leads to an estimate of the systematic uncertainty of $11.9^{+4.1}_{-3.7}$~km~s$^{-1}$. These values are also consistent with the simulations from the gas and sky emission line subtraction residuals (see above). Based on these results, we include an additive error of 11 km s$^{-1}$ for S0-2 RV measurements with OSIRIS and NIFS. This error is smaller than the RV uncertainties for NIRSPEC, NIRC2, and SINFONI, so we do not include it for those instruments. 

\begin{figure}
\begin{center}
\includegraphics[scale=0.4]{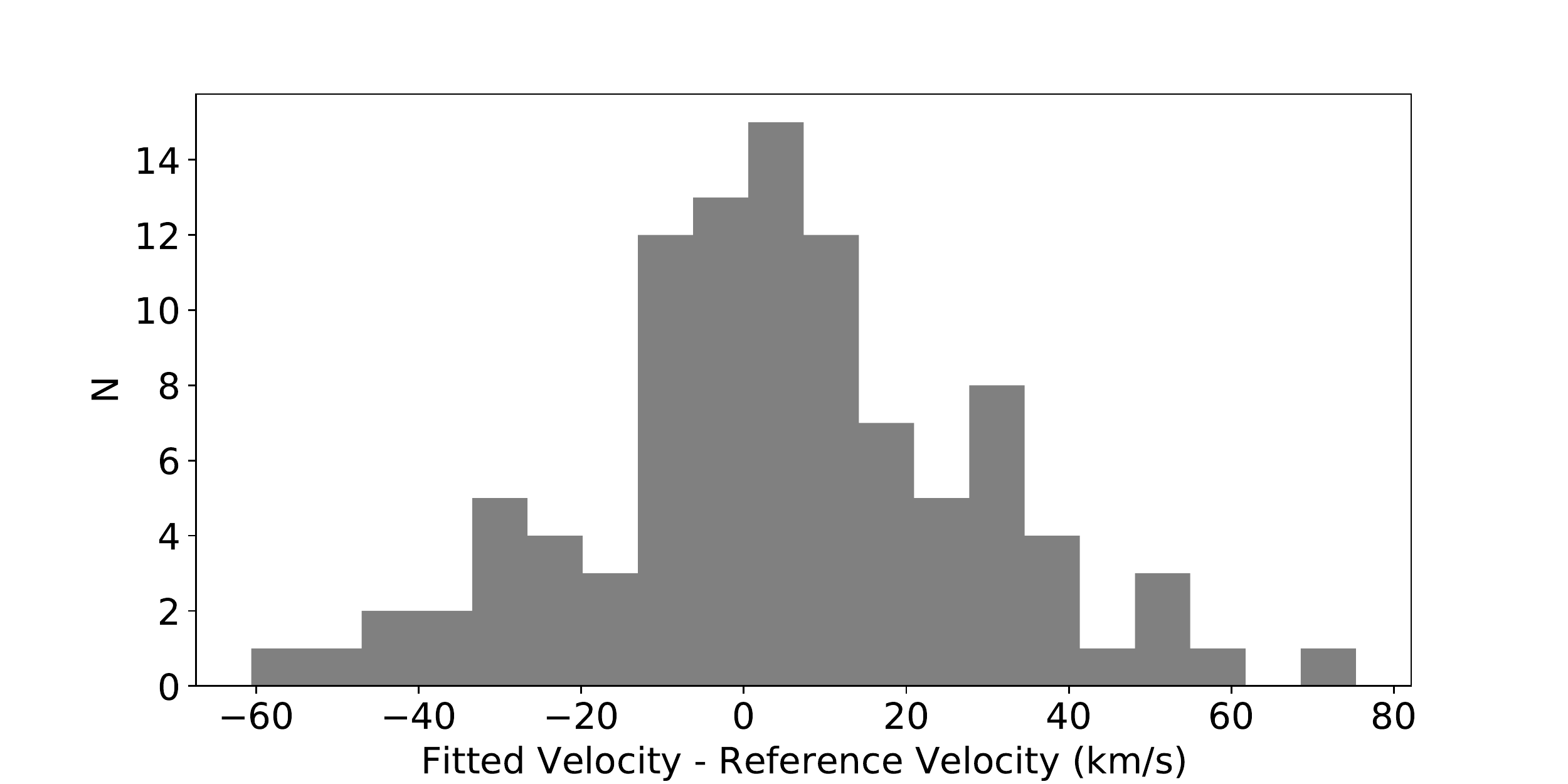}
\end{center}
\caption{Histogram of the velocity offsets from fits to simulated data compared to their reference values. The goal of these simulations are to evaluate the effect of residuals from background subtraction on the spectral features of S0-2 and how it affects the fitted velocities. These simulations suggest that the RV uncertainties of S0-2 should be larger than the fitted uncertainties. In this simulation, the uncertainty should be larger by about 14~km~s$^{-1}$ added in quadrature to the fitted uncertainties.}
\label{fig:sky_line_simulations}
\end{figure}

\begin{figure}
\begin{center}
\includegraphics[scale=0.4]{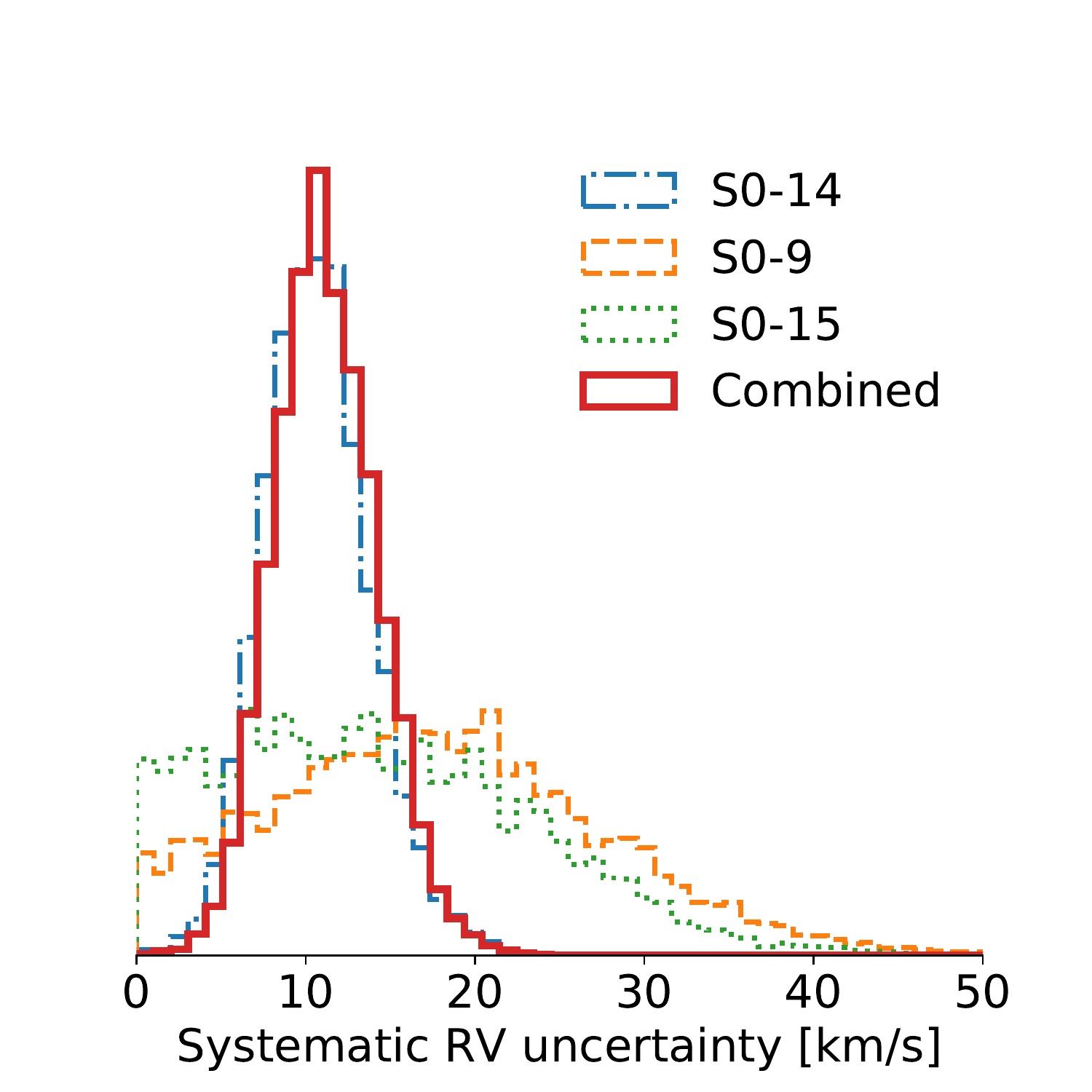}
\end{center}
\caption{Posterior probability distribution for the fitted additive systematic RV uncertainty obtained from independent fits of the three stars S0-14 (blue), S0-9 (orange) and S0-15 (green). The combined fit from these three stars (red) show that an additive uncertainty of 11 km s$^{-1}$ is preferred by the model. 
\label{fig:rv_uncertainty}
}
\end{figure}

We also examine two additional sources of systematic uncertainties in the RV measurements: uncertainty in the wavelength solution and optical fringing in the NIRC2 spectra. The wavelength solutions for OSIRIS and NIFS are derived from Ar, Ne, Xe arclamp lines, while the wavelength solution for NIRC2 is derived using the OH skylines. We measure the accuracy of the wavelength solution by comparing the observed centroid of the OH lines from observations of the sky to their vacuum wavelength values. We estimate the systematic error in the radial velocity using the standard deviation of these differences. Using this measure, the wavelength solutions for OSIRIS and NIFS have uncertainties less than 2 km s$^{-1}$. NIRC2 shows larger variations between different observations, with wavelength solution offsets as high as 26 km s$^{-1}$. We therefore include an additive uncertainty from the wavelength solution of 2 km s$^{-1}$ for OSIRIS and NIFS data and 26 km s$^{-1}$ for NIRC2 data. 
The NIRC2 observed spectra also exhibited fringing that can not be corrected. Fringing, likely as a result of optical interference patterns in the filters, creates quasi-periodic flux variations in the NIRC2 observed spectra. This can cause apparent shifts in the spectral features. We have attempted to estimate the effect of the fringing by examining measurements made on two consecutive nights in 2003. Between these nights, the measurements disagreed by about 70 km s$^{-1}$.  We thus know that the fringing can cause offsets in RV as large as 70 km s$^{-1}$. We account for possible systematic offset in all NIRC2 RVs by introducing an offset parameter into the orbit fit. This parameter represents a global RV offset in the NIRC2 RV and is fitted simultaneously with other model parameters. Based on the Bayesian information criteria, such an offset very significantly improves the fit compared to offsets applied to the other spectroscopic instruments (See Section \ref{sec:systematics}). 

New spectroscopic observations are reported in Table \ref{tab:new_spec_obs}. Table \ref{tab:rv_measurements} presents the RV measurements both before ($v_z$) and after correction for the local standard of rest velocity ($v_{lsr}$). The RV uncertainty ($\sigma_{v_z}$) includes the additive error for each epoch of observation. We also include literature measurements from \cite{Gillessen:2017} used in the orbit fitting.

\subsection{Imaging and astrometric measurements}\label{sec:astro}

The sky positional measurements of S0-2 are made using 2 instruments: speckle imaging with NIRC on Keck I (1995-2005) and AO imaging with NIRC2 on Keck II (2005-2018). The data reduction and point source detection methods are described in detail in \cite{Ghez:2008,Yelda:2010}. 
Here, we summarize the data and methods used to place the measurements of stellar positions in a common reference frame. Table \ref{tab:new_astro_obs} present new astrometric observations and Table \ref{tab:astro_measurements} presents the astrometric measurements used in the S0-2 orbit fit. We also transform the positions into separation and position angle (defined to increase East from North) from the origin of the reference frame in each epoch. Using the separation and angle is not straightforward (due to the fit for drift in the reference frame), so we use the coordinates in all our fitting procedures.

{\scriptsize
\begin{longtable}{llcrrccccrcc}
\caption{New Astrometric Observations}\\
\label{tab:new_astro_obs}\\
\toprule
Date  & Date  & Filter    & $N_{\textrm{frames}}$    & $N_{\textrm{frames}}$    & $t_{\textrm{int}}$ & $N_{\textrm{coadds}}$	& FWHM & Strehl Ratio & $N_{\textrm{stars}}$  & $m_{\textrm{lim}}$    & $\sigma_{\textrm{pos}}$   \\
 (UT) & (MJD) &  &  Obtained & Used &  (s) &  & (mas) & & & (mag) & (mas) \\
\midrule
\endhead
\midrule
\multicolumn{3}{r}{{Continued on next page}} \\
\midrule
\endfoot

\bottomrule
\endlastfoot
2017-05-07  & 57880.558 & $H$   & 140   & 99    & 7.4   & 4 & 56.38 & 0.18  & 1591  & 21.07 & 0.07  \\
2017-08-13  & 57978.276 & $H$   & 80    & 41    & 7.4   & 6 & 58.05 & 0.16  & 1413  & 21.71 & 0.12  \\
2017-08-23,24,26    & 57990.266 & $H$   & 101   & 45    & 7.4   & 4 & 65.09 & 0.14  & 1201  & 20.30 & 0.09  \\
2018-03-17  & 58194.635 & $H$   & 35    & 27    & 7.4   & 4 & 59.51 & 0.20  & 1466  & 20.60 & 0.06  \\
2018-03-22  & 58199.620 & $H$   & 50    & 40    & 7.4   & 4 & 82.06 & 0.10  & 936   & 19.62 & 0.10  \\
2018-03-30  & 58207.629 & $H$   & 47    & 21    & 7.4   & 4 & 66.06 & 0.15  & 1118  & 19.91 & 0.10  \\
2018-05-19  & 58257.545 & $H$   & 9     & 7     & 7.4   & 4 & 98.11 & 0.08  & 408   & 18.15 & 0.40  \\
2018-05-24  & 58262.529 & $H$   & 46    & 30    & 7.4   & 4 & 83.19 & 0.09  & 845   & 19.57 & 0.11  \\
2018-09-03  & 58364.259 & $H$   & 84    & 46    & 7.4   & 4 & 75.68 & 0.10  & 1055  & 20.72 & 0.06
\end{longtable}}

{\footnotesize
\begin{longtable}{rrrrrrrrrr}
\caption{\textbf{S0-2 Astrometric Measurements.} Col. 1: Date of observation, Col. 2: Modified-Julian date, Col. 3 \& 4: Separation from center of reference frame in right ascension and declination, Col. 5 \& 6: Separation and position angles transformed from $\Delta$ R.A. and $\Delta$ D.EC.. The full table (46 entries) is provided in Data S2.}\\
\label{tab:astro_measurements}\\
\toprule
Date & MJD & $\Delta$ R.A. & $\sigma_{\Delta R.A.}$ & $\Delta$ Dec.  & $\sigma_{\Delta Dec.}$  & Separation & $\sigma_{sep}$ & Pos. Angle & $\sigma_{angle}$ \\
 (UT) & (Days) & (arcsec) & (arcsec) & (arcsec) & (arcsec) & (arcsec) & (arcsec) & (deg) & (deg) \\
\midrule
\endhead
\midrule
\multicolumn{3}{r}{{Continued on next page}} \\
\midrule
\endfoot

\bottomrule
\endlastfoot
1995-06-10 & 49878.40621 & 0.04354 & 0.00139 & 0.16901 & 0.00206 & 0.17453 & 0.00204 & 345.553 &  0.475 \\ 
 1996-06-26 & 50260.37913 & 0.05331 & 0.00322 & 0.15518 & 0.00353 & 0.16408 & 0.00347 & 341.041 &  1.125 \\ 
 1997-05-14 & 50582.45453 & 0.05881 & 0.00111 & 0.14064 & 0.00141 & 0.15244 & 0.00138 & 337.305 &  0.442 \\ 
 1999-05-02 & 51300.48994 & 0.06826 & 0.00077 & 0.09692 & 0.00055 & 0.11855 & 0.00063 & 324.844 &  0.342 \\ 
 1999-07-24 & 51383.31410 & 0.06869 & 0.00059 & 0.09142 & 0.00065 & 0.11436 & 0.00063 & 323.080 &  0.305 \\ 
 2000-04-21 & 51655.57339 & 0.07057 & 0.00192 & 0.06569 & 0.00374 & 0.09642 & 0.00290 & 312.948 &  1.816 \\ 
 2000-05-19 & 51683.47321 & 0.06812 & 0.00058 & 0.06498 & 0.00079 & 0.09414 & 0.00069 & 313.649 &  0.419 \\ 
 2000-07-19 & 51744.26340 & 0.06575 & 0.00096 & 0.05922 & 0.00165 & 0.08849 & 0.00133 & 312.010 &  0.901 \\ 
 2000-10-18 & 51835.18844 & 0.06465 & 0.00191 & 0.05104 & 0.00167 & 0.08237 & 0.00181 & 308.291 &  1.216 \\ 

 ... & ... & ...  & ... & ... & ... & ...  & ... & ... & ... \\ 
\end{longtable}}

\subsubsection{Speckle imaging}

Speckle imaging consists of very short exposure images ($t\_exp= 0.1 s$) designed to be shorter than the atmospheric turbulence time scale. The individual images are combined and post-processed to produce a deep image for each epoch of observation using a speckle holography technique; this process can reconstruct images that are at the diffraction-limit of the telescope and is described in \cite{Schoedel:2016,Boehle:2016}, with additional improvements described in \cite{Jia:2018}. The speckle images have a field of view of $\sim5^{\prime\prime}\times5^{\prime\prime}$ and have a magnitude limit for detection of stars of $K=16.6$ mag as defined by the brightness above which 90\% of the stars are detected. In total, 27 epochs of speckle data were re-derived and used in this analysis. 

\subsubsection{Adaptive optics imaging}

Two types of adaptive optics imaging data were obtained with NIRC2 on Keck II for this work: i) imaging of the central 10"$\times$10" region around the supermassive black hole, which is used to obtain the astrometric position of S0-2 \cite{Jia:2018}, ii) imaging of a wider 22''$\times$22'' region to obtain the positions of maser stars (near-infrared visible stars which also have accurate astrometry from radio observations) used to construct the reference frame \cite{Sakai:2018}. For the central 10"$\times$10" observation, we use observations from the Kp filter from 2005-2017 and the H filter in 2018. We switched to a shorter wavelength filter in 2018 to avoid astrometric biases due to the infrared source associated with Sgr A* itself, which is very red \cite{Witzel:2018}. The central AO data is 2.4 magnitudes deeper than with speckle imaging, with a limiting magnitude of $Kp = 19.0$ mag averaged across the 10"$\times$10" field. In total we use 18 AO measurements for S0-2, with 9 new measurements from 2017-2018. 

\subsubsection{Reference frame construction}
\label{sec:reference_frame}

All the stars in the NIRC2 field of view are moving; thus establishing a common astrometric reference frame for observations over 20 years is difficult. 
We have developed a Sgr A*-rest frame in which the SMBH is assumed to be at the origin and at rest. This reference frame is constructed from observations of maser-emitting stars at radio wavelengths, where both Sgr A* and the masers are very bright \cite{Reid:2007}, and in the near-infrared, where only the masers have clean counterparts.
By matching the positions and velocities of the seven masers (observed between 2005-2017), we are able to place the near-infrared observations into the Sgr A*-rest frame. Details of this analysis can be found in \cite{Sakai:2018}. Overall, the reference frame localizes the expected Sgr A* position in the near infrared to 0.645 mas (dominated by distortion uncertainties) and is stable at the level of 0.009 mas yr$^{-1}$.

\subsubsection{Alignment of 24 years of data}

We place the speckle and AO astrometric observations of the central region into to a common reference frame using a subset of the sample of secondary astrometric standards described above (see \ref{sec:reference_frame}). 
This process, which places stars such as S0-2 in the Sgr A*-radio rest reference frame, is described detail in \cite{Yelda:2010}, with improvements in the methodology described in \cite{Jia:2018}. 
The resulting positional measurements of S0-2 typically have an uncertainty of 1.1 mas for the speckle points and 0.26 mas for the AO points. The uncertainties on S0-2 are derived from a combination of the positional measurement error, a magnitude-dependent additive error, transformation errors, and errors in the Sgr A*-rest reference frame. 

\subsubsection{Astrometric confusion with stars and Sgr A*}

There are two main causes of source confusion for S0-2, which can bias its astrometric position: confusion with other stars and confusion with Sgr A*, the near-infrared counterpart of the SMBH. Due to the high stellar density of the region, there are several years where another star is very close to S0-2 in projection. This can result in a bias to the measured position of S0-2 if the nearby source is not detected \cite{Ghez:2008}. To avoid this bias, we have excluded the following epochs from the fit: 1998.251, 1998.366, 1998.505, 1998.590, 1998.771, 2006.336, 2006.470, 2006.541, 2007.374, 2007.612, 2013.318, 2013.550, 2014.380, 2014.596, and 2015.606. Confusion with Sgr A* when S0-2 is very close can also have similar astrometric biases. 
We exclude speckle K-band observations in 2002.309, 2002.391, and 2002.547 from the first observed closest approach of S0-2 to Sgr A*. For 2018, we observed the closest approach using the H-band filter instead of the Kp-band filter. Sgr A* is fainter relative to S0-2 in H-band because it is a very red source. This allows us reduce its potential astrometric bias and to exclude fewer epochs for the most recent closest approach. We exclude one epoch, 2018.673, because S0-2 was substantially brighter than it was in other epochs in 2018, indicating that Sgr A* was unusually bright that night and thus likely to bias the position of S0-2.

\subsection{Orbital fit}
The astrometric and radial velocity measurements presented in the previous sections are simultaneously combined to fit an orbit. This section discusses our orbital fit: the orbital modeling, the likelihood, a brief description of the sampler, and the statistical tools for model comparison.

\subsubsection{Modeling of the observations}
The modeling of the astrometric and radial velocity measurements is described in two parts: (i) the orbital dynamics, and (ii) the modeling of the observables.

\subsubsection{Orbital dynamics}
The orbital dynamics of the star S0-2 is determined by integrating the post-Newtonian equations of motion derived from a Schwarzschild metric in harmonic coordinates \cite{Will:2017}
\begin{equation}\label{eq:eom}
    \frac{d^2 \bm R}{dt^2}=-\frac{GM}{R^3}\bm R + \frac{G M }{c^2 R^3} \left(4\frac{G M}{R}-v^2 \right)\bm{R}  + 4 \frac{G M (\bm R \cdot \bm V)}{c^2 R^3} \bm{V} \, ,
\end{equation}
where $\bm R$ is the position of S0-2 with respect to the BH, $\bm V$ its velocity, $R=\left|\bm R\right|$ and $V=\left|\bm V\right|$.
The black hole spin and the quadrupole moment that arise respectively at the 1.5 and 2 post-Newtonian orders in the equations of motion \cite{Will:2017} can be neglected.

The initial conditions used to integrate these equations of motion are computed from a set of orbital parameters: the period $P$, the eccentricity $e$, the time of closest approach $T_0$, the inclination $i$, the argument of periastron $\omega$ and the longitude of ascending node $\Omega$. A detailed description of these orbital elements can be found in \cite{Ghez:2005}. 

Any orbital modeling beyond a pure Newtonian 2-body interaction will lead to a time dependence of these orbital parameters, requiring these parameters to be considered as oscillating elements. In particular, the post-Newtonian perturbation from Eq.~(\ref{eq:eom}) induces a time variation of the orbital elements $P$, $e$, $\omega$ and $T_0$ \cite{Brumberg:1991}. We use the J2000 epoch ($t_\mathrm{J2000}$) as a convention for reporting the initial conditions. The set of 6 orbital parameters are therefore transformed into a cartesian initial position and velocity for the epoch J2000 and used to integrate (forward and backward) the equations of motion. The coordinate system used in this integration is centered on the BH. The Z-axis of the coordinate system is defined by the vector pointing from the Solar System to the Galactic Center, and the X and Y axes are defined such that the X-Y plane is parallel to the plane of the sky, with the X-axis pointing East and the Y-axis pointing North. The units used in our calculations are astronomical units (au) and Julian years, defined as 365.25 Julian days. 

The transformation between the oscillating elements and the initial conditions is a classic Newtonian transformation: the eccentric anomaly $E$ is computed by solving the Kepler equation
\begin{equation}
    E-e\sin E=n\left(t_\mathrm{J2000}-T_0\right) \, ,
\end{equation}
where $n=2\pi/P$. 
This leads to the position and velocity of the star in its orbital plane 
\begin{eqnarray}
    \xi       & =& a\left(\cos E-e\right) \, , \\
    \eta      & =& a\sqrt{1-e^2}\sin E \, ,\\
    \dot \xi  & =& -na \frac{\sin E}{1-e\cos E}\, , \\
    \dot \eta & =&  na \sqrt{1-e^2} \frac{\cos E}{1-e\cos E}\, , 
\end{eqnarray}
where $a=\left(GM/n^2\right)^{1/3}$ is the semi-major axis, $\xi$ is the projection along the major axis, and $\eta$ is the projection along the minor axis. The initial conditions in our coordinate system are obtained after applying the three Euler rotations:
\begin{eqnarray}
    X&= B\xi+G\eta \, , &\qquad V_X= B\dot \xi+G \dot \eta \, ,\\
    Y&= A\xi+F\eta \, ,&\qquad V_Y=  A\dot \xi+F \dot \eta \, ,\\
    Z&= C\xi+H\eta \, ,&\qquad V_Z=  C\dot \xi+H \dot \eta \, ,    \label{eq:vz}
\end{eqnarray}
where $A$, $B$, $C$, $F$, $G$, $H$ are the classic Thiele-Innes constants:
\begin{eqnarray}
 A & = & \cos\Omega  \cos\omega - \sin\Omega  \sin\omega  \cos i   \, , \\
 B  & = & \sin\Omega  \cos\omega + \cos\Omega  \sin\omega  \cos i  \, , \\
 C  & = &  \sin\omega  \sin i  \, ,  \\
 F  & = & -\cos\Omega  \sin\omega - \sin\Omega  \cos\omega  \cos i \, ,  \\
 G  & = & -\sin\Omega  \sin\omega + \cos\Omega  \cos\omega  \cos i \, , \\
 H & = &  \cos\omega \sin i\, .
\end{eqnarray}
These initial conditions are then used to integrate the equations of motion to provide $\bm R(t)=\left(X(t),Y(t),Z(t)\right)$ and $\dot {\bm R}(t)=\left(V_X(t),V_Y(t),V_Z(t)\right)$.

\subsubsection{Modeling of the observables}\label{sec:model}
\paragraph{Light propagation:}
The R\"omer time delay is the first effect that needs to be taken into account when modeling the observables. This time delay is due to the fact that the speed of light is finite and thus the signal from the star takes a certain amount of time to propagate through S0-2's orbit in the Z-direction, producing a modulation of the propagation time of the signal between S0-2 and Earth. The light propagation time is obtained by solving the equation (see e.g. \cite{Damour:1986})
\begin{equation}
    t_\textrm{\tiny obs}-t_\textrm{\tiny em}=\frac{Z(t_\textrm{\tiny em})}{c} \, , 
\end{equation}
where $t_\textrm{\tiny obs}$ is the epoch of observation, $t_\textrm{\tiny em}$ is the epoch of emission, and $Z(t)$ is the component of S0-2's orbit parallel to the line of sight. In practice, this equation can efficiently be solved iteratively (e.g. \cite{Hees:2014}) using the following iteration scheme: $t^{(i+1)}_\textrm{\tiny em}=t_\textrm{\tiny obs}-Z(t^{(i)}_\textrm{\tiny em})/c$, starting with $t^{(0)}_\textrm{\tiny em}=t_\textrm{\tiny obs}$. For our purposes, only one iteration is required:
\begin{equation}\label{eq:romer}
    t_\textrm{\tiny em}= t_\textrm{\tiny obs}-\frac{Z(t_\textrm{\tiny obs})}{c} \, .
\end{equation}
This leads to a modulation of the light propagation time between -0.5 days at closest approach and 7.5 days at apoastron. A second iteration would lead to a correction of $<$ 20 minutes, which is negligible at our level of accuracy. We also neglect the Shapiro time delay, which yields a maximum correction of $\sim$5 minutes.

\paragraph{Astrometric observable:}
The astrometric observations are given in terms of angular positions $x$ and $y$, which are modeled as
\begin{eqnarray}
    x(t_\textrm{\tiny obs}) &=  -\frac{X(t_\textrm{\tiny em})}{R_0} + x_0 + v_{x_0}\left(t_\textrm{\tiny obs}-t_\textrm{\tiny J2000}\right)\, ,\\
    y(t_\textrm{\tiny obs}) &=  \phantom{-}\frac{Y(t_\textrm{\tiny em})}{R_0} + y_0 + v_{y_0}\left(t_\textrm{\tiny obs}-t_\textrm{\tiny J2000}\right)\, ,   
\end{eqnarray}
where $R_0$ is the line-of-sight distance to the Galactic center and $x_0$, $y_0$, $v_{x_0}$ and $v_{y_0}$ model a 2D offset and linear drift of the gravitational center of mass with respect to the center of our reference frame. These four parameters are included to model systematics that appear at the level of the construction of the reference frame \cite{Sakai:2018}. We neglect the gravitational light deflection of the SMBH, which produces an effect on the order of 20 $\mu$as on the astrometry at closest approach \cite{Grould:2017}, smaller than our  observational uncertainty.

\paragraph{Spectroscopic observable:}
The spectroscopic observable $\mathcal V$ is given by Eq.~(\ref{eq:spectro}),
\begin{equation}
    \mathcal V=\frac{\lambda_\textrm{\tiny obs}-\lambda_\textrm{\tiny em}}{\lambda_\textrm{\tiny em}}c=c\left(\frac{\nu_\textrm{\tiny em}}{\nu_\textrm{\tiny obs}}-1\right)\, .
\end{equation}
Using a regular post-Newtonian expression for the frequency shift (e.g. \cite{Blanchet:2001,Hees:2014}), the spectroscopic observable becomes
\begin{equation}\label{eq:S_c}
    \frac{\mathcal V}{c}= \frac{1-\frac{U_\textrm{\tiny obs}}{c^2}-\frac{V^2_\textrm{\tiny obs}}{2c^2}}{1-\frac{U_\textrm{\tiny em}}{c^2}-\frac{V^2_\textrm{\tiny em}}{2c^2}} \times \frac{1+\frac{1}{c}\bm N\cdot\bm V_\textrm{\tiny em}}{1+\frac{1}{c}\bm N \cdot \bm V_\textrm{\tiny obs}}-1+\mathcal O(1/c^3)\, , 
\end{equation}
where $U_\textrm{\tiny em/obs}$ is the gravitational potential at the emission/observation of the light signal, $V_\mathrm{ em/obs}=\left|\bm V_\mathrm{ em/obs}\right|$ is the norm of the velocity of the emitter/observer at the emission/reception of the signal, and $\bm N=\frac{\bm X_\textrm{\tiny em}-\bm X_\textrm{\tiny obs}}{\left|\bm X_\textrm{\tiny em}-\bm X_\textrm{\tiny obs}\right|}$ is a unit vector pointing in the direction of the line-of-sight. In Eq.~(\ref{eq:S_c}),  the derivative of the Shapiro time delay is neglected (terms of the order of $\mathcal O(1/c^3)$). The maximal contribution from this term arises at closest approach and remains below 5 km/s \cite{Grould:2017}, below the current measurement uncertainty.

Eq.~(\ref{eq:S_c}) can also be written as    
\begin{eqnarray}
    RV(t_\textrm{\tiny obs})= \mathcal V +\bm N \cdot \bm V_\mathrm{obs}&= \bm N \cdot \bm V_\mathrm{em}+\frac{1}{c}\left(\bm N \cdot \bm V_\mathrm{obs}\right)^2-\frac{1}{c}(\bm N \cdot \bm V_\mathrm{obs})(\bm N\cdot\bm V_\mathrm{em})\\
    &+\frac{U_\mathrm{em}+V^2_\mathrm{em}/2}{c}-\frac{U_\mathrm{obs}+V^2_\mathrm{obs}/2}{c}+\mathcal O(1/c^2)\, ,\nonumber
\end{eqnarray}
where $RV$ is the observed radial velocity corrected for the Velocity of the Local Standard of Rest ($VLSR=\bm N\cdot\bm V_\mathrm{obs}$). All $\bm V_\mathrm{em}$ are evaluated at $t=t_\mathrm{em}$, such that the R\"omer time delay is taken into account. The first term on the right hand side of this equation $\bm N\cdot\bm V_\mathrm{em}$ is the standard Newtonian line-of-sight velocity of the star S0-2, which is $V_Z(t)$ from Eq.~(\ref{eq:vz}). The second term $\left(\bm N\cdot\bm V_\mathrm{obs}\right)^2/c$ is a second order term proportional to the square of the VLSR. The contribution from that term to the radial velocity remains below $\sim$ 30 m/s so is neglected. The $(\bm N\cdot\bm V_\mathrm{obs})(\bm N\cdot\bm V_\mathrm{em})/c$ term is a cross term that remains below 0.1 km/s and can  be neglected as well. The last terms comprise the relativistic contributions to the redshift. The $U_\mathrm{obs}/c$ term is the gravitational redshift contribution related to the observer and $V^2_\mathrm{obs}/2c$ is the transverse Doppler shift predicted by special relativity. In our case, these two terms comprise several contributions, the main ones being from the Galactic potential, from the potential of the Sun, of the Earth and of the Moon. The order of magnitude of all these terms is below $\sim$ 0.1 km/s \cite{Alexander:2005} and can safely be neglected as well. The last terms are the second-order  transverse Doppler effect from special relativity and the gravitational redshift from the star S0-2. The combination of these two terms is the signal we are seeking to measure.

Dropping all negligible terms, we model the RV as:
\begin{equation}
    RV(t_\mathrm{obs})=V_Z(t_\mathrm{em})+\frac{V^2_\mathrm{em}(t_\mathrm{em})}{2c} + \frac{GM}{c R(t_\mathrm{em})}+v_{z_0}\, , 
\end{equation}
where $v_{z_0}$ is a constant velocity offset that accounts for possible systematic effects in the radial velocity measurement or in the VLSR correction. The second term of this equation is the relativistic transverse Doppler shift predicted by special relativity, while the third term is the gravitational redshift predicted by GR.  The signatures produced by the second and third contributions are highly correlated.  Indeed, using a Newtonian orbital model, one gets
\begin{eqnarray}
	\left[RV\right]_\mathrm{spec} &= \frac{V^2_\mathrm{em}(t_\mathrm{em})}{2c}\approx \frac{GM}{cR(t_\mathrm{em})}-\frac{GM}{2ac}=\left[RV\right]_\mathrm{grav}-\frac{GM}{2ac}\, ,
\end{eqnarray}
where $a$ is S0-2's semi-major axis, $\left[RV\right]_\mathrm{spec}$ is the contribution from special relativity to the RV and $\left[RV\right]_\mathrm{grav}$ is the contribution from general relativity, i.e. the gravitational redshift. Since an offset $v_{z_0}$ is fitted to our data, the constant term $\frac{GM}{2ac}$ is unobservable and both signals from special relativity and from the gravitational redshift are completely degenerate and the data is only sensitive to the sum of the two. 

To quantify possible deviations from the predicted relativistic signal, we introduce a dimensionless parameter $\Upsilon$ whose value is 0 for a purely Newtonian model and 1 in GR (see \cite{Zucker:2006}). The expression for the radial velocity used in our orbital fit is given by
\begin{equation}\label{eq:RV1}
    RV(t_\mathrm{obs})=V_Z(t_\mathrm{em})+\Upsilon \left[\frac{V^2_\mathrm{em}(t_\mathrm{em})}{2c} + \frac{GM}{cR(t_\mathrm{em})}\right] +v_{z_0}\, .
\end{equation}

Alternatively, we can assume that special relativity is correct and only search for a deviation from the gravitational redshift prediction. Such a deviation is usually parametrized by a parameter $\alpha$ whose value is 0 in GR (see Eq.~(6) from \cite{Will:2014}) and the radial velocity are then modeled as
\begin{equation}\label{eq:RV2}
    RV(t_\mathrm{obs})=V_Z(t_\mathrm{em})+\frac{V^2_\mathrm{em}(t_\mathrm{em})}{2c} + (1+\alpha) \frac{GM}{cR(t_\mathrm{em})} +v_{z_0}\, .
\end{equation}
All our estimations of the $\Upsilon$ parameter can be translated into an estimation of the $\alpha$ parameter through: 
\begin{equation}\label{eq:alpha}
    \alpha=2(\Upsilon-1)
\end{equation}
and the uncertainty on $\alpha$ is given by $\sigma_\alpha=2\sigma_\Upsilon$.

\subsubsection{Summary of the model}
To summarize, our modeling is based on the integration of the first post-Newtonian equations of motion, includes the R\"omer time delay and the 2nd order transverse Doppler shift  as well as the gravitational redshift. This model is parametrized by 6 orbital parameters for S0-2 and 8 global parameters: the gravitational parameter $GM$ of the BH, the distance to the Galactic Center $R_0$, the 2-D position and velocity of the BH $x_0$, $y_0$, $v_{x_0}$, $v_{y_0}$, an offset for the radial velocity $v_{z_0}$ and a parameter that encodes deviations from General Relativity at the level of the redshift $\Upsilon$.

The observable depends on the SMBH gravitational parameter $GM$, which is the parameter that is fitted (and not the mass directly). We express the SMBH $GM$  in units of the Sun's gravitational parameter $GM_\odot$ where $M_\odot$ is the mass of the Sun and whose value is precisely measured from planetary ephemerides (see table 8 from \cite{Folkner:2014}.)

\subsection{Bayesian samplers and software}\label{sec:bayesian}

Parameter exploration was done using the Nested Sampling package MultiNest \cite{Feroz:2008,Feroz:2009}. The analysis was also done independently using the STAN modeling language and NUTS sampler \cite{STAN}. The MultiNest sampler used in this analysis was preferred because it allowed for efficient calculation of the Bayesian evidence and has been successfully used in previous GC orbit analyses \cite{Meyer:2012,Boehle:2016,Hees:2017}.  The STAN implementation of our analysis was primarily used to confirm our results.  

\subsection{Model selection and information criteria}
We base model selection on two criteria. The first is the Bayesian evidence
\begin{equation}
\mathcal{E}_k \equiv \mathcal{P}(\{d_j\} \mid \mathrm{Model}_k) = \int \mathcal{P}_k(\{d_j\} \mid \theta)\mathcal{P}_k(\theta) \mathrm{d}\theta
\end{equation}
that is a direct output of the nested sampling algorithm.  Here, the subscript $k$ denotes that the probability is conditioned on $\mathrm{Model}_k$ being true ($\mathcal{P}_k(\dots \mid \dots) \equiv \mathcal{P}(\dots \mid \dots, \mathrm{Model}_k)$), $\{d_j\}$ represents the dataset, $\theta$ the estimated parameters, $\mathcal{P}_k(\{d_j\} \mid \theta)$ is the likelihood and $\mathcal{P}_k(\theta)$ the prior. The evidence is used to infer the probability of the different models via Bayes theorem ($\mathcal{P}(\mathrm{Model}_k \mid \{d_j\}) \propto \mathcal{E}_k \mathcal{P}(\mathrm{Model}_k)$) assuming that the a priori probability, $\mathcal{P}(\mathrm{Model}_k)$, is even over all the models under consideration. Sometimes, this assumption can artificially bias probabilities toward fine-tuned models \cite{Brewer09, Jaynes03} and may give inconsistent results when the true model is not included in the comparison \cite{Yao17}. The ratio of evidences under the assumption of uniform priors is known as the ``Bayes factor'' or odd ratio: $\mathcal{E}_0/\mathcal{E}_1\propto \mathcal{P}(\mathrm{Model}_0 \mid \{d_j\})/\mathcal{P}(\mathrm{Model}_1 \mid \{d_j\})$. The logarithm of the ratio of evidences is often compared to roughly judge the strength of one model over another with a log ratio ($\log(\mathcal{E}_0/\mathcal{E}_1)$) value under $1$ considered "barely mentioning", $1$ to $2.5$ being ``positive'', $2.5$ to $5$ having ``strong evidence'', and greater than $5$ having ``very strong evidence'' of one model over another \cite{KassRaftery95,jeffreys61, Jeffreys35}.

We also use the expected log probability density (elpd) as an additional model selection criteria \cite{Gelman14}:
\begin{equation}
\label{eq:elpd}
\mathrm{elpd}^{(k)} \equiv \int \mathcal{P}(d^{*}) \log \mathcal{P}_k(d^{*} \mid \{d_j\}) \mathrm{d} d^{*}
\end{equation}
where $\mathcal{P}_k(d^{*} \mid \{d_i\}) = \int \mathcal{P}_k(d^{*} \mid \theta)\mathcal{P}_k(\theta \mid \{d_j\}) \mathrm{d}\theta$ is the probability of observing $d^{*}$ assuming $\mathrm{Model}_k$ and $\mathcal{P}(d^{*})$ is the actual, unknown, data distribution and $d^*$ being a measurement that does not belong to $\{d_j\}$. Several inference criteria have been developed based on approximations to Equation \ref{eq:elpd}, such as the Akaike Information Criteria (AIC) \cite{Akaike73}, Deviance information criterion (DIC) \cite{DIC}, and Watanabe Akaike information criterion (WAIC) \cite{Watanabe07}. We use leave-one-out cross validation,
\begin{equation}
\label{eq:loo-cv}
\widehat{\mathrm{elpd}}_{\mathrm{loo-cv}}^{(k)} \equiv \sum_i \log \mathcal{P}_k(d_i \mid \{d_j\}_{j \neq i})
\end{equation}
which has been shown to be, asymptotically, a good approximation to the elpd \cite{Watanbe10}. In principle, the probability of observations $d_i$ given the dataset that excludes $d_i$ ($\mathcal{P}_k(d_i \mid \{d_j\}_{j \neq i}) = \int \mathrm{d}\theta \mathcal{P}_k(d_i \mid \{d_j\}_{j \neq i}, \theta)  \mathcal{P}_k(\theta \mid \{d_j\}_{j \neq i})$) needs to be reevaluated for each data point since the posterior probability, $\mathcal{P}_k(\theta \mid \{d_j\}_{j \neq i})$, is unique for each data set, $\{d_j\}_{j \neq i}$.  We avoid this by reweighting the posterior probability given the full data set ($\mathcal{P}_k(\theta \mid \{d_j\})$) to match the distribution $\mathcal{P}_k(d_i \mid \{d_j\}_{j \neq i})$.  Using this approximation, Equation \ref{eq:loo-cv} becomes \cite{Vehtari17}:
\begin{equation}
\label{eq:approx-loo-cv}
\widehat{\mathrm{elpd}}_{\mathrm{loo-cv}}^{(k)} \approx -\sum_i \log \left[{\sum_l \frac{w_l}{\mathcal{P}_k(d_i \mid \{d_j\}_{j \neq i}, \theta_l)}}\right]
\end{equation}
where $\{\theta_k\}$ and $\{w_k\}$ are deviates, and corresponding weights, of the posterior $\mathcal{P}_k(\theta \mid \{d_j\})$.  The inverse sum in Equation \ref{eq:approx-loo-cv} is usually numerically unstable because infrequent deviates will correspond to low $\mathcal{P}_k(d_i \mid \{d_j\}_{j \neq i}, \theta_k)$ values and thus be weighted higher \cite{Vehtari17}.  This is avoided if a nested sampling algorithm, such as MultiNest, is used to sample from the posterior.  In this case, the inverse divergence behavior is avoided because nested sampling weights are proportional to the likelihood, $\mathcal{P}_k(\{d_j\} \mid \theta)$ \cite{Feroz:2008,Feroz:2009}. 

As a summary, in this analysis a selection for a more complex model is decided when both the difference of the evidence $\mathcal E$ and the elpd are larger than 2.5 in favor of the more complex model.

\subsubsection{Likelihoods}
We consider several likelihoods in our analysis to capture different sources of errors.

For the first one, we assume the astrometric positions ($\{x_i\}$, $\{y_i\}$) and radial velocity ($\{\mathrm{RV}_i\}$) to be normally distributed about the astrometric ($\{x(t_{\mathrm{astro}, i})\}$, $\{y(t_{\mathrm{astro}, i})\}$) and radial velocity ($\mathrm{RV}(t_{\mathrm{RV}, i})$) predicted values and with their dispersions equal to their uncertainties:
\begin{eqnarray}
x_i & \sim & \mathcal{N}(x(t_{\mathrm{astro}, i}), \sigma_{x_i}^2) \label{eq:like_1a}\\
y_i & \sim & \mathcal{N}(y(t_{\mathrm{astro}, i}), \sigma_{y_i}^2) \\
\mathrm{RV}_i & \sim & \mathcal{N}(\mathrm{RV}(t_{\mathrm{RV}, i}), \sigma_{\mathrm{RV}_i}^2)\label{eq:like_1b}
\end{eqnarray}
Here we define $x \sim \mathcal{N}(\mu, \sigma)$ to mean that variable $x$ is normally distributed about $\mu$ with a dispersion $\sigma$.

To determine whether we have under-estimated the uncertainties, we also explore likelihoods that include an additional additive error for the astrometry ($\sigma_{\mathrm{astro}}$) and the radial velocities ($\sigma_{\mathrm{inst}}$):
\begin{eqnarray}
x_i & \sim & \mathcal{N}(x(t_{\mathrm{astro}, i}), \sigma_{x_i}^2 + \sigma_{\mathrm{astro}}^2) \label{eq:like_2a}\\
y_i & \sim & \mathcal{N}(y(t_{\mathrm{astro}, i}), \sigma_{y_i}^2 + \sigma_{\mathrm{astro}}^2) \\
\mathrm{RV}_i & \sim & \mathcal{N}(\mathrm{RV}(t_{\mathrm{RV}, i}), \sigma_{\mathrm{RV}_i}^2 + \sigma_{\mathrm{instr}}^2)\label{eq:like_2b}
\end{eqnarray}
(where $\sigma_{\mathrm{instr}}$ is different of each instrument: NIRSPEC, NIRC 2, Osiris Kbb, Osiris Kn3, NIFS, SINFONI and IRCS). When using this likelihood, the additional systematic uncertainties are fitted simultaneously with the model parameters.

To account for potential correlations in the uncertainty of astrometric measurements, we also consider a likelihood with separate covariance matrices for the astrometric positions (${\bm x} \equiv \{x_i\}$ and ${\bm y} \equiv \{y_i\}$) corresponding to times ${\bm t}_{\mathrm{astro}} \equiv \{t_{\mathrm{astro}, i}\}$:
\begin{eqnarray}
{\bm x} & \sim & \mathcal{N}(x({\bm t}_{\mathrm{astro}}), {\bm \Sigma}_{x}) \label{eq:like_3a}\\
{\bm y} & \sim & \mathcal{N}(y({\bm t}_{\mathrm{astro}}), {\bm \Sigma}_{y}) \\
\mathrm{RV}_i & \sim & \mathcal{N}(\mathrm{RV}(t_{\mathrm{RV}, i}), \sigma_{\mathrm{RV}_i}^2)\label{eq:like_3b}
\end{eqnarray}
where  ${\bm x} \sim \mathcal{N}({\bm \mu}, {\bm \Sigma})$ denotes that the vector ${\bm x}$ is normally distributed around the vector ${\bm \mu}$ with a covariance matrix of ${\bm \Sigma}$.  We model the covariance matrices by a single correlation matrix, ${\bm \rho}$, defined by $\left[{\bm \Sigma_x}\right]_{i j} = \sigma_{x_i} \sigma_{x_j} {\bm \rho}_{i j}$ and $\left[{\bm \Sigma_y}\right]_{i j} = \sigma_{y_i} \sigma_{y_j} {\bm \rho}_{i j}$ where the covariance matrix is given by
\begin{eqnarray}
\left[{\bm \rho}\right]_{i j} & = & (1-p) \delta_{i j} + p \exp{\left[-\frac{|d_{i j}|}{\Lambda}\right]} \label{eq:cor_d}
\end{eqnarray}
where $d_{ij}$ is the 2D projected distance between point $i$ and point $j$ ($d_{ij} = \sqrt{(x_i - x_j)^2 + (y_i - y_j)^2}$).  This matrix introduces a correlation length scale $\Lambda$ characteristic of the correlation and a mixing parameter $p$, both of which will be fitted simultaneously with the model parameters. 

\subsubsection{Priors}
In the orbit fitting, we used uniform priors on all fitted parameters. While this choice is common, it has been shown that uniform priors  can potentially bias estimated parameters \cite{Lucy:2014,Kosmo:2018}. With this in mind, we used simulated data to understand the impact of our fitting procedure by identifying possible biases in the estimated parameters and assessing the accuracy of the confidence intervals obtained in our analysis.

\begin{figure}
\begin{center}
\includegraphics[scale=0.75]{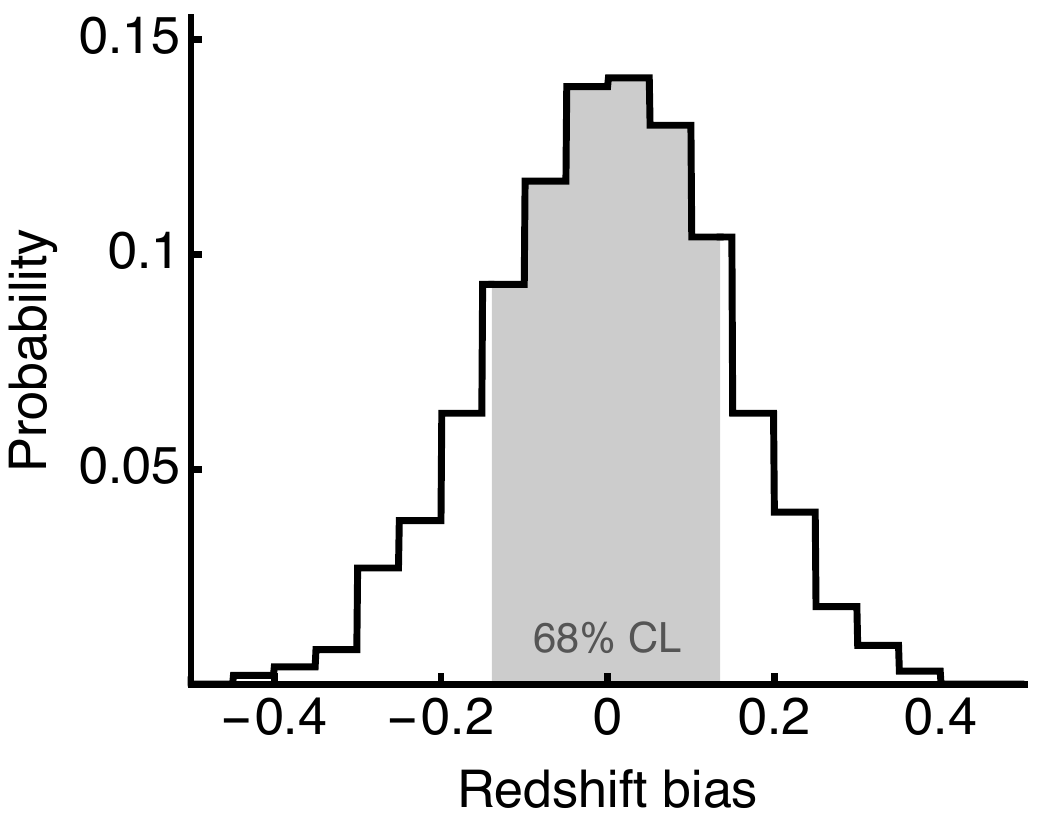}
\end{center}
\caption{Distribution of the bias values on the redshift parameter. This distribution is obtained by considering fits of 1000 mock datasets generated using the same epochs of measurements as our real dataset. The specific bias factor is defined as the difference between the estimated redshift parameter and the input redshift parameter. This distribution shows that the procedure used in our fit does not induce any substantial bias on the estimated redshift parameter.
\label{fig:bias}
}
\end{figure}

Here, we summarize the methodology to test for fitting biases (for a full description, see \cite{Kosmo:2018}). We generated 1000 mock datasets by simulating measurements at epochs corresponding to our observations. Each mock dataset was generated by drawing a random measurement from a normal distribution distributed about an assumed true value, with a dispersion taken to be the actual measurement error at that epoch. We then fit these 1000 mock datasets using the same procedure that is used to fit the real data. For these 1000 fits, we computed the redshift bias, or the difference between the estimated redshift parameter and the input redshift parameter. The distribution of the 1000 bias values is shown in Fig.~\ref{fig:bias}. This distribution is centered around zero and indicates the bias on the redshift parameter is negligible.

We next consider the accuracy of our confidence intervals. In the classical definition of a confidence interval, for a sufficiently large number of experiments, the confidence interval inferred from each experiment will contain, or cover, the universally ``true'' value a prescribed fraction of the time (confidence level $\times$ 100\%) \cite{Neyman:1937}. For example, by this definition, given 100 possible observed (or randomly drawn) datasets, a 68\% confidence interval requires that about 68 out of 100 fits produce a confidence interval that covers the true value. Thus, the statistical efficiency, defined as the ratio of effective coverage (the experimentally determined percentage of datasets in which the inferred confidence or credible interval covers the true value) to stated coverage (68\% for a 1-$\sigma$ confidence interval), is a powerful performance diagnostic that can be used to investigate the accuracy of calculated confidence or credible intervals \cite{Kosmo:2018}. By definition, a statistical efficiency of one would indicate exact coverage. The statistical efficiency for the redshift parameter determined from the 1000 mock datasets is $1.002 \pm 0.02$ (or $0.682 \pm 0.015$ coverage for a 68 \% confidence interval). This shows that the confidence interval on the redshift parameter we derive represents a robust estimate of the statistical uncertainty.

The bias analysis shows that using uniform priors in this analysis does not lead to any substantial bias in the estimation of the redshift parameter. In addition, the statistical efficiency demonstrates that the confidence intervals used in this analysis are well defined and have close to exact coverage.

\subsection{Accounting for instrumental systematic effects on the RV measurements in the orbital fit}
\label{sec:systematics}
We assessed two potential sources of systematics in the RV measurements: (i) systematic RV offsets between the different instruments and (ii) possible additive uncertainties for the different instruments.

The RV measurements used in our analysis have been performed by seven different instruments: NIRSPEC, NIRC2, OSIRIS Kbb, OSIRIS Kn3 at the Keck Observatory, NIFS with Gemini, SINFONI at the VLT and with IRCS at SUBARU. Since the different instruments work differently and are sensitive to different systematics, it is important to cross-validate the different datasets. In order to do so, we added 7 parameters to our orbital fit: one offset per instrument. We performed 8 different fits: one reference fit with all the offsets forced to 0 and 7 where we fitted for one offset at a time simultaneously with the model parameters. We then used the model selection criteria described in section~\ref{sec:bayesian} to assess which offsets, if any, are significant. The results from these fits are presented in Table~\ref{tab:offset_RV}. Using the threshold on the information criteria presented in section~\ref{sec:bayesian} ($\Delta \mathcal E >5$ and $\Delta$ elpd $>$5), we conclude that the NIRC2 dataset presents a significant offset. This conclusion, which is obtained based purely  on statistical arguments, is reinforced by the fact that NIRC2 measurements are affected by fringing, as explained with more details in Section~\ref{sec:NIRC2RV}. For these two reasons, in all our orbital fits, we include an offset for the NIRC2 dataset that is fitted simultaneously with the other model parameters.

\begin{table}[htb]
 \caption{Results from orbital fits that include an offset per spectrograph instrument. The third and fourth columns are the differences between the log-evidence and the elpd of the fit with the offset and the reference fit where no offset is considered. A more complex model is adopted when both differences are larger than 5. The measurements show strong evidence for a NIRC 2 RV offset.}
	\label{tab:offset_RV} 
	\centering
	\begin{tabular}{c| c  | c c}
	    Instrument  & Estimated    & $\Delta \log \mathcal E$   & $\Delta$ elpd   \\
        with offset &offset [km/s]&  \\
	\hline
	 NIRSPEC & -2.4  $\pm$ 100.1 & -1.6 & -3.8 \\
	 NIRC 2  & 80.6  $\pm$  19.1 &  \textbf{6.8} & \textbf{8.4}  \\
	 Kbb     & -8.9  $\pm$   7.2 & -2.5 & 0.1 \\
	 Kn3     &  4.0  $\pm$   4.4 & -3.3 & -1.5 \\	
	 VLT     &-16.6  $\pm$   6.5 & 0.1 & 2.9 \\		  
	 NIFS     & 17.1  $\pm$   8.3 & -1.0 & 1.9 \\
     SUBARU  &  -11.0 $\pm$ 9.1 &  -1.9 & -1.3
	\end{tabular}
\end{table}

In addition to considering instrumental offsets in the RV dataset, we have also considered additional possible instrumental systematic uncertainties. We use the likelihood defined by Eqs.~(\ref{eq:like_2a}-\ref{eq:like_2b}) and to introduce 7 additional parameters: one systematic uncertainty for each instrument. The methodology is the same as the one followed for the instrumental offsets: we performed different fits by considering one additional systematic uncertainty  at a time and assessed the significance of each instrument systematic uncertainty by using our model selection criteria described in Section~\ref{sec:bayesian}. The results of these fits are presented in Table~\ref{tab:sigma_RV}. This analysis does not show any evidence for any additional systematic uncertainty. This means that our RV uncertainties are not underestimated and in the following, no additional systematic uncertainty in the RV is included.

\begin{table}[htb]
 \caption{Same as Table \ref{tab:offset_RV}, but for additional systematic uncertainty for each spectrograph instrument. No additional RV systematic uncertainty is suggested by the measurements.}
	\label{tab:sigma_RV} 
	\centering
	\begin{tabular}{c |  c| c c}
	    Instrument with  & Estimated    & $\Delta \log Ev$   & $\Delta elpd$ \\
        sys. uncertainty & uncertainty [km/s] &  \\
	\hline
	 NIRSPEC & 84.9 $\pm$ 56.1   & 0.1  & -0.2   \\
	 NIRC 2  & 91.0 $\pm$ 92.2 & -1.4  & -2.8   \\
	 Kbb     &  12.0 $\pm$ 9.6  & -2.3  &  -0.9   \\
	 Kn3     &   6.4 $\pm$ 4.1  & -3.0  &  0.0   \\	
	 VLT    &   6.6 $\pm$ 5.0  & -3.2  &  -0.3  \\		  
	 NIFS     &  11.5 $\pm$ 9.3  & -2.5  & -0.4 \\
     SUBARU   &  36.6 $\pm$ 19.0  & -0.1  & 0.8 \\
	\end{tabular}
\end{table}

\subsection{Accounting for systematic effects on the astrometric measurements and analysis of correlation within the astrometric dataset at the level of the orbital fit}\label{sec:dom}
At the level of the orbital fitting, we have assessed our astrometric dataset by considering three effects: (i) an additional possible systematic uncertainty on the astrometric measurements (ii) a possible offset between the different instruments and (iii) the correlation within the astrometric measurements.

\subsubsection{Possible additional systematic uncertainties for the astrometric measurements}
For the study of the systematic uncertainty, we followed the same methodology as the one presented in the previous section for the RV measurements. We used the likelihood defined by Eqs.~(\ref{eq:like_2a}-\ref{eq:like_2b}) and considered two additional systematic uncertainties in the orbital fit: one for the Speckle measurements and one for the AO measurements. We compared three different fits: a reference fit where no additional systematic uncertainty is considered, one fit where the Speckle systematic uncertainty is fitted and one fit where the AO systematic uncertainty is fitted. We used our model selection criteria described in section \ref{sec:bayesian} to identify if these additional systematic uncertainties are relevant. The results of these fits are presented in Table~\ref{tab:sigma_astro}. The difference in our information criteria are below the threshold presented in section \ref{sec:bayesian} which leads to the conclusion that no additional systematic uncertainty needs to be added to our astrometric dataset.

\begin{table}[htb]
 \caption{Same as Table \ref{tab:offset_RV}, but for systematic uncertainty for each type of astrometric measurements. No additional astrometric systematic uncertainty (or rescaling of the two datasets) is suggested by the measurements.}
	\label{tab:sigma_astro} 
	\centering
	\begin{tabular}{c| c | c c}
	\hline
	    Instrument with  & Estimated   & $\Delta \log Ev$   & $\Delta elpd$ \\
    sys. uncertainty &uncertainty [mas] & \\\hline
	 Speckle & 0.5 $\pm$ 0.2 & -0.7 & 1.3 \\
	 AO & 0.07 $\pm$ 0.05 &-3.4 &-0.3 
	\end{tabular}
\end{table}

\subsubsection{Possible additional offset between the different astrometric instruments}
We also assessed the possibility of an offset between the Speckle and AO measurements. We added a 2D offset to the fit and compared the log-evidence and elpd of a fit that includes these additional 2 parameters to a fit where no offset is considered. Tab.~\ref{tab:offset_astro} shows the result of these fits and shows that no additional systematic offset between the two instruments need to be considered

\begin{table}[htb]
 \caption{Same as Table \ref{tab:offset_RV}, but for a 2-D offset between the Speckle and AO measurements. No astrometric offset is suggested by the measurements.}
	\label{tab:offset_astro} 
	\centering
	\begin{tabular}{c| c c | c c}
	      & Estimated  & & $\Delta \log Ev$   & $\Delta elpd$ \\
        &X  offset  [mas]&Y offset  [mas]  & \\\hline
	 x  & 0.49 $\pm$ 0.30 & -0.98 $\pm$ 0.32 & 0.03 &2.0   \\
	\end{tabular}
\end{table}

\subsubsection{Correlation within the astrometric measurements}
The second effect assessed regarding our astrometric dataset is the presence of correlation in the measurements. Such correlations are expected because of detected and undetected source confusions and because of the presence of correlated systematic effects that arise during the construction of the reference frame. Assuming all the measurements to be statistically independent will therefore lead to uncertainties that are overoptimistic for the estimated parameters. In order to study correlations within our dataset, we have used two different methods: (i) we modeled the correlation and we fitted for the related parameters simultaneously with the model parameters and (ii) we used a Monte Carlo approach where we kept only one measurement per correlation length scale. 

\begin{figure}
\begin{center}
\includegraphics[scale=0.5]{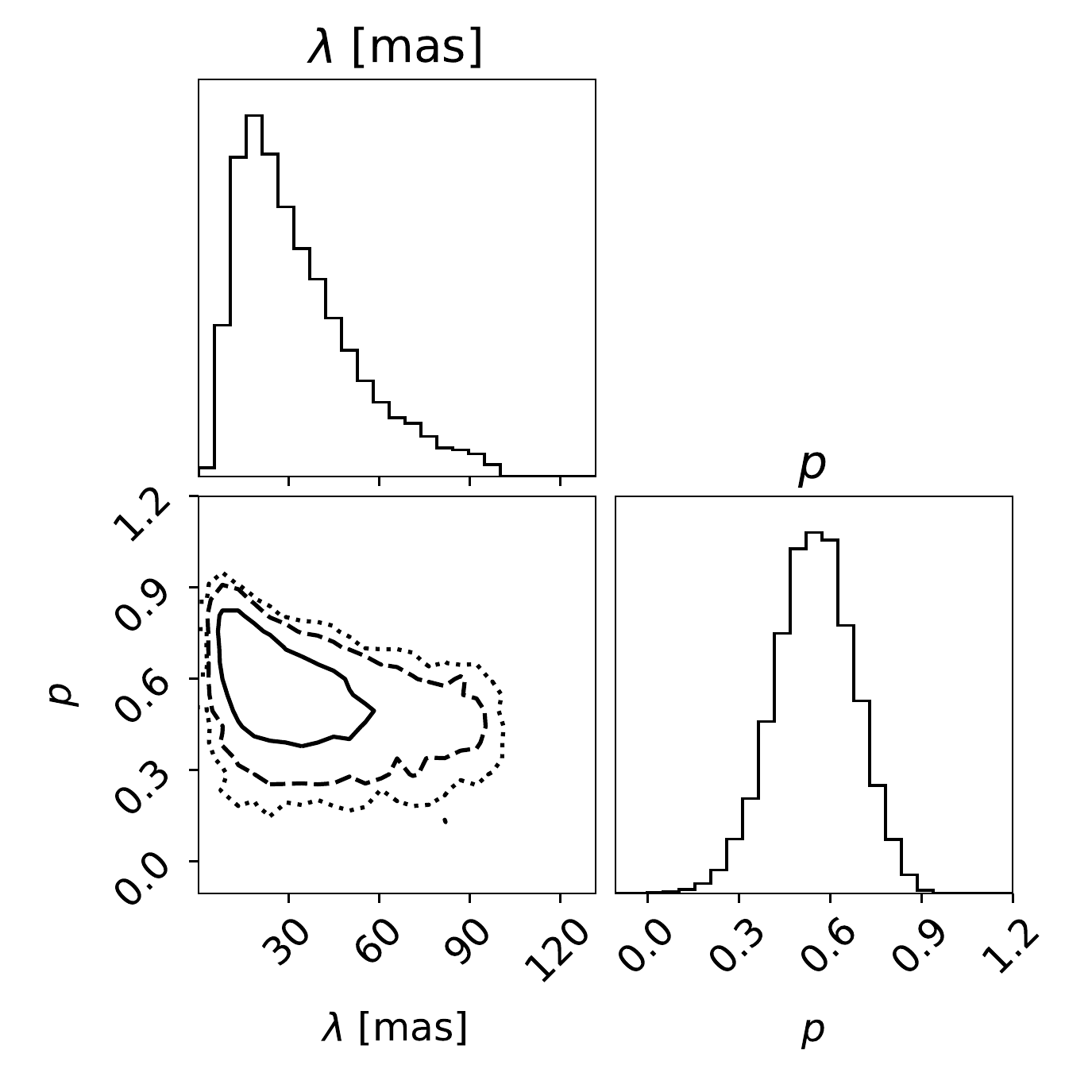}
\end{center}
\caption{Posterior probability distribution function for the two parameters that parametrized the correlation matrix: $\Lambda$ a correlation length scale and $p$ a mixing parameter. The marginalized 68\% confidence interval for the correlation length scale is $\Lambda=28^{+24.6}_{-13.6}$ mas, which corresponds roughly to half the Keck diffraction limit.
\label{fig:astro_post_cor}
}
\end{figure}

For the first method, we used the likelihood defined by Eqs.~(\ref{eq:like_3a}-\ref{eq:like_3b}). This likelihood introduces a correlation matrix between the astrometric measurements. We considered the form for the correlation matrix given by Eq.~(\ref{eq:cor_d}), which assumes that the correlation within the astrometric dataset is related to the 2-D projected distance between measurements.  This likelihood introduces two additional parameters that are fitted simultaneously with the model parameters: a correlation length scale $\Lambda$ and a mixing parameter $p$. We used our model selection criteria described in Section~\ref{sec:bayesian} to compare orbital fits that consider this correlation with a fit that uses the regular uncorrelated likelihood and the difference in the log-evidence is of 7.3 while the difference of the elpd is of 13.5, showing a strong  evidence in favor of the model that includes correlations.  In this analysis, we therefore use this correlation matrix and fit for the two parameters $\Lambda$ and $p$ simultaneously with the model parameters. Fig.~\ref{fig:astro_post_cor} shows the posterior distribution for these two parameters. The fitted correlation length scale is around 30 mas, which corresponds to half the Keck diffraction limit.

\begin{figure}
\begin{center}
\includegraphics[scale=0.5]{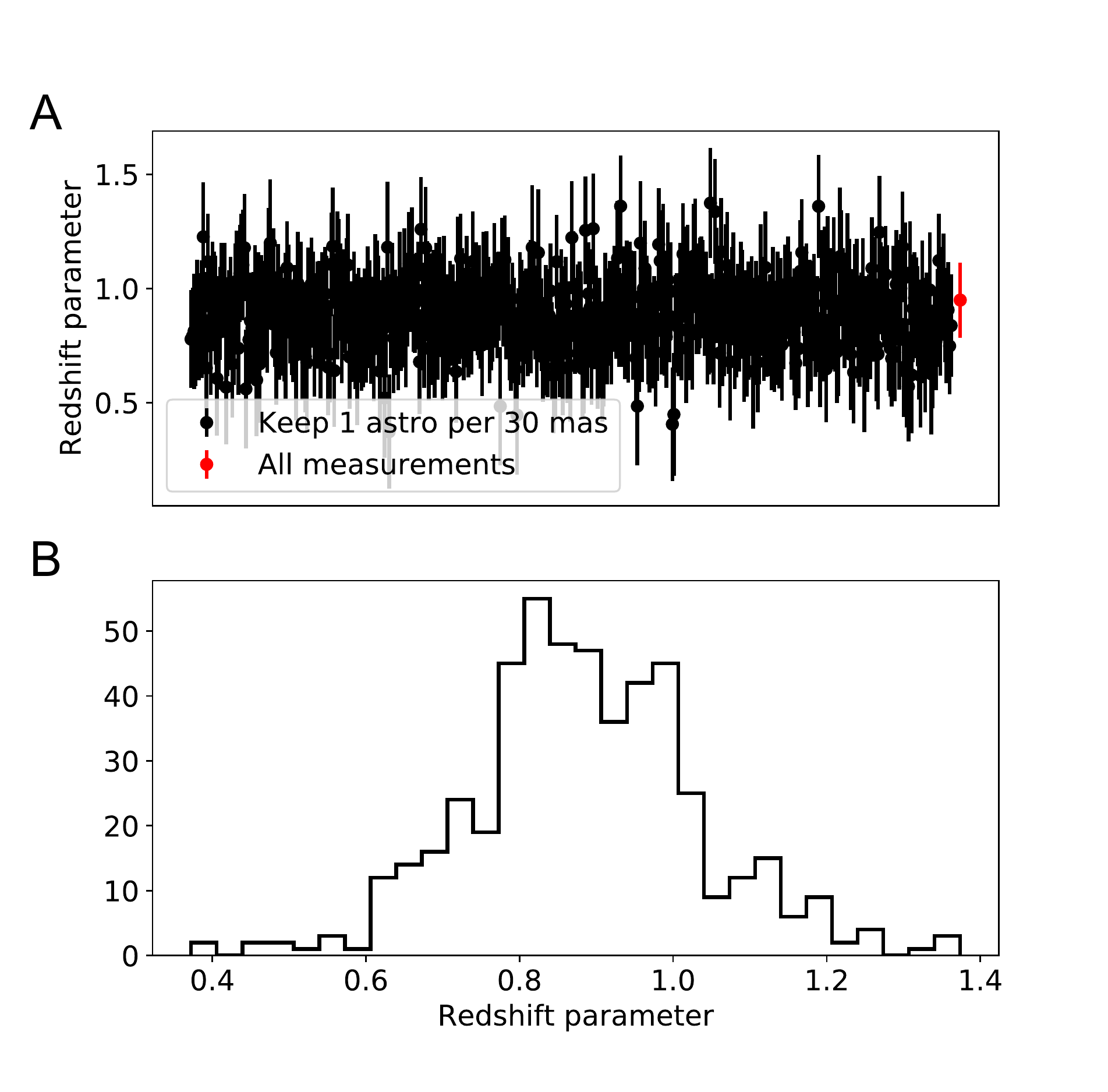}
\end{center}
\caption{Estimation of the redshift parameter for 500 datasets generated by choosing randomly one astrometric measurement per correlation length scale of 30 mas. The top panel shows the 1$\sigma$ estimated values for the redshift parameter for these 500 fits (and the red errobar indicates the estimation obtained using all the astrometric measurements and assuming their error to be statistically independent). The bottom panel shows the distribution of the 500 redshift parameter best-fitting values.
\label{fig:redshift_dop}
}
\end{figure}

In order to validate the analysis presented in the previous section, we used a second method to assess the correlations in our dataset. In this second method, we used a Monte Carlo approach and generated 500 mock astrometric datasets by choosing randomly one observation per 2D projected length scale of 30 mas. We then performed 500 fits by considering that the observations of these datasets are independent. As a result, we obtained a distribution of estimated parameters. Fig.~\ref{fig:redshift_dop} shows the distribution of the redshift parameter $\Upsilon$ for these 500 fits. The mean of these estimated redshift parameters is 0.88 and their dispersion, which is an estimation of the uncertainty due to correlations is 0.13. The total uncertainty, which is the quadratic sum of the statistical uncertainty obtained when considering all the astrometric measurements as independent (which is given by 0.89 $\pm$ 0.13) and the correlation uncertainty is given by $\sqrt{0.13^2+0.13^2}=0.18$. The results obtained with this method agrees within 1 $\sigma$ with the results obtained by the first method (fitting directly for the correlation simultaneously with the model parameters) that leads to 0.88 $\pm$ 0.16.

This result gives confidence in both methods. In our analysis, we include the correlation in our model and fit for the two parameters $\Lambda$ and $p$ simultaneously with all other model parameters.

\subsubsection{Systematic effects arising from the construction of the astrometric absolute reference frame}

The Sgr A*-radio rest astrometric reference frame is defined by the positions of seven masers (\cite{Sakai:2018}). The small number of reference stars may lead to systematic uncertainties in the construction of the reference frame. In order to estimate these systematics, we have undertaken a jackknife resampling method. The details of the jackknife method are described in \cite{Boehle:2016} and \cite{Sakai:2018}. We construct seven reference frames by dropping one different maser for each of them. Each  is applied to the cross-epoch alignment following the same  methodology as described in Section~\ref{sec:astro}. As a result, 7 sets of S0-2's positions are derived (see Table~\ref{tab:jackknife}).  We used these 7 sets of S0-2's astrometric measurements combined with the RV measurements to fit for S0-2's orbit following the same methodology as above. 

\begin{figure}
\begin{center}
\includegraphics[scale=0.5]{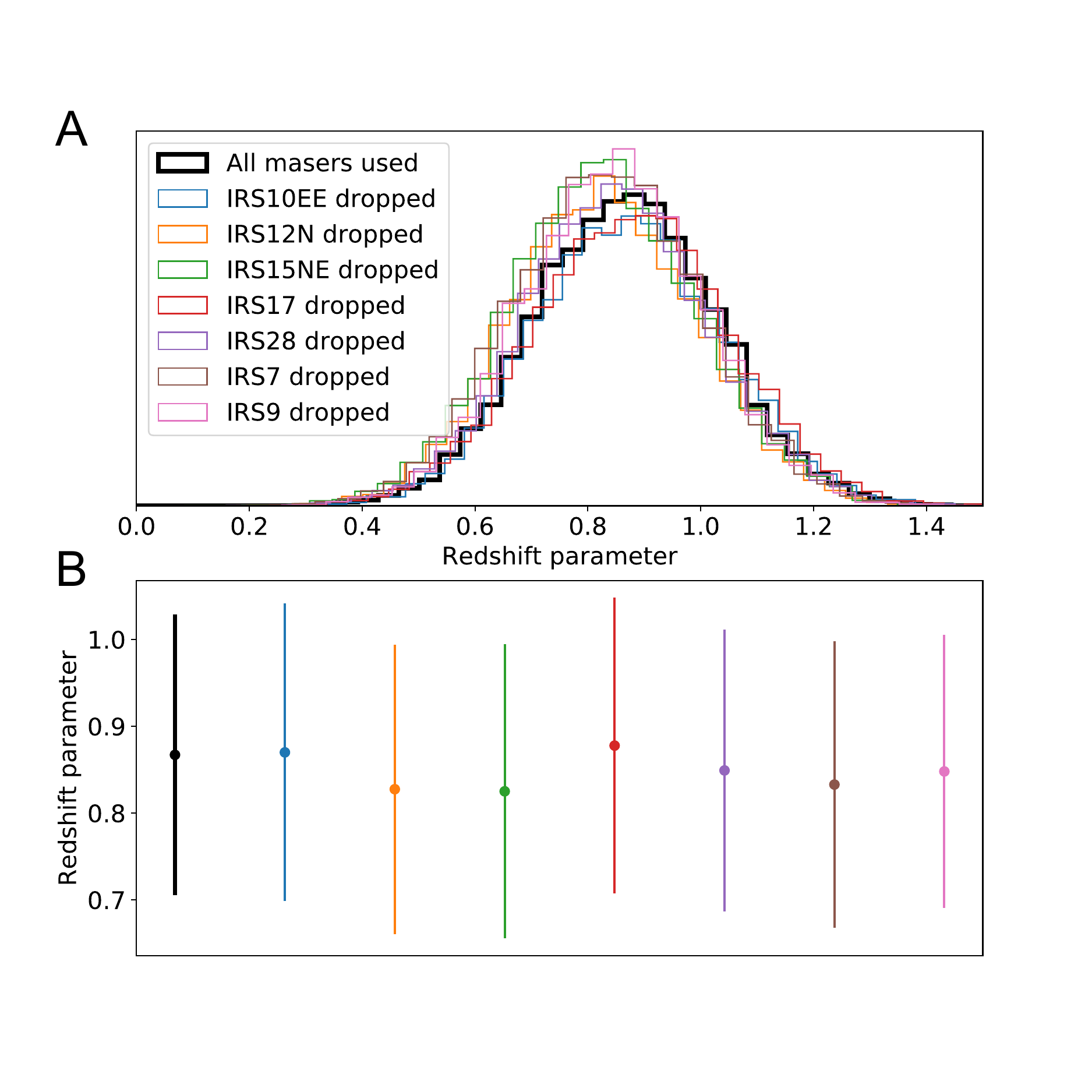}
\end{center}
\caption{\textbf{A:} Posterior distribution for the redshift parameter for 8 different fits: 7 for which the astrometric positions of S0-2 have been derived by constructing a reference frame where one maser is dropped at a time and one where all 7 masers are used in the construction of the reference frame. The legend gives the name of the maser dropped. \textbf{B:} estimations of the redshift parameter and its 68\% confidence interval for each of the eight fits.
\label{fig:DOM}
}
\end{figure}

The resulting estimations of the redshift parameter are presented in Fig.~\ref{fig:DOM} and in Table~\ref{tab:DOM}. These estimations can then be used to infer the systematic uncertainty due to the construction of our reference frame by evaluating
\begin{equation}
	\sigma^2_\mathrm{sys}=\frac{n-1}{n}\sum_{i=1}^n\left(x_{n-1,i}-\bar x_{n-1}\right)^2 \, ,
\end{equation}
where $n$ is the sample size (in this case 7), $x_{n-1,i}$ is the redshift estimator derived by excluding the $i$th maser and $\bar x_{n-1}$ is the average of these subsets. The estimations of the redshift parameter from Table~\ref{tab:DOM} give an estimation of the systematic uncertainty of 0.0466. This systematic uncertainty is independent of the statistical uncertainty determined in the orbital fit and needs to be added in quadrature. The systematic uncertainties for the other parameters are obtained in the same way and are presented in Table ~\ref{tab:fit_results}. The systematic uncertainty is larger than the statistical one only for the Black Hole position and velocity parameters.

\begin{table}[htb]
 \caption{Estimation of the redshift parameters (68\% confidence interval of the marginalized posterior) for 8 different fits: seven for which the astrometric positions of S0-2 have been derived by constructing a reference frame where one maser is dropped at a time and one where all 7 masers are used in the construction of the reference frame. Col. 1: Maser dropped from analysis, Col. 2 \& 3: Maser location (offset from the location of Sgr A* in right ascension and declination), Col. 4: estimate of the gravitational redshift parameter}
	\label{tab:DOM} 
	\centering
	\begin{tabular}{lccc}
	\hline
	Maser dropped   & $\Delta$ R.A. (arcsec) & $\Delta$ DEC (arcsec)  & Estimation of the    \\
      &&  & redshift parameter  \\\hline
	None  & &  &$0.88^{\phantom{-}0.16}_{-0.16}$         \\\hline
    IRS10EE & 7.7  & 4.2   &$   0.87^{\phantom{-}0.17}_{-0.17}$     \\
    IRS12N & -3.3 & -6.9 & $   0.83^{\phantom{-}0.17}_{-0.16}$    \\
    IRS15NE & 1.2 & 11.3 & $   0.83^{\phantom{-}0.17}_{-0.16}$     \\
    IRS17 & 13.1 & 5.6 & $   0.88^{\phantom{-}0.17}_{-0.17}$      \\
    IRS28 & 10.5 & -5.8 & $   0.85^{\phantom{-}0.16}_{-0.16}$     \\
    IRS7  & 0.0 & 5.5 & $   0.83^{\phantom{-}0.17}_{-0.17}$    \\
   	IRS9  & 5.7 & -6.3 & $   0.85^{\phantom{-}0.16}_{-0.16}$     \\
   	\hline
	\end{tabular}
\end{table}

\begin{longtable}{lccccc}
\caption{\textbf{S0-2 Astrometric Measurements used in Jackknife Analysis}. This table contains the astrometric position of S0-2 in the seven reference frames from maser jackknife analysis. Col. 1: Date observation, Col. 2: Modified Julian date, Col.3 \& 4: Offset from the origin of reference frame in right ascension and declination, Col. 5 \& 6: Uncertainty in the offset. Full table (368 entries) in Supplement D4.}\\
\label{tab:jackknife}\\
\toprule
Date &	MJD	&	$\Delta$R.A. & $\Delta$ DEC & $\Delta$ R.A. Error	&	$\Delta$ DEC Error	\\
(Year) &   & (arcsec)  & (arcsec) & (arcsec) & (arcsec) \\
\midrule
\endhead
\midrule

\multicolumn{6}{r}{{Continued on next page}} \\
\midrule
\endfoot

\bottomrule
\endlastfoot
All Seven Masers     &  &  &  &  &  \\ 
1995.439 & 49878.406215 &  0.04350 &  0.16900 &  0.00141 &  0.00206 \\ 
 1996.485 & 50260.379132 &  0.05312 &  0.15539 &  0.00308 &  0.00336 \\ 
 1997.367 & 50582.454528 &  0.05862 &  0.14074 &  0.00115 &  0.00137 \\ 
 1999.333 & 51300.489935 &  0.06819 &  0.09725 &  0.00072 &  0.00062 \\ 
 1999.559 & 51383.314098 &  0.06867 &  0.09164 &  0.00059 &  0.00062 \\ 
 2000.305 & 51655.573395 &  0.07057 &  0.06574 &  0.00188 &  0.00369 \\ 
 2000.381 & 51683.473210 &  0.06809 &  0.06509 &  0.00060 &  0.00075 \\ 
 2000.548 & 51744.263398 &  0.06572 &  0.05924 &  0.00086 &  0.00161 \\ 
 2000.797 & 51835.188435 &  0.06466 &  0.05116 &  0.00181 &  0.00177 \\ 
 2001.351 & 52037.589269 &  0.05773 &  0.02777 &  0.00099 &  0.00078 \\ 
 ... & ... & ... & ... & ... & ... \\

\end{longtable}

\section{Supplemental Text}
\subsection{Results and discussion}
\label{sec:discussion}
Here we summarize the orbital fitting methodology described in detail above.  We use a Gaussian likelihood for the RV observations and include an offset for the NIRC2 measurements in the fit. The likelihood for the astrometric part is given by a multivariate distribution characterized by a correlation matrix that is parametrized by a length scale $\Lambda$ and a mixing parameter $p$, see Eq.~(\ref{eq:cor_d}). The fit includes  6 orbital parameters for S0-2 and 7 parameters related to the SMBH: the $GM$, the distance to the GC $R_0$, the linear drift of the SMBH $x_0$, $y_0$, $v_{x_0}$, $v_{y_0}$, a RV offset $v_{z_0}$. This amounts for 16 fitted parameters. Different fits using different models were done in this analysis:

\begin{minipage}[h]{\textwidth}
\begin{enumerate}
    \item a fit using a purely Newtonian modeling.
    \item a fit using a purely GR modeling.
    \item a fit using a modeling that include the correction from special relativity but no effects from GR.
    \item a fit using a pure GR modeling and including the presence of an extended mass modeling a possible population of compact objects near the SMBH, such as neutron stars and stellar mass black holes.
    \item a fit introducing an additional parameter $\Upsilon$ to encode deviation from special and general relativity at the level of the redshift as described in Eq.~(\ref{eq:RV1}).
    \item a fit similar to the previous one but including the extended mass.
    \item a fit introducing an additional parameter $\alpha$ to encode a deviation from General Relativity at the level of the gravitational redshift as descirbed in Eq.~(\ref{eq:RV2}).
\end{enumerate}
\end{minipage}
\smallskip

Table~\ref{tab:fit_ev} lists the log evidence and the elpd for these 7 fits. These quantities are useful to assess the models favored by the measurements (see  section~\ref{sec:bayesian}). The chains sampling the posterior probability distribution from the fits 1, 2 and 5 are available as Data S3.


\begin{table}[htb]
 \caption{Bayesian evidence and elpd difference from Newtonian model for the 7 fits discussed in this section. }
	\label{tab:fit_ev} 
	\centering
	\begin{tabular}{c c  | c c}
	    Fit  & Model    & $ \Delta \log \mathcal E$   & $\Delta$ elpd   \\
	\hline
	 1 & Newton                  & 0           & 0     \\
	 2 & GR correct              &  10.68 & 10.09     \\
	 3 & special rel. only       & 6.45            & 5.88     \\
	 4 & GR correct + Extended Mass         & 7.79  & 9.26   \\	
	 5 & redshift test with $\Upsilon$ & 8.46 & 9.08 \\		  
	 6 & redshift test + Extended Mass      &  4.88 & 8.36   \\
     7 & redshift test with $\alpha$  & 8.46 & 9.08 		  
	\end{tabular}
\end{table}

First, the difference of the log-evidence and of the elpd between the fits 1 and 2 are 10.68 and 10.09, respectively (see table~\ref{tab:fit_ev}), indicating that the data favors highly the GR modeling over the Newtonian modeling (see section~\ref{sec:bayesian}). In other words, the relativistic modeling has a Bayes factor of 10.68, which means that it is 43,650 more likely given the measurements. This demonstrates  that the relativistic modeling is favored with respect to the Newtonian modeling. Similarly, the difference in maximum of the log-likelihood between the two fits is  10.4.

To assess the significance of the corrections from General Relativity, we compared fit 2 (GR model) to fit 3, which includes only corrections from special relativity but not General Relativity. The difference of log-evidence and elpd between these two fits is 4.23 and of 4.21, respectively, indicating a clear detection of the general relativistic model over a model that includes special relativity only. In terms of Bayes factor, this means that the pure GR modeling is 70 times more likely than the modeling using special relativity given the measurements. 

\begin{figure}
\begin{center}
\includegraphics[scale=0.5]{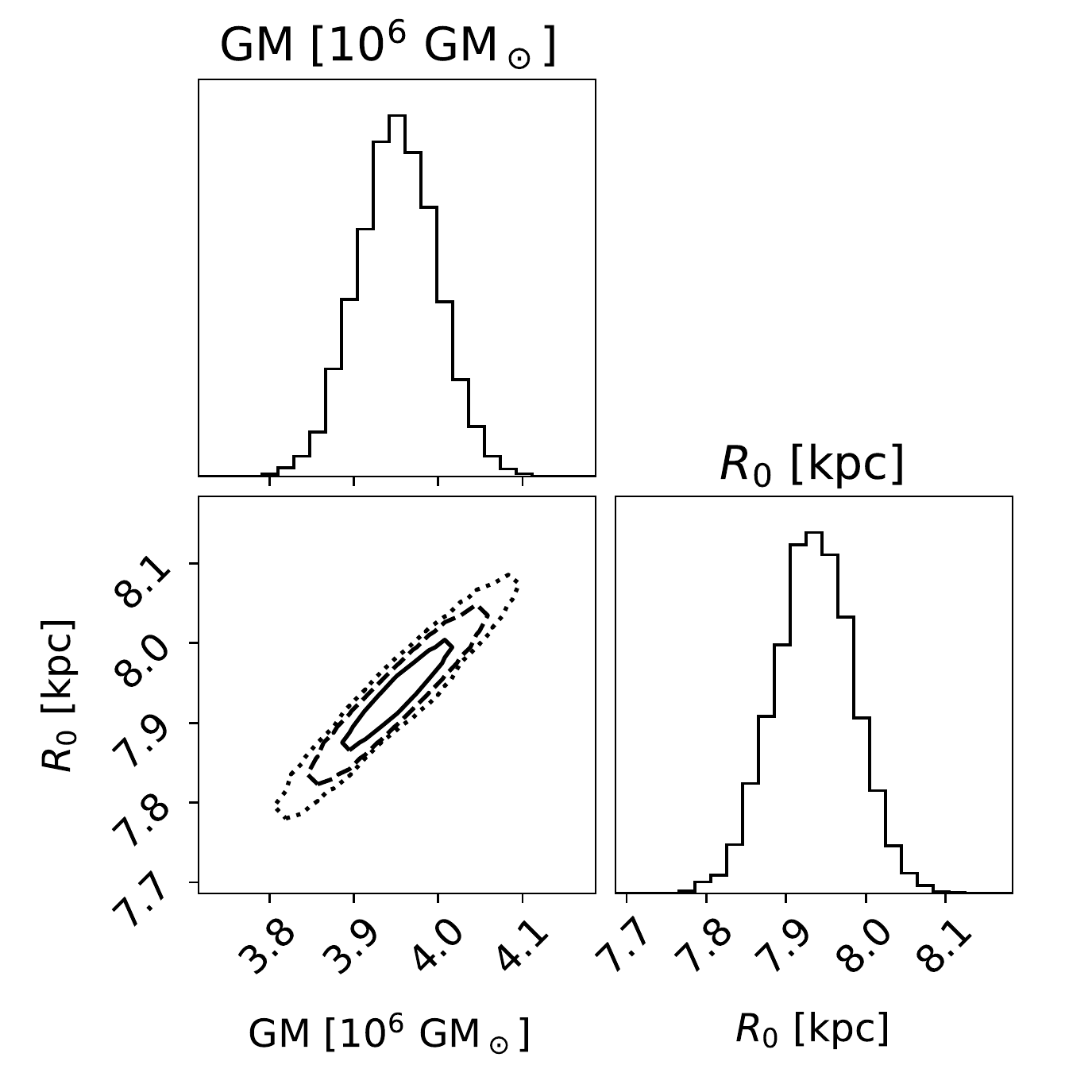}
\end{center}
\caption{Posterior pdf for the global parameters $GM$ and $R_0$ for a fit that assumes General Relativity to be correct (Fit 2 in Section \ref{sec:discussion}).
\label{fig:corner_M_R}
}
\end{figure}

The results from fit 2 include the estimation of the SMBH mass and $R_0$. The 1D marginalized 68 \% confidence interval on these two parameters are: $M_\mathrm{BH}=\left(3.964 \pm 0.047 \pm 0.026 \right) \times 10^{6} M_\odot$ and $R_0= 7,946\pm 50 \pm 32$ pc, where the first uncertainty is the statistical uncertainty and the second uncertainty is the systematic $\sigma$ from the jackknife analysis. Figure \ref{fig:corner_M_R} presents the posterior distribution for these two parameters. The estimated BH mass and $R_0$ differs at the level of 2$\sigma$ with those measured by the GRAVITY collaboration \cite{Gravity:2018}. Their systematic uncertainty was given only for the redshift parameters and not for the other parameters \cite{Gravity:2018}. 

\begin{figure}
\begin{center}
\includegraphics[scale=0.5]{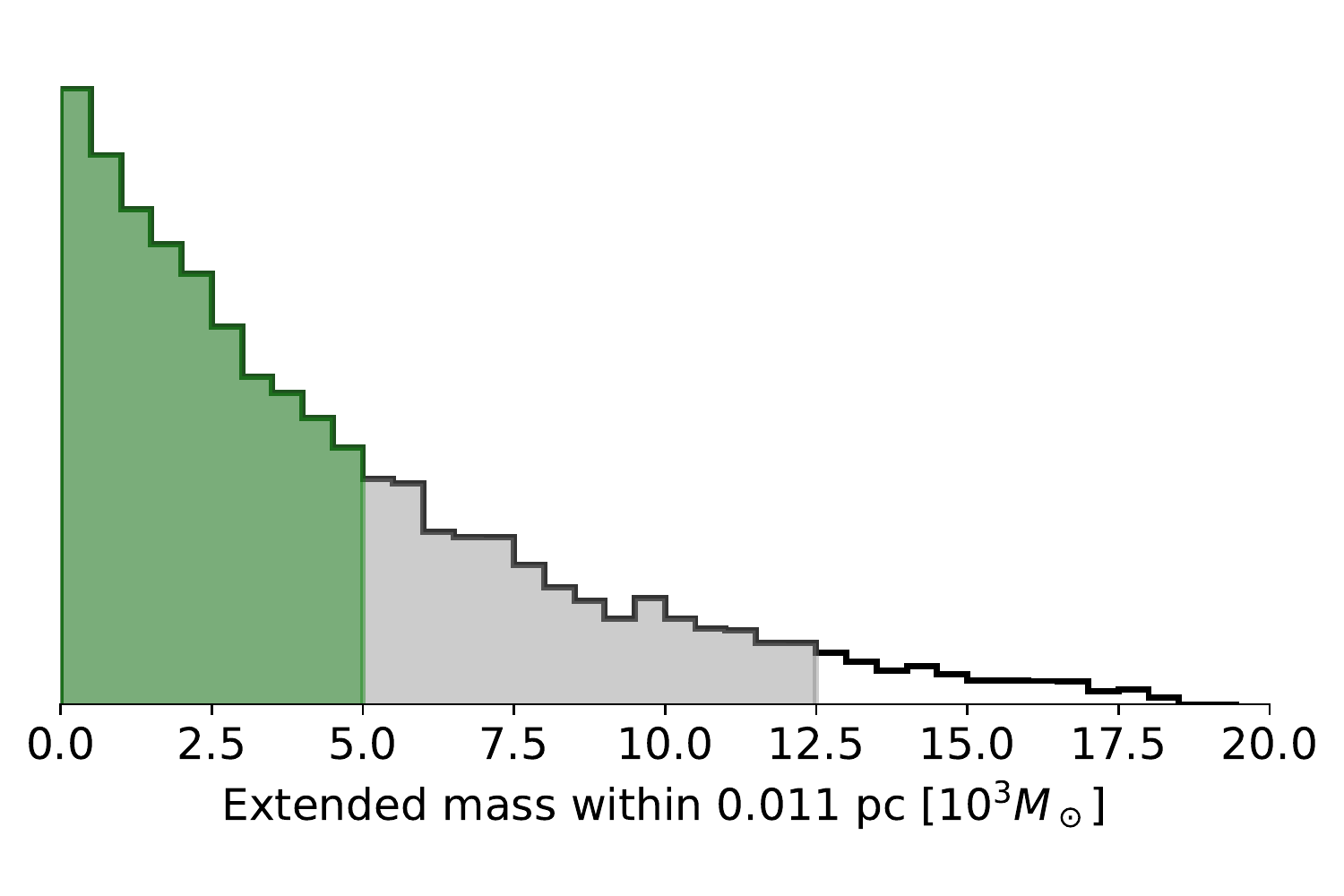}
\end{center}
\caption{Posterior pdf for the extended mass parameter $M_0$ from Eq.~(\ref{eq:EM}). This fit assumes a power law slope for the density profile of any extended mass, with index $\gamma = 1.5$ ($M_0$ is not significantly sensitive to this value, see \cite{Boehle:2016}). No significant extended mass has been detected, but this analysis has resulted provides a improved limit compared to previous works \cite{Boehle:2016}. 
\label{fig:EM}
}
\end{figure}

Fit 4 includes the extended mass in the context of General Relativity. The extended mass distribution is modeled by a power-law as in \cite{Boehle:2016}:
\begin{equation}\label{eq:EM}
    M_\mathrm{ext}(r)=M_0 \left(\frac{r}{r_0}\right)^{3-\gamma} \, ,
\end{equation}
where $\gamma$ is a powerlaw parameter, $M_0$ is the extended mass enclosed inside $r_0$ and where $r_0$ has been taken as 0.011 pc such that it encloses S0-2 apoapse. The comparison of the elpd and Bayesian evidence from fits 2 and 4 on table~\ref{tab:fit_ev} indicates that the model with extended mass is not significantly preferred over a point mass. While this means we have not detected extended mass, an upper limit can be inferred on $M_0$. The astrophysical interpretation of this parameter includes a possible population of compact objects near the SMBH, such as neutron stars and stellar mass black holes.  The posterior distribution for the extended mass enclosed within S0-2's orbit is presented in figure~\ref{fig:EM} and the 68\% (respectively 95\%) upper confidence limit is $M_0<5.5 \times 10^3\  M_\odot \ (< 12.7\times 10^3\ M_\odot)$.

\begin{figure}
\begin{center}
\includegraphics[scale=0.32]{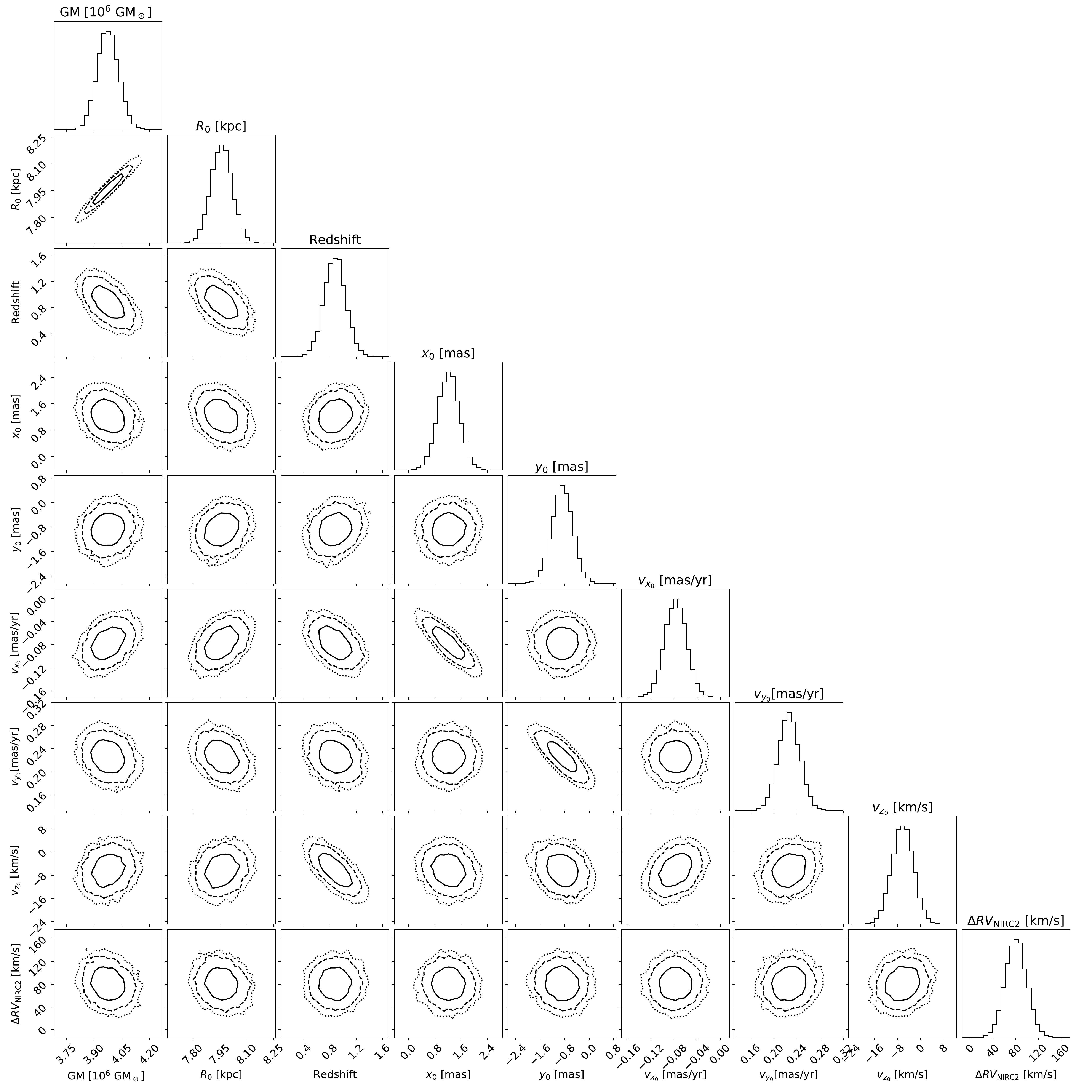}
\end{center}
\caption{Posterior pdf for the global black hole parameters for Fit 5 (the parameterized relativistic redshift fit). The posterior pdf for the parameters that characterized the correlation matrix is presented in Fig.~\ref{fig:astro_post_cor}. The continuous, dashed and dotted contours represent the 68, 95 and 99 \% credible area.
\label{fig:corner_glob}
}
\end{figure}

Fit 5 is our canonical solution presented in the main text. Here we discuss it in further detail. The estimated parameters for this fit are presented in Table~\ref{tab:fit_results}. The posterior pdf for the global parameters is presented in Fig.~\ref{fig:corner_glob} while the posterior pdf for the coefficients of the astrometric correlation matrix is presented in Fig.~\ref{fig:astro_post_cor}. 

\begin{figure}
\begin{center}
\includegraphics[scale=0.5]{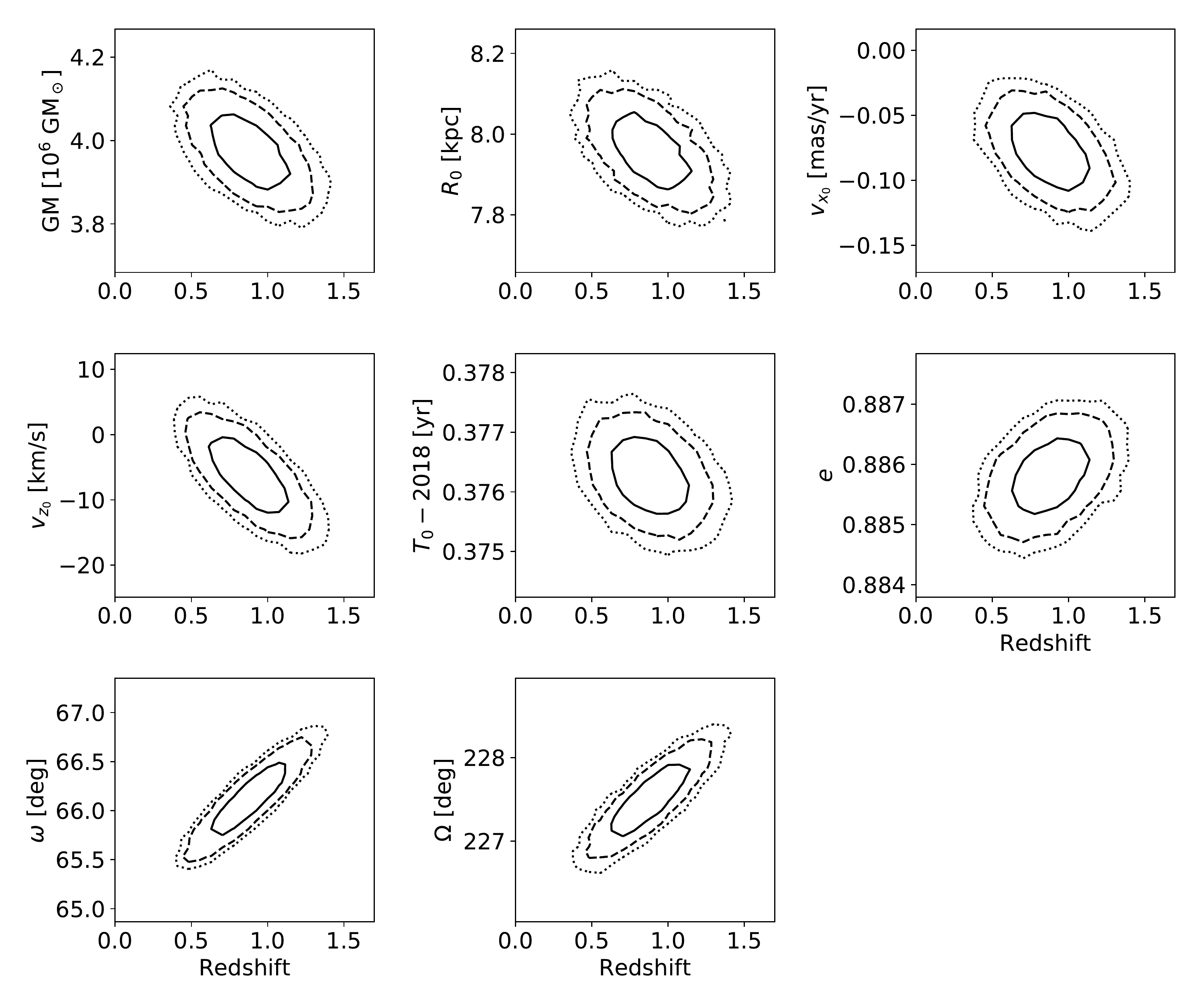}
\end{center}
\caption{Joint posterior probability distributions for the redshift parameter and other fitted parameters. We include parameters that are the most correlated with the redshift. The $GM$ parameter shows a strong correlation since the redshift signal is proportional to this parameter. The $v_{z_0}$ parameter impacts directly the RV, similarly to the redshift parameter. S0-2's orbital parameters  determine the distance between S0-2 and the central SMBH and hence how deep S0-2 is in the gravitational potential. The continuous, dashed and dotted contours represent the 68, 95 and 99 \% credible area.
\label{fig:redshift_correlations}
}
\end{figure}

The redshift parameter is correlated with several other fitted parameters. Fig.~\ref{fig:redshift_correlations} presents some joint-posterior distributions of the parameters that show the greatest correlation with the redshift parameter.

\begin{figure}
\begin{center}
\includegraphics[scale=0.5]{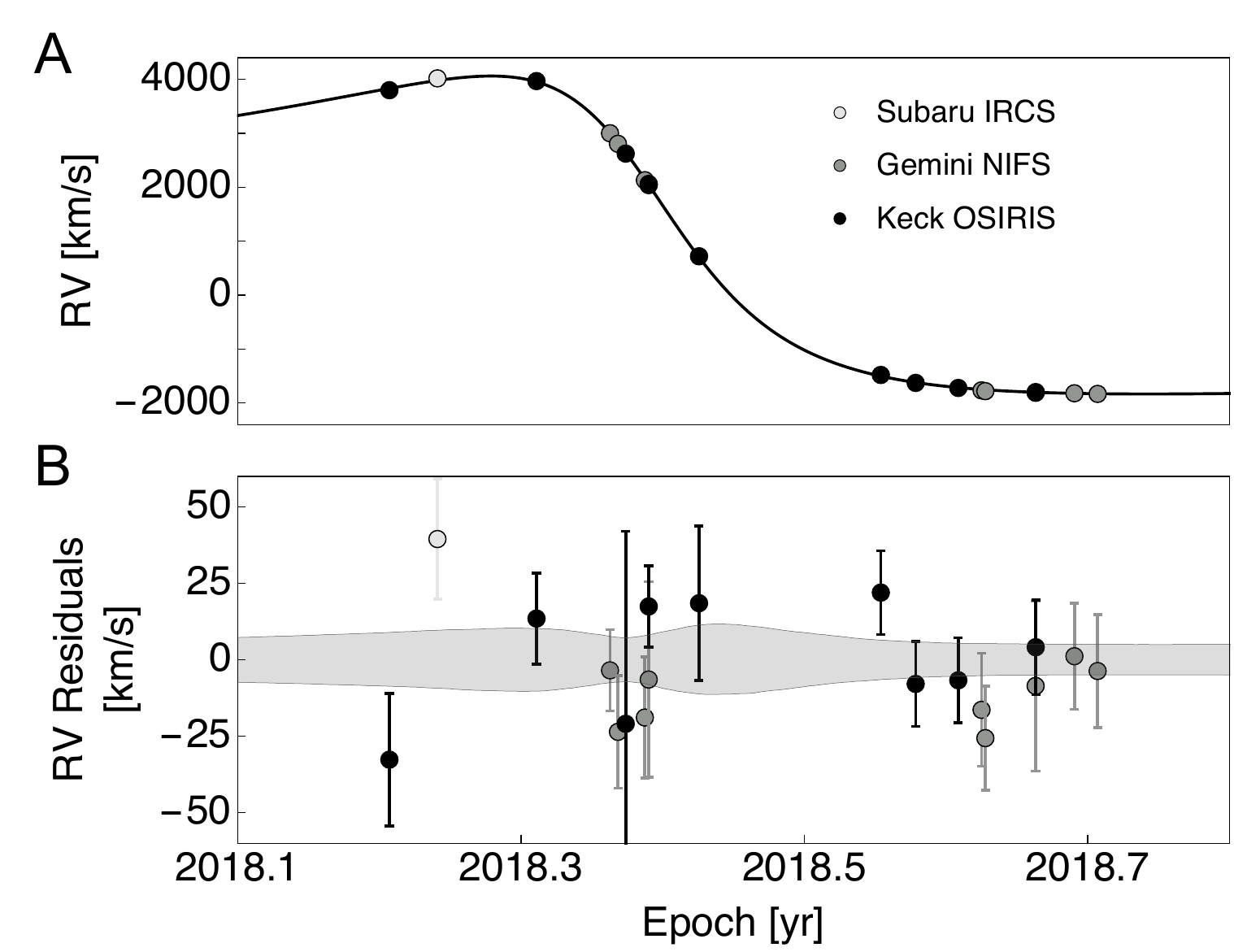}
\end{center}
\caption{\textbf{A:} RV model (line) and observations (circles) as a function of time, zoom of Figure~\ref{fig:res} B. \textbf{B:} RV residuals and  model uncertainty (68\% confidence limit) as a function of time, zoom of Figure~\ref{fig:res} C. 
\label{fig:RV_res}
}
\end{figure}

\begin{figure}
\begin{center}
\includegraphics[scale=0.5]{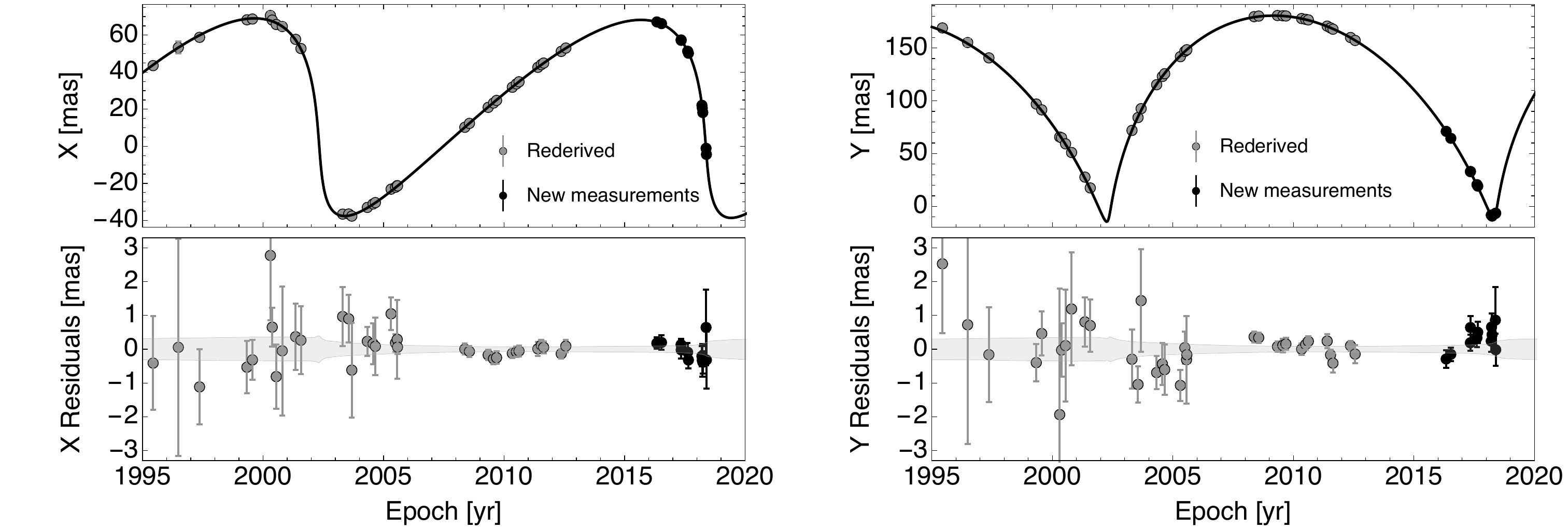}
\end{center}
\caption{\textbf{Top:} Astrometric model (line) and observations (circles) as a function of time. \textbf{Bottom} Astrometric residuals and  model uncertainty (68\% confidence limit) as a function of time. 
\label{fig:astro_res}
}
\end{figure}

\begin{figure}
\begin{center}
\includegraphics[scale=0.6]{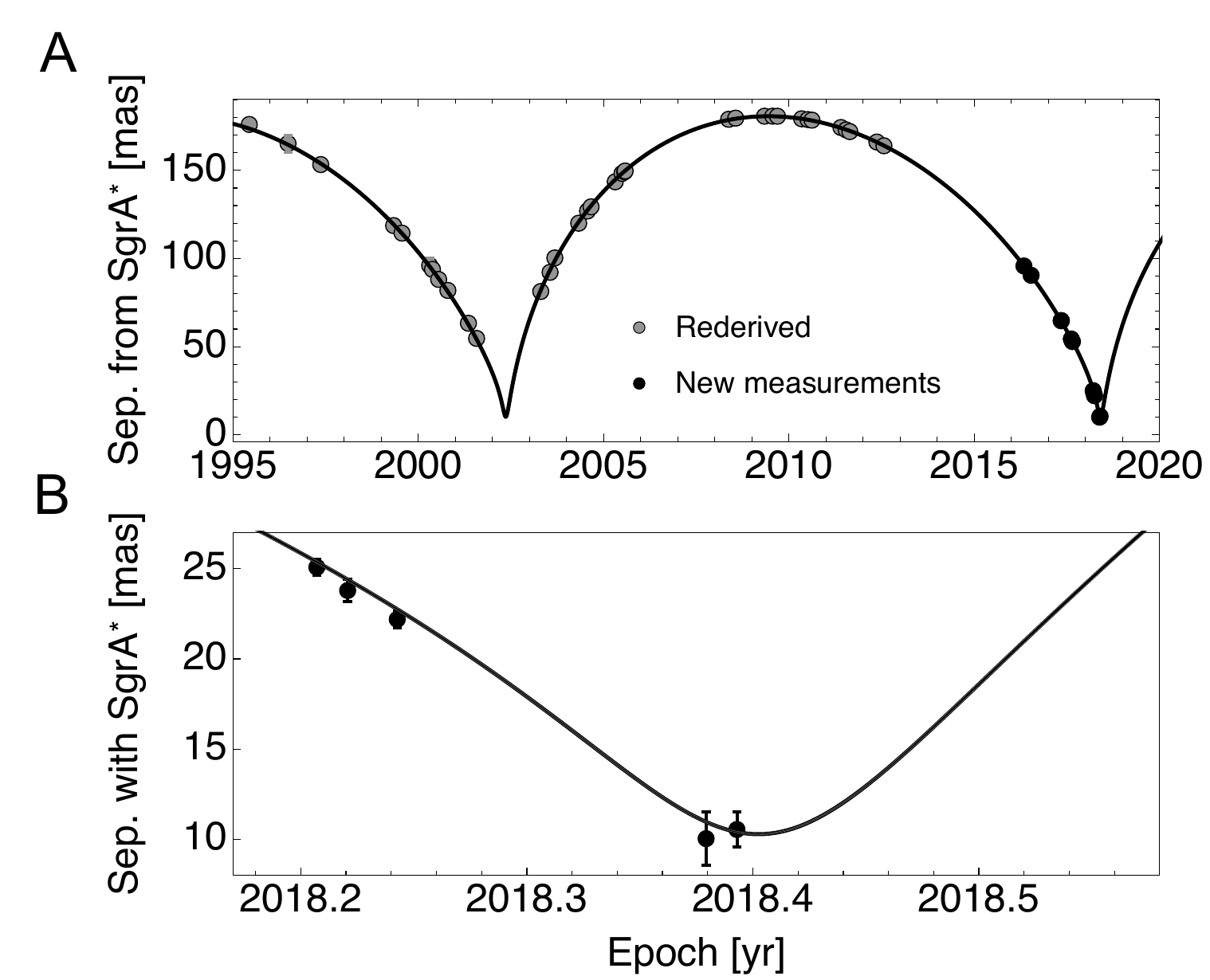}
\end{center}
\caption{Angular separation (best fitting model and measurements) between S0-2 and the SMBH center of mass as a function of time. The bottom panel is a zoom around the 2018 closest approach.
\label{fig:sep}
}
\end{figure}

The full RV residuals are presented in the main text. The 2018 RV residuals that are presented in Fig.~\ref{fig:RV_res}. The astrometric measurements, best fit model and residuals are presented in Fig.~\ref{fig:astro_res}. The separation between Sgr A* and S0-2 is presented in Fig.~\ref{fig:sep} and allows us to compare our astrometric measurements with the one from the GRAVITY collaboration \cite{Gravity:2018}. Our model 
predicts an angular separation of 22.6 mas at the end of March, which seems in agreement with their figure 1 (although no uncertainty is given in \cite{Gravity:2018}). 

The results presented in Table~\ref{tab:fit_results} have been obtained assuming that the extended mass resulting from massive non-luminous object such as stellar remnants is negligible. We tested the impact of adding an additional parameter to the fit to include an the hypothetical presence of such an extended mass (see fit 6). We use a power law density-profile modeling for the extended mass, see Eq.~(\ref{eq:EM}). Adding this parameter to the fit does not change the redshift parameter estimate which remains $\Upsilon=0.87\pm 0.16\pm 0.047$.

On the other hand, if we use a Newtonian modeling of the orbital dynamics instead of integrating the post-Newtonian equations of motion (i.e. not including the precession of the periapsis), we obtain a slightly lower estimate of the redshift parameter of $\Upsilon=0.78\pm 0.16\pm 0.047$. This behavior is similar to the one observed in \cite{Gravity:2018}.

The NIRC2 offset that is added to our model does not change the redshift estimate significantly. If we do not include this offset, the estimate of the redshift parameter is 0.86 $\pm$ 0.16 $\pm$ 0.047.

\begin{figure}
\begin{center}
\includegraphics[scale=0.5]{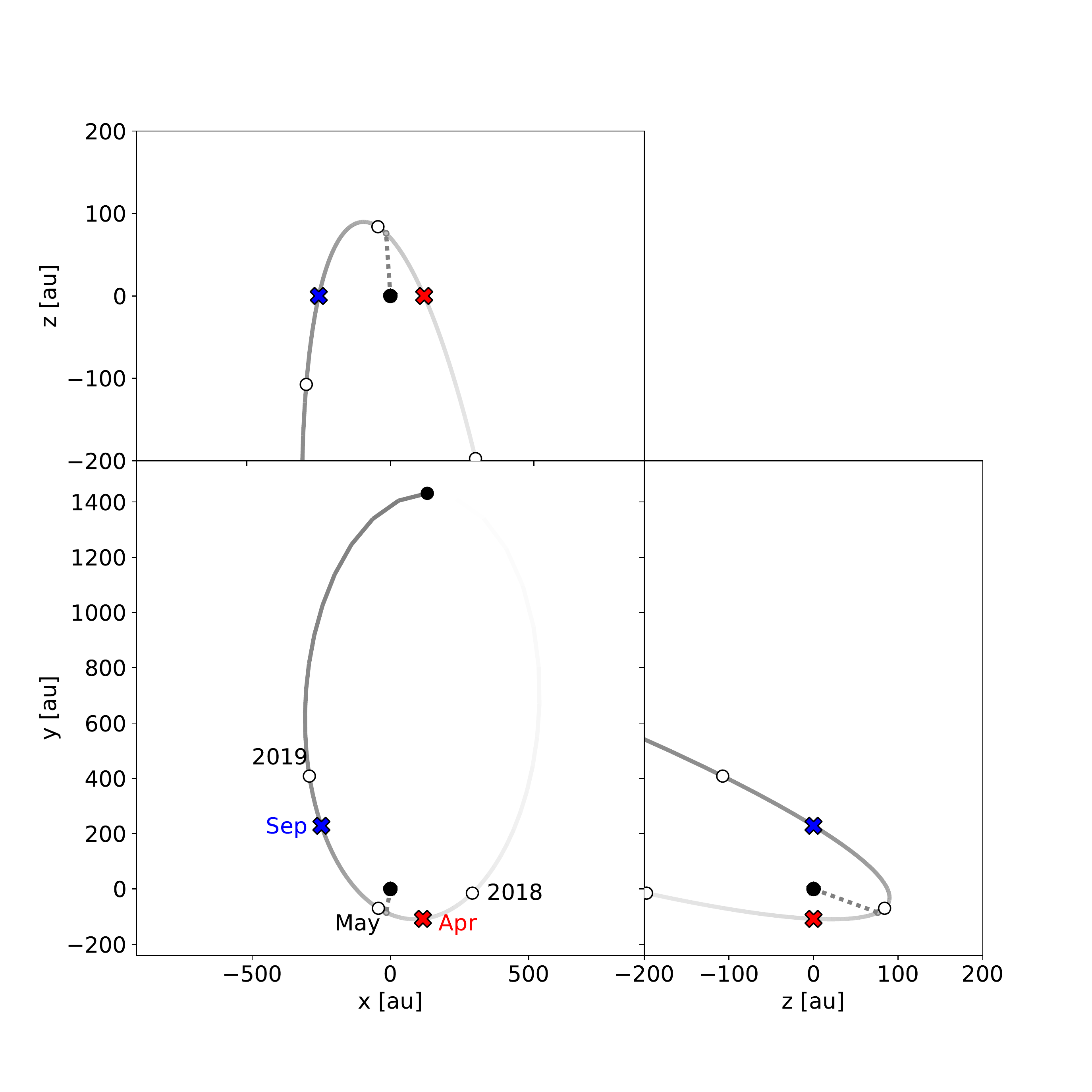}
\end{center}
\caption{The orbit of S0-2 in different projections, centered on the supermassive black hole (black circle). The solid line indicates the orbital motion, with the darker portions more recent in time. The X-Y projection is on the plane of the sky. Z is along the line of sight with positive Z directed away from Earth and negative Z toward Earth. The orbit of S0-2 is tilted toward the Earth, with its closest approach to Sgr A* lying behind the black hole in this projection. Also labeled are the location along the orbit of RV maximum (red cross in April) and RV minimum (blue cross in Sept). These two points, along with the 2D closest approach (white circle in May) provide the most sensitive observations for the redshift parameter estimation. The 3D closest approach position (grey dotted line) is very close the projected 2D closest approach. Also marked are the location of S0-2 in Jan, 1 2018 and Jan 1, 2019. This figure was produced with the aid of the Rebound code \cite{Rein:2012}.
\label{fig:geometry}
}
\end{figure}

To aid in understanding the geometry of the orbit of S0-2, we show the orbit using different projections in Figure \ref{fig:geometry}.

The last fit (fit 7) is similar to fit 5 (parameterized redshift model), but here we only allow the gravitational redshift to vary. This is parametrized by the conventional parameter $\alpha$ (see Eq.~(\ref{eq:RV2}), or the section 2.3.1. from \cite{Will:2014}). The resulting estimation of the $\alpha$ parameter is given by $\alpha = -0.24\pm 0.32\pm 0.047$ in agreement with Eq.~(\ref{eq:alpha}) above. This parametrization provides constraints on theories of modified gravity. Several modified theories of gravitation violate the equivalence principle so can be characterized with this parametrization (see e.g.~\cite{Hees:2019arxiv}). 

One example of a theory that violates the equivalence principle around a black hole is the quadratic Einstein-Gauss-Bonnet theory where the scalar field is non-minimally coupled to the standard matter fields \cite{doneva:2018aa,silva:2018aa,antoniou:2018aa}. The quadratic Einstein-Gauss-Bonnet theory is a tensor-scalar theory with a specific coupling between the scalar field and the Gauss-Bonnet invariant. These recently studied theories are (amongst others) motivated by string theory and by unification scenarios \cite{doneva:2018aa,silva:2018aa,antoniou:2018aa}. These theories can present large deviations from General Relativity around black holes while they will behave exactly as GR around non compact objects, such as those in our solar system \cite{doneva:2018aa,silva:2018aa,antoniou:2018aa}. Black holes would acquire a large scalar charge that characterizes the $1/r$ behavior of the scalar field at large distances \cite{doneva:2018aa,silva:2018aa,antoniou:2018aa}.

If this hypothetical scalar field is non-miniminally coupled to the standard matter fields, it will produce a violation of the Einstein equivalence principle. A well-studied prototype for the Lagrangian modeling the interaction between the scalar field and the standard matter is given by the Lagrangian from Damour and Donoghue \cite{damour:2010zr,damour:2010ve}. In this modeling, the scalar field presents a dilatonic coupling to the standard matter fields
that leads to a dependency of the constants of nature (the fine structure constant, the mass of the fermions and the quantum chromodynamic energy scale)  to the scalar field. Such a coupling is known to break the equivalence principle and leads to violations of the universality of free fall (see e.g. \cite{damour:2010zr}) and of the gravitational redshift \cite{damour:1999fk}. It can be shown (see e.g. \cite{damour:1999fk,Hees:2019arxiv}) that the deviation from the gravitational redshift in such a theory will be parametrized by the $\alpha$ parameter introduced in Eq.~(\ref{eq:RV2}). This parameter is directly  related to the hypothetical scalar charge of the SMBH and to the dilatonic coefficients that characterize the coupling between the scalar field and the standard matter fields \cite{Hees:2019arxiv}. This example illustrates how the current constraint on the gravitational redshift of S0-2 around Sgr A* can probe some effects beyond Solar System tests (because the scalar field will vanish around non compact objects) or with gravitational waves (because violation of the gravitational redshift is currently not testable with gravitational wave measurements).

\subsection{Adaptive scheduling tool}
In order to maximize the use of telescope time in 2018 and to enhance our chance of a successful detection of the relativistic redshift, we developed an adaptive scheduling tool to help us to plan our observations. This tool is fully described in \cite{Hees:2019} and here we  briefly outline its principles and results regarding the redshift measurement.

This tool determines what type of measurements and which nights of observation are optimal to measure a given parameter. This scheduling tool (or cadence tool) is based on the computation of the Fisher matrix using the assumption that the posterior probability density functions are Gaussian. This method is computationally fast to consider a large number of possible measurements to find the optimal ones to measure a signal.

The Fisher matrix is given by
\begin{equation}
    \bm F = \bm P^T \cdot \bm P \, ,
\end{equation}
where $\bm P$ is the matrix of the partial derivatives of the model with respect to the parameters:
\begin{equation}
    \left[\bm P \right]_{ij}=\frac{1}{\sigma_i}\frac{\partial M(t_i, \bm p)}{\partial p_j} \, ,
\end{equation}
where $M(t_i,\bm p)$ is the expression of the model of our observables (in practice, we have two types of observables: the astrometry and the RV), $\bm p$ is the set of fitted parameters and $\sigma_i$ corresponds to the uncertainty of the $i$th measurement. In practice, we used the modeling presented in section~\ref{sec:model} for which we computed the partial derivatives analytically. An estimation of the uncertainties on the estimated parameters resulting from a fit of the model $M$ using a set of observation $\left\{\left(t_i,O_i,\sigma_i\right)\right\}$, where $t_i$ is the epoch, $O_i$ the measurement and $\sigma_i$ its uncertainty, can be determined from the covariance matrix $\bm \Sigma$ which is the inverse of the Fisher matrix $\bm\Sigma=\bm F^{-1}$. In particular, the uncertainties of the estimated parameters are given by the diagonal of this matrix: $\sigma_{p_i}^2=\Sigma_{ii}$. This covariance matrix depends on (i) the measurements dataset (the timing of the measurements and their uncertainties) and (ii) on the set of parameters that are fitted (i.e. $\bm p$).

The cadence tool iteratively determines the future optimal observations (the optimal scheduling and the optimal type of measurement) in a given observational window assuming that we already have past measurements available. More precisely, we start the procedure with existing data and search within a given window for the measurement that most increases the information entropy related to a given parameter. Once this optimal measurement is found, it is added to the set of ``existing'' data and we iteratively search for the next most important measurement. This procedure furnished a sequence of optimal measurements that maximizes the signal to noise ratio of a given parameter. This sequence of measurements depends on the parameter that we are trying to optimize, on the set of fitted parameters (because of correlation), on the past measurements, on the expected uncertainty for the future observations, and on the future observational window that is considered.

In practice, the optimization procedure relies on a brute force method in the sense that we compute the covariance matrix for every additional measurement included in our future observational window. However, it remains sufficiently fast for two reasons: (i)  it requires only one evaluation of the model and of its partial derivatives for the past measurements and over the full future observational window and (ii) instead of inverting the full Fisher matrix after each additional measurement, we update the covariance matrix by adding one observation. In other words, if $\bm\Sigma^{(n)}$ is the covariance matrix related to a set of $n$ measurements, the covariance matrix for a set which includes an additional measurement at time $t_k$ will be given by
\begin{equation}
    \bm\Sigma^{(n+1)} = \bm\Sigma^{(n)} +\bm U^{(n)}_{(k)}\, ,
\end{equation}
where $\bm U^{(k)}$ is the update matrix which depends on the measurement at time $t_k$ and is given by
\begin{equation}
    \bm U^{(n)}_{(k)}= - \frac{\left(\bm\Sigma^{(n)} \cdot \tilde {\bm P}_{(k)} \right)\cdot \left(\bm\Sigma^{(n)} \cdot \tilde {\bm P}_{(k)} \right)^T }{1+\tilde {\bm P}_{(k)}^T\cdot\bm \Sigma^{(n)}\cdot \tilde {\bm P}_{(k)}}
\end{equation}
where $\tilde {\bm P}_{(k)}$ is a column vector containing the partial derivatives of the measurement at time $t_k$ with respect to the fitted parameters $\bm p$: $\left[\tilde {\bm P}_{(k)}\right]_i =\frac{1}{\sigma_k}\frac{\partial M(t_k,\bm p)}{\partial p_i}$. In practice, since we are optimizing on the uncertainty of one single parameter, only one term of the update matrix needs to be computed and this avoids the need to invert the full Fisher matrix a large number of times.

\begin{figure}
\begin{center}
\includegraphics[scale=0.75]{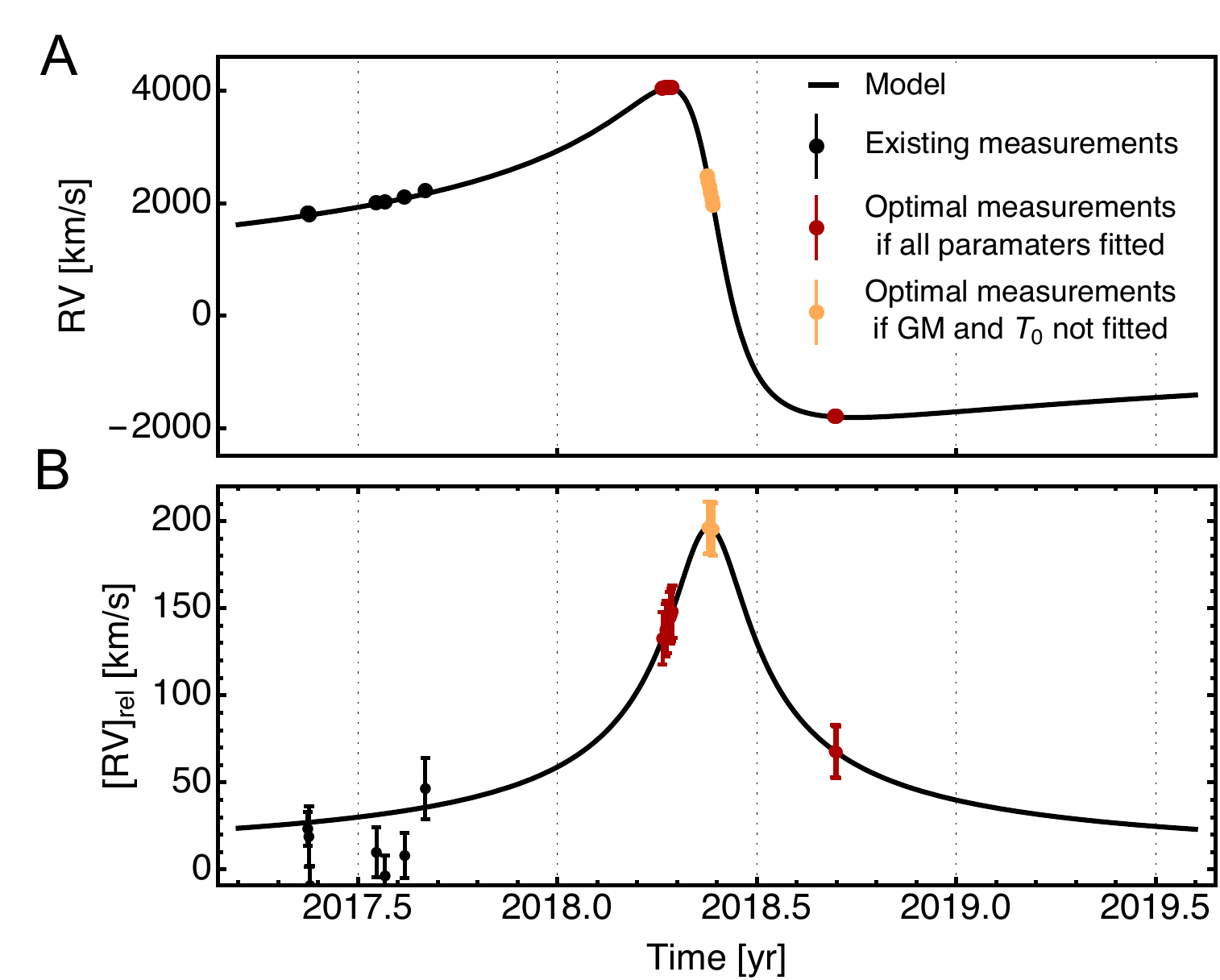}
\end{center}
\caption{Representation of the RV measurements sequence in 2018 optimized to detect the relativistic redshift (see Table~\ref{tab:cadence}). The top panel represents the optimal RV measurement to detect the relativistic redshift. The black curve is the RV model. The bottom panel shows the relativistic contribution from the same measurements with the redshift signal. In these simulations, all measurements prior to 2018 are used (black points). The red points show the optimal measurements sequence under the assumption that all the model parameters are fitted simultaneously. The optimal RV measurements are located at the RV turning points and not at the maximum of the redshift because of correlations with $T_0$ and $GM$. The observations in orange are the optimal measurements under the assumption that $T_0$ and $GM$ are not fitted. These optimal measurements are located at the maximum of the redshift signal.
\label{fig:cadence_all}
}
\end{figure}

This adaptive scheduling tool has been used in the planning of our measurements in 2018.  We ran this tool by considering the measurements before 2018 as ``past observations'' and searched for the optimal sequence of observations in 2018 in order to measure the redshift parameter $\Upsilon$. The result was the sequence presented in Table \ref{tab:cadence}, under the assumption that  new RV uncertainties would be 20 km/s and  new astrometric uncertainties 0.9 mas. For this simulation, we assumed that all the model parameters are fitted. Correlations between the parameters are substantial.  9 out of the 12 optimal measurements are RV observations. This makes sense considering that the redshift impacts directly the RV but not the astrometry. In addition, the adaptive scheduling does not favor nights at the maximum of the redshift signal. Rather, it favors measurements that are at the RV turning points (i.e. the maximum and minimum of the RV curves, see Fig.~\ref{fig:cadence_all}). This is due to correlations between the redshift parameter and the other model parameters, in particular with $T_0$ and with the SMBH $GM$. These correlations are maximal exactly when the redshift signal is maximal, therefore that epoch is not optimal in order to measure $\Upsilon$. In other words, the redshift signal can easily be absorbed by a small change in $T_0$ or in the SMBH $GM$. To demonstrate the effect of correlations, we ran a test case and searched for the optimal measurements if we fit for all the model parameters except for $T_0$ and $GM$. Fig.~\ref{fig:cadence_all} shows the 12 optimal observations that are all located close to the maximal of the redshift signal. 

In conclusion, our analysis has shown that, due to correlations, the optimal measurements to detect the relativistic redshift are RV measurements taken at the epochs corresponding to the maximum and minimum of the RV and not at the maximum of the relativistic signal. This conclusion is highly sensitive to the assumed uncertainties for future measurements (in this analysis 20 \kms for RV and 0.9 mas for the astrometry). The optimal measurements sequence is also highly dependent on the parameter we are optimizing for (here the redshift parameters). Another optimal sequence of measurements would be obtained for optimal measurement of other parameters, such as $R_0$. 

\begin{table}[htb]
 \caption{Results of the adaptive scheduling tool: optimal measurements sequence in 2018 in order to measure the relativistic redshift. In this analysis, we are fitting all the model parameters. We assumed 2018 RV uncertainty of 20 \kms and astrometric uncertainty of 0.9 mas.}
	\label{tab:cadence} 
	\centering
	\begin{tabular}{c c c   }
	\hline
	  Order of importance &  Epoch &   Type of observation  \\\hline
 1 &  2018.286 & RV \\
 2 & 2018.283  & RV \\
 3 & 2018.700  & RV \\
 4 & 2018.264  & RV \\
 5 & 2018.275  & RV \\
 6 & 2018.699  & RV \\
 7 & 2018.272  & RV \\
 8 & 2018.700  & astro \\
 9 & 2018.288  & RV \\
 10 & 2018.696  & RV \\
 11 & 2018.546  & astro \\
 12 & 2018.699  & astro \\
 \hline
	\end{tabular}
\end{table}

\clearpage
\begin{itemize}
    \item Data S1: \textbf{S0-2 RV measurements}. Col. 1: Date of observation, Col 2: Julian Year, Col. 3 Modified Julian Date, Col 4: Measured radial velocity [\kms], Col. 5: Radial velocity uncertainty [\kms], Col. 6: Radial velocity corrected for local standard of rest velocity [\kms]. 
    \item Data S1: \textbf{S0-2 Astrometric Measurements.}  Col. 1: Date of observation, Col. 2: Modified-Julian date, Col. 3 \& 4: Separation from center of reference frame in right ascension and declination [arcsec], Col. 5 \& 6: Separation and position angles transformed from $\Delta$ R.A. and $\Delta$ D.EC. [arsecond, deg.].
    \item Data S3: \textbf{Nested-sampling Chains.} Contains 3 files. chains\_Newton.txt: chain corresponding to a fit using a Newtonian modeling (i.e. no advance of the periastron, no Romer time-delay and no relativistic redshift included). chains\_GR.txt: chain corresponding to a fit using a relativistic modeling (i.e. the Romer time delay, the relativistic redshift and the 1 post-Newtonian equations of motion are included in the fit). chains\_redshift: chain corresponding to a fit where the redshift parameter is free (i.e. the Romer time delay and the 1 post-Newtonian equations of motion are included and the relativistic redshift is included and fitted for). Col 1: weight, Col 2: the SMBH gravitational parameter (GM) [1E6 solar GM], Col 3: R0 [kpc], Col 4: Redshift parameter [-], Col 5: x0 [mas], Col 6: y0 [mas],Col 7: vx0 [mas/yr], Col 8: vy0 [mas/yr], Col 9: vz0 [\kms], Col 10: S0-2's orbital period [yr], Col 11: S0-2's time of closest approach T0 [yr], Col 12: S0-2's eccentricity [-], Col 13: S0-2's orbital inclination [deg], Col 14: S0-2's orbital argument of the periastron [deg], Col 15: S0-2's orbital longitude of the ascending node  [deg], Col 16: NIRC2 radial velocity offset [\kms], Col 17: astrometric correlation length [mas], Col 18: astrometric correlation mixing parameter p [-].
    \item Data S4: \textbf{S0-2 Astrometric Measurements used in Jackknife Analysis}. This table contains the astrometric position of S0-2 in the seven reference frames from maser jackknife analysis. Col. 1: Date observation, Col. 2: Modified Julian date, Col.3 \& 4: Offset from the origin of reference frame in right ascension and declination, Col. 5 \& 6: Uncertainty in the offset.
\end{itemize}

\end{document}